\documentclass[epjST]{svjour}
\usepackage{graphics}
\usepackage{epsfig}

\begin{document}
\title{Electromagnetic Excitation of Nucleon Resonances}
\author{L. Tiator\inst{1}\fnmsep\thanks{\email{tiator@kph.uni-mainz.de}},
D. Drechsel\inst{1}\fnmsep\thanks{\email{drechsel@kph.uni-mainz.de}},
S.S. Kamalov\inst{2}\fnmsep\thanks{\email{kamalov@theor.jinr.ru}} and
M. Vanderhaeghen\inst{1}\fnmsep\thanks{\email{marcvdh@kph.uni-mainz.de}} }
\institute{Institut f\"ur Kernphysik, Johannes Gutenberg-Universit\"at, D-55099
Mainz, Germany \and Laboratory of Theoretical Physics, JINR Dubna, 141980
Moscow Region, Russia}
\abstract{Recent progress on the extraction of electromagnetic properties of
nucleon resonance excitation through pion photo- and electroproduction is
reviewed. Cross section data measured at MAMI, ELSA, and CEBAF are analyzed and
compared to the analysis of other groups. On this basis, we derive longitudinal
and transverse transition form factors for most of the four-star nucleon
resonances. Furthermore, we discuss how the transition form factors can be used
to obtain empirical transverse charge densities. Contour plots of the thus
derived densities are shown for the Delta, Roper, $S_{11}$, and $D_{13}$
nucleon resonances.}
\maketitle

\section{Introduction}
\label{section:1}

\begin{figure}
\begin{center}
\includegraphics[height=10cm]{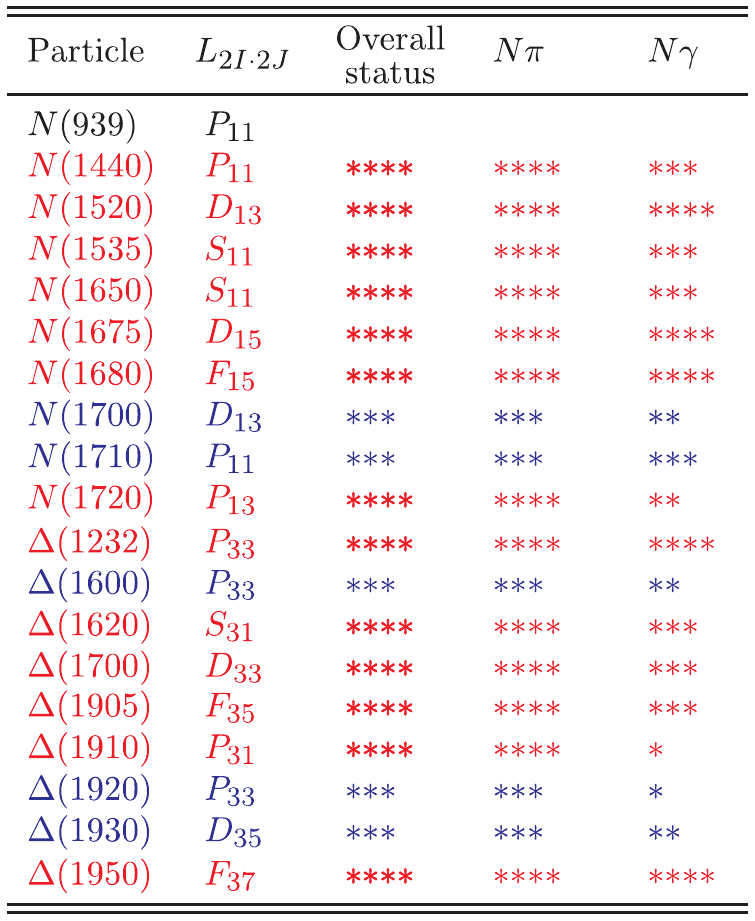}
\vspace{3mm} \caption{$N$ and $\Delta$ resonances with overall status of three
and four stars below 2 GeV. Taken in part from {\em Review of Particle
Physics}~\cite{PDG10} } \label{fig:nstars}
\end{center}
\end{figure}

\begin{figure}
\begin{center}
\includegraphics[width=12cm]{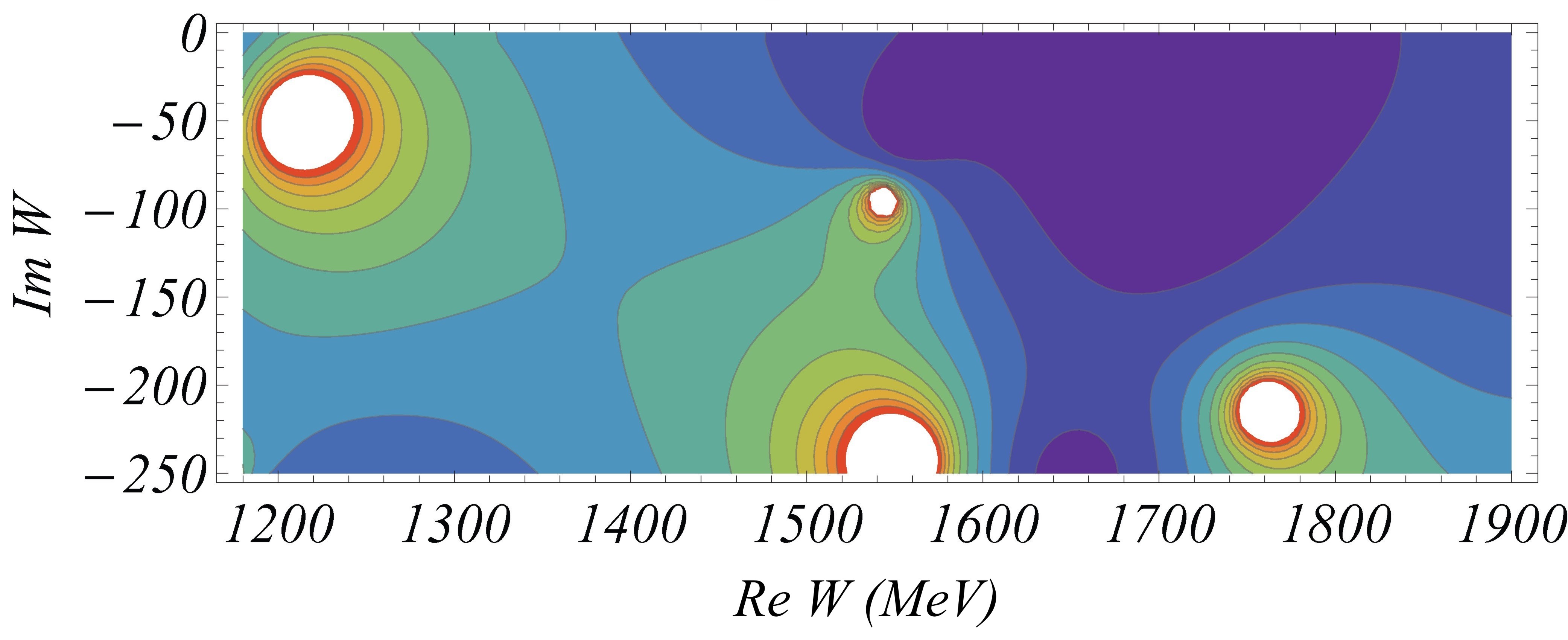}
\vspace{3mm} \caption{Contour plot of $|T_{\pi N}(W)|$ for the $P_{33}$ partial
wave in the Dubna-Mainz-Taipei model~\cite{Chen:2007cy,Tiator:2010rp}. The
light and dark regions show poles and zeroes, respectively, in the first
unphysical sheet of $W$. Three of the four poles seen in this plot belong to
the $P_{33}(1232)$, $P_{33}(1600)$, and $P_{33}(1920)$ resonances listed in
PDG. The fourth pole at $(1554 - 243i)$~MeV has no noticeable influence for
$\pi N$ in the physical region. } \label{fig:p33poles}
\end{center}
\end{figure}

\begin{figure}
\begin{center}
\includegraphics[width=12cm]{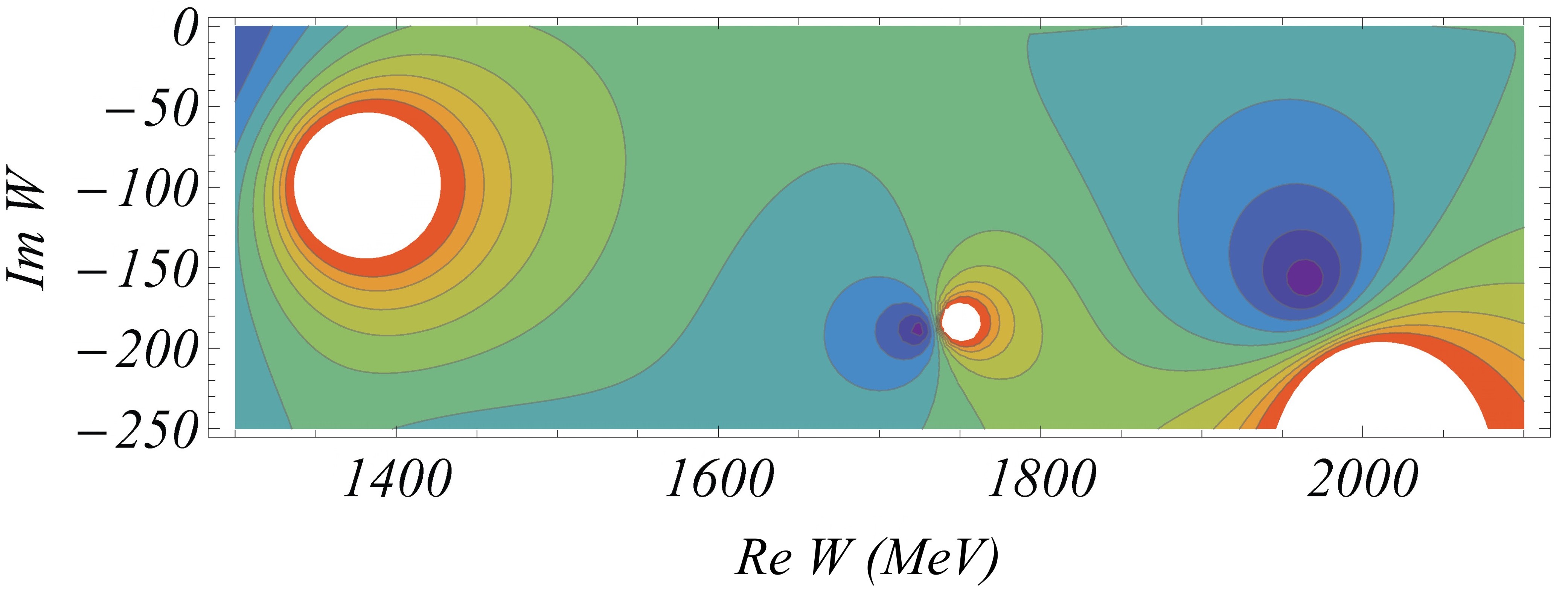}
\vspace{3mm} \caption{Same as in Fig.~\ref{fig:p33poles} for the $P_{11}$
partial wave. The three poles seen in this plot belong to the $P_{11}(1440)$,
$P_{11}(1710)$, and $P_{11}(2100)$ resonances listed in PDG. }
\label{fig:p11poles}
\end{center}
\end{figure}

Our knowledge about the excitation spectrum of the nucleon was originally
provided by elastic pion-nucleon scattering. All the resonances listed in the
Particle Data Tables (PDG)~\cite{PDG10} were identified by partial-wave
analyses of this process with both Breit-Wigner and pole extraction techniques.
In PDG the nucleon resonances $N^*$ and $\Delta$ are given a status from
one-star for a `poor evidence of existing' up to four-star for a `certain
existence', where `properties are at least fairly well explored'. In
Fig.~\ref{fig:nstars} we show part of the PDG summary table including only
three- and four-star resonances up to a mass of 2~GeV. But out of 13 four-star
resonances, only 5 are listed with also four stars in the $N\gamma$ channel.
For others the e.m. properties are at best only approximately known. From
partial wave analyses we know the resonance masses, widths, pole positions, and
branching ratios into the $\pi N$ and $\pi\pi N$ channels. These are reliable
parameters for the four-star resonances, with only few exceptions. In
particular, there remains some doubt about the structure of two prominent
resonances, the Roper $P_{11}(1440)$, which appears unusually broad, and the
$S_{11}(1535)$ whose pole position can not be uniquely determined because its
closeness to the $\eta N$ threshold. In
Figs.~\ref{fig:p33poles}-\ref{fig:p11poles} we have mapped the $P_{33}$ and
$P_{11}$ partial waves in the complex energy plane, as derived from an analysis
within the Dubna-Mainz-Taipei dynamical meson-exchange
model~\cite{Chen:2007cy,Tiator:2010rp}. In both cases, the plots show 3-4 poles
for imaginary values of $W$ from zero down to about $-250$~MeV. The residues of
the poles are proportional to the size of the white disks in the figures.
However, one should keep in mind that the influence of a pole on the physical
region decreases rapidly with the distance from the real axis.

On the basis of these relatively firm grounds, additional information can be
obtained for the electromagnetic (e.m.) $\gamma N N^*$ couplings through meson
photo- and electroproduction. By far, the main source for resonance structure
is pion production, but in some cases with a small $\pi N$ branching ratio,
also $\eta$, $\pi\pi$, $K$, $\rho$, $\omega$, etc. production can give valuable
information. A typical such example is the $S_{11}(1535)$ resonance, which is
located close to the $\eta N$ threshold and strongly coupled to this channel.

Here we will concentrate on the analysis of pion photo- and electroproduction.
The $\gamma N$ couplings can be given in terms of electric, magnetic, and
Coulomb transition moments or, alternatively, as helicity amplitudes. The PDG
lists only the helicity amplitudes $A_{1/2}$ and $A_{3/2}$ for real photons.
These amplitudes are real numbers, which are determined at the Breit-Wigner
position of a given resonance. In electroproduction, an additional longitudinal
amplitude $S_{1/2}(Q^2)$ can be determined and all three amplitudes become
functions of $Q^2$. In particular for the $N\to\Delta$ transition, a set of 3
Sachs form factors ($G_E^*$, $G_M^*$, $G_C^*$) is often introduced and, most
recently, also a set of covariant Dirac form factors ($F_1^{N N^*}, F_2^{N
N^*}$, $F_3^{N N^*}$) has been used, especially for comparison with lattice QCD
and as a source to derive transverse transition densities in the light-front
frame. A general derivation of $N N^*$ transition form factors for arbitrary
spin and parity states can be found in the paper by Devenish, Eisenschitz, and
K\"orner~\cite{Devenish:1975jd}.

Until about 10 years ago, information on the transition form factors (FF) at
finite $Q^2$ was very scarce and, in particular, nonexistent for the
longitudinal form factor. The best knowledge we had was on the magnetic
$N\to\Delta(1232)$ form factor $G_M^*(Q^2)$, which was fairly well known up to
about $Q^2=10~{\rm {GeV}}^2$. This is the only $N N^*$ transition form factor
which can be directly measured in inclusive electron
scattering~\cite{Ash67,Bartel,Baetzner,Alder:1972di,Stein,Jon73,Stoler:1993yk,Stuart:1996zs}
because of the very strong $M1$ transition to the isolated $\Delta(1232)$
resonance.

The new era started with $\pi^0$ electroproduction experiments at
Mainz~\cite{Pos01,Elsner06,Stave06}, Bonn~\cite{Got00,Ban03}, and
Bates~\cite{Mer01,Sparveris:2004jn} in search for the small $E2$ and $C2$
transition form factors for $N\to\Delta$ and the determination of the $E/M$ and
$S/M$ ratios.

\begin{figure}
\begin{center}
\includegraphics[width=12cm]{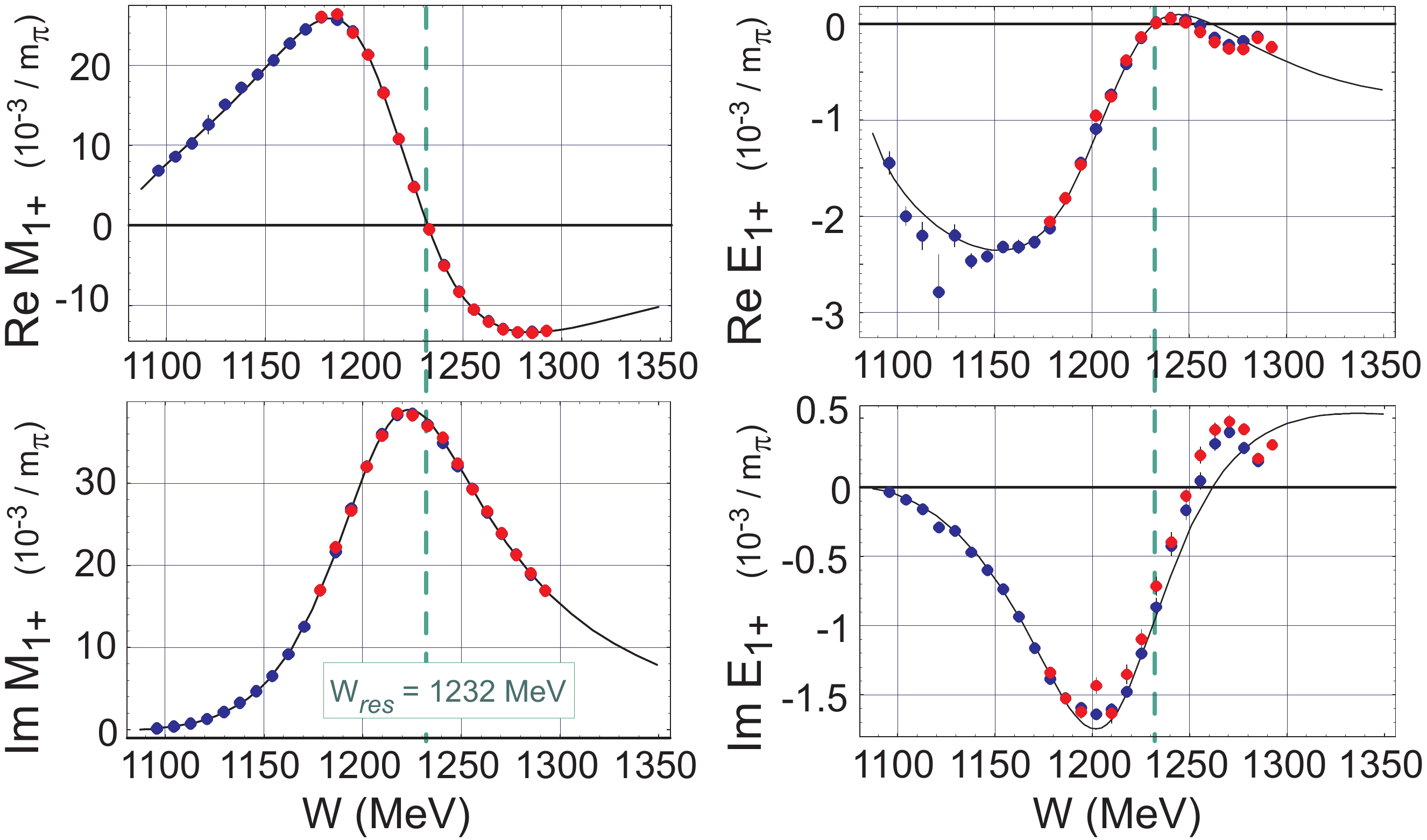} \vspace{3mm} \caption{Pion
photoproduction multipoles $M_{1+}$ and $E_{1+}$ in the $P_{33}$ channel. The
red and blue data points are mostly overlapping and are obtained from the
analyses of Beck et al.~\cite{Beck97,Beck00} and Hanstein et
al.~\cite{HDT97,HDT98}, respectively. The solid lines show the energy-dependent
dispersion theoretical analysis of Hanstein et al.~\cite{HDT98}.}
\label{fig:p33multipoles}
\end{center}
\end{figure}

Already in the mid-nineties, the resonant multipoles (FFs at $Q^2=0$) and in
particular the $E/M$ ratio were investigated in experimental programs at MAMI
in Mainz~\cite{Beck97} and LEGS in Brookhaven~\cite{Blanpied:1997zz}. In
Fig.~\ref{fig:p33multipoles} we display the $P_{33}$ amplitudes as derived from
unpolarized differential cross sections and photon beam asymmetries measured
with MAMI B for both $p(\gamma,\pi^0)p$ and $p(\gamma,\pi^+)n$. The shown
results are based on a model-independent partial wave analysis as performed by
the Mainz group during the years 1995-2000. With the well justified assumptions
of the Watson theorem (due to unitarity) and the neglect of non-Born D and
higher wave contributions, the measurement of only 2 observables for each
channel was sufficient to determine the partial waves without any model input.
From the $E_{1+}^{3/2}$ and $M_{1+}^{3/2}$ multipoles, the following
$\Delta(1232)$ resonance properties were
obtained~\cite{Beck97,Beck00,HDT97,HDT98}:
\begin{itemize}
\item e.m. transition moments and $R_{EM}$ ratio at $W_{\rm{res}}=1232$~MeV
\begin{eqnarray}
\mu_{N\Delta}  &=& \quad(3.46 \pm 0.03)\, \mu_N\,,\label{eq:muNDelta}\\
Q_{N\Delta}    &=& -(0.0846 \pm 0.0033)\, \mbox{fm}^2\,,\\
R_{EM}         &=& -(2.5 \pm 0.1_{stat.} \pm 0.2_{syst.})\,\%
\end{eqnarray}
\item pole position and residues at $W_{\rm{pole}}=M_p - i/2\,\Gamma_p$
\begin{eqnarray}
M_p        &=&  (1212 \pm 1)\,\mbox{MeV}\,,\\
\Gamma_p   &=&  (99 \pm 2)\,\mbox{MeV}\,,\\
r(M1)      &=&   21.16 \cdot e^{-i\,27.5^\circ}\, 10^{-3}/m_\pi\,,\\
r(E2)      &=&    1.23 \cdot e^{-i\,154.7^\circ}\, 10^{-3}/m_\pi\,,\\
R_\Delta=\frac{r(E2)}{r(M1)} &=& -0.035-0.046\,i\,.
\end{eqnarray}
\end{itemize}

During the last decade, a big step forward was made at
JLab~\cite{Fro99,Lav04,Joo02,Joo03,Joo04,Kelly05,Egi06,Ung06,Park:2007tn,Villano:2009sn}
by a series of experiments measuring electroproduction of $\pi^0$ and $\pi^+$
on the proton. Most of these experiments did not use polarization degrees of
freedom, except for the longitudinal and transverse polarizations of the
virtual photon in electroproduction, which are accessible in experiments with
large (azimuthal) angular coverage. However, especially in the $\Delta(1232)$
region, some experiments were performed also with polarized electrons and
polarized targets. In Hall A, Kelly et al.~\cite{Kelly05} even performed an
almost complete experiment, which yielded 16 unpolarized and recoil
polarization observables at $Q^2=1.0$~GeV$^2$.

In parallel with the ongoing experiments, several theoretical groups developed
models and analysis techniques, which were applied to the data. The
model-inde\-pen\-dent GWU/SAID analysis~\cite{Arndt:1990ej,Arndt:1995ak} mostly
analyzed the pion photoproduction data and improved the values of the photon
couplings over the years. In addition, this group analyzed also the $N\Delta$
transition form factors~\cite{Arndt:2006ym}. However, the model-independent
ansatz used in the SAID analysis could not determine the $Q^2$ dependence of
the transition form factors in a reliable way because of rather limited
experimental input from polarization observables. Further coupled channels
analyses were performed by the Giessen
group~\cite{Penner:2002md,Shklyar:2006xw} and by the Bonn-Gatchina
group~\cite{Anisovich:2005tf,Anisovich:2009zy}. Transition amplitudes were also
determined in the framework of dynamically generated resonances by coupling to
meson-baryon channels, e.g., for the $S_{11}(1535)$ resonance by the J\"ulich
group~\cite{Doring:2009yv,Doring:2009uc,Huang:2010vb}. In addition, the nature
of $N^*$ resonances and their physical meaning was investigated within chiral
effective field theory~\cite{Djukanovic:2007bw,Capstick:2007tv,Gegelia:2009py},
and in particular the $\Delta(1232)$ magnetic form factor and the $E/M$ ratio
was evaluated by different
approaches~\cite{Hemmert:1997ye,Gail:2005gz,Pascalutsa:2005vq,Pascalutsa:2006up}.
Furthermore, dynamical models as the Dubna-Mainz-Taipei (DMT)
model~\cite{DMT:2001,KY99} and the Sato-Lee model~\cite{SL01} as well as its
extensions by the EBAC group~\cite{JuliaDiaz:2009ww,Suzuki:2010yn} were used to
analyze photo- and electroproduction in the framework of bare and dressed
nucleon resonances. However, most successful concerning the general
applicability to the higher resonances, were the unitary isobar models of the
Mainz group (MAID
model)~\cite{Maid98,MAID07,Tiator:2003uu,Tiator:2003xr,Tiator:2009mt} and of
the JLab group~\cite{Aznauryan:2008pe,Azn09,Aznauryan:2011ub} who used
dispersion relations as an additional constraint to reduce the model dependence
due to incomplete experimental input.

With our unitary isobar model MAID, we analyzed all available electroproduction
data in order to determine the transition form factors for all four-star
resonances below $W=1.8$~GeV. In most cases we could obtain both single-$Q^2$
and $Q^2$-dependent transition form factors for the proton target. In the case
of the neutron, the parametrization of the $Q^2$ dependence had to take a
simpler form because of the much smaller world database. Already in
2003~\cite{Tiator:2003uu} we obtained transverse transition form factors for
the $\Delta(1232),\,P_{11}(1440),\,S_{11}(1535),\,D_{13}(1520)$, and
$F_{15}(1680)$ as well as longitudinal form factors for the
$\Delta(1232),\,P_{11}(1440)$, and $S_{11}(1535)$ by using unpolarized $\pi^0$
electroproduction data in the range of $Q^2=0.4-1.5~GeV^2$~\cite{Joo02,Lav04}.

The main motivation for exploring the nucleon transition form factors is to
obtain a precise knowledge of the nucleon excitation spectrum, which provides
--- together with the elastic form factors --- a complete description of the
nucleon's electromagnetic structure. This structure can be compared with QCD
inspired quark models and, in recent years, more and more also with lattice QCD
calculations~\cite{Alexandrou:2010uk,Alexandrou:2009hs,Lin:2008qv,Braun:2009jy}.
Moreover, the nucleon transition form factors provide an essential input for
dispersive calculations of both sum rules and two-photon corrections to
electron
scattering~\cite{Drechsel:2004ki,Pasquini:2004nq,Pasquini:2006yi,Pasquini:2007fw,Pasquini:2011ek}.

Finally, the precise e.m. FF data yield the information to map out the quark
charge densities in a baryon. A proper interpretation of such densities was
found by looking at the baryon in the light-front frame. This procedure yields
the spatial distribution of the quark charges in the plane transverse to the
line-of-sight. Along these lines, the transverse charge densities of the quarks
were mapped out for both the nucleon~\cite{Miller:2007uy,Carlson:2007xd} and
the deuteron~\cite{Carlson:2008zc} on the basis of empirical FF data. If
precise transition FF data are available, the same technique can also be
applied to map out the transition densities for nucleon resonance excitation.
The resulting density plots reveal the spatial distribution of the quark
charges inducing the excitation of a particular resonance and provide an
immediate view at its multipole structure. Using the empirical information on
the $N \to N^*$ transition form factors from the MAID analysis~\cite{MAID07},
the transition charge densities were mapped out for the transitions $N\to
\Delta(1232)$~\cite{Carlson:2007xd} and $N \to P_{11}(1440)$~\cite{Tiator2009}.
Most recently we have extended this method to the quark transition charge
densities inducing the e.m. excitations of the $S_{11}(1535)$ and
$D_{13}(1520)$ resonances~\cite{Tiator:2009mt}.

\section{Electromagnetic couplings and transition form factors}

The transverse photon couplings  $A_{1/2}$ and $A_{3/2}$ listed by the PDG are
related to the helicity amplitudes of pion photoproduction, $A_{\ell\pm}$ and
$B_{\ell\pm}$, as follows~\cite{PDG92}:
\begin{eqnarray}
A_{\ell\pm} &=& \mp \alpha C_{\pi N}A_{1/2}\,,\\
B_{\ell\pm} &=& \mp \frac{4\alpha }{\sqrt{(2J-1)(2J+3)}} C_{\pi N}A_{3/2}\,,
\end{eqnarray}
where
\begin{equation}
\alpha\equiv
\sqrt{\frac{1}{\pi}\frac{k}{q}\frac{1}{(2J+1)}\frac{M_N}{M_R}\frac{\Gamma_\pi}{\Gamma^2}}\,.
\end{equation}
Here $k$ and $q$ are the photon and pion c.m. momenta at $W=M_R$, $J$ is the
angular momentum, $M_R$ the mass, $\Gamma$ the full width, and $\Gamma_\pi$ the
$N\pi$ partial width of the resonance. Furthermore, $M_N$ is the nucleon mass
and $C_{\pi N}$ is the Clebsch Gordan coefficient for the decay of the
resonance into the relevant $N\pi$ charge state. This coefficient is
$\sqrt{3/2}$ for isospin $I=3/2$ and $-1/\sqrt{3}$ for $I=1/2$. The helicity
amplitudes are related to the usual electric and magnetic multipoles by
\begin{eqnarray}
A_{\ell+} &=& \frac{1}{2}[(\ell+2)E_{\ell+} +\ell M_{\ell+}]\,, \\
B_{\ell+} &=& E_{\ell+} - M_{\ell+}\,, \\
A_{\ell+1,-} &=& \frac{1}{2}[(\ell+2)M_{\ell+1,-} -\ell E_{\ell+1,-}]\,, \\
B_{\ell+1,-} &=& E_{\ell+1,-} + M_{\ell+1,-}\,.
\end{eqnarray}

The $N\gamma$ partial width $\Gamma_\gamma$ is then given by
\begin{equation}
\Gamma_\gamma=\frac{k^2}{\pi}\frac{2M_N}{(2J+1)M_R}[|A_{1/2}|^2+|A_{3/2}|^2]\,.
\end{equation}

However, this formulation as explicitly given, e.g., in PDG(1992)~\cite{PDG92},
needs some additional interpretation. In order to obtain a pure resonance
coupling, the pion photoproduction amplitudes must be separated in resonance
and background parts. This separation is in general a model-dependent issue.
Mathematically, the cleanest way is to separate the amplitudes at the pole
position and to identify the resonance part by the residues in each helicity
channel. The first steps to study the e.m. $\Delta(1232)$ excitation at the
pole position were made by Hanstein et al.~\cite{HDT98} and Workman et
al.~\cite{Workman:1998tv}. Very recently also the
Bonn-Gatchina~\cite{Anisovich:2009zy} and the EBAC~\cite{Suzuki:2010yn} groups
have started to derive the photon couplings for higher resonances at the
$t-$matrix pole. Of course, the pole values arising from complex residues will
eventually lead to complex photon couplings and complex form factors.

Here and in most publications as well as in all the listings of PDG, the
resonance-background separation has been performed in a Breit-Wigner formalism.
A typical ansatz can be found in an early publication of the SAID
group~\cite{Arndt:1990ej}.

Starting from the photon couplings, the transverse transition form factors in
the helicity basis can be straightforwardly introduced by defining the
functions $A_{1/2}(Q^2)$ and $A_{3/2}(Q^2)$. The FFs of the transverse e.m.
multipoles follow from the above equations in an analogous way. In addition, a
third form factor $S_{1/2}(Q^2)$ appears due the longitudinal photon field in
the same partial wave. The details and conventions used will be given in the
next section where we describe our MAID ansatz for obtaining the form factors.

\section{The MAID ansatz}

In the spirit of a dynamical approach to pion photo- and electroproduction, the
$t$-matrix of the unitary isobar model is set up by the ansatz
\begin{equation}
t_{\gamma\pi}(W)=t_{\gamma\pi}^B(W) + t_{\gamma\pi}^{R}(W)\,,\label{eq:DM}
\end{equation}
with a background and a resonance $t$-matrix, each of them constructed in a
unitary way. Of course, this ansatz is not unique. However, it is a very
important prerequisite to clearly separate resonance and background amplitudes
within a Breit-Wigner concept also for higher and overlapping resonances.

For a specific partial wave $\alpha = \{j,l,\ldots\}$, the background
$t$-matrix is set up by a potential multiplied by the pion-nucleon scattering
amplitude in accordance with the K-matrix approximation,
\begin{equation}
 t^{B,\alpha}_{\gamma\pi}(W,Q^2)=v^{B,\alpha}_{\gamma\pi}(W,Q^2)\,[1+it_{\pi N}^{\alpha}(W) ]\, ,
\label{eq:Kmatrix}
\end{equation}
where only the on-shell part of pion-nucleon rescattering is maintained and the
off-shell part from pion-loop contributions is neglected. Whereas this
approximation would fail near the threshold for
$\gamma,\pi^0$~\cite{Bernard:1991rt,Kamalov:2001qg}, it is well justified in
the resonance region because the main contribution from pion-loop effects is
absorbed by the nucleon resonance dressing.

The background  potential $v_{\gamma\pi}^{B,\alpha}(W,Q^2)$ is described by
Born terms obtained with an energy-dependent mixing of
pseudovector-pseudoscalar $\pi NN$ coupling and $t$-chan\-nel vector meson
exchanges. The mixing para\-meters and coupling constants are determined by an
analysis of nonresonant multipoles in the appropriate energy
regions~\cite{Maid98}. In the latest version MAID2007~\cite{MAID07}, the $S$,
$P$, $D$, and $F$ waves of the background contributions are unitarized as
explained above, with the pion-nucleon elastic scattering amplitudes,
$t^{\alpha}_{\pi N}=[\eta_{\alpha} \exp(2i\delta_{\alpha})-1]/2i$, described by
phase shifts $\delta_{\alpha}$ and the inelasticity parameters $\eta_{\alpha}$
taken from the GWU/SAID analysis~\cite{Arndt:1995ak}.

For the resonance contributions we follow Ref.~\cite{Maid98} and assume
Breit-Wigner forms for the resonance shape,
\begin{eqnarray}
t_{\gamma\pi}^{R,\alpha}(W,Q^2) = {\bar{\cal A}}_{\alpha}^R(W,Q^2)\,
\frac{f_{\gamma N}(W)\Gamma_{tot}(W)\,M_R\,f_{\pi N}(W)}{M_R^2-W^2-iM_R\,
\Gamma_{tot}(W)} \,e^{i\phi_R(W,Q^2)}\,, \label{eq:BW}
\end{eqnarray}
where $f_{\pi N}(W)$ is the usual Breit-Wigner factor describing the decay of a
resonance with total width $\Gamma_{tot}(W)$, partial $\pi N$ width
$\Gamma_{\pi N}(W)$, and spin $j$,
\begin{equation}
f_{\pi N}(W)=C_{\pi N}\left[\frac{1}{(2j+1)\pi}\frac{\kappa(W)}{q(W)}
\frac{M_N}{M_R}\frac{\Gamma_{\pi N}(W)}{\Gamma_{\rm
{tot}}^2(W)}\right]^{1/2}\,, \label{eq:fpin}
\end{equation}
with
\begin{eqnarray}
k(W,Q^2)&=&\frac{\sqrt{(Q^2+(W+M_N)^2)(Q^2+(W-M_N)^2)}}{2W}\,,\\
q(W)&=&\frac{\sqrt{(W^2-(M_N+m_\pi)^2)(W^2-(M_N-m_\pi)^2)}}{2W}\,,\\
\kappa(W)&=&\frac{W^2-M_N^2}{2W}\,,
\end{eqnarray}
and $C_{\pi N}=\sqrt{3/2}$ and $-1/\sqrt{3}$ for isospin $\frac {3}{2}$ and
$\frac {1}{2}$, respectively. The energy dependence of the partial width is
given by
\begin{equation}
\Gamma_{\pi N}(W)=\beta_{\pi}\,\Gamma_R\,\left(\frac{q(W)}{q_R}
\right)^{2l+1}\,\left(\frac{X^2_R+q_R^2}{X^2_R+q^2(W)}\right)^{\ell} \,
\frac{M_R}{W}\,, \label{eq:Gampin}
\end{equation}
with $q_R=q(M_R)$, $\Gamma_R=\Gamma_{\rm{tot}}(M_R)$, $X_R$ a damping parameter
(cut-off), and $\beta_{\pi}$ the single-pion branching ratio. The expression
for the total width $\Gamma_{\rm{tot}}$ is given in Ref.~\cite{Maid98}. The
$\gamma NN^*$ vertex is assumed to have the following dependence on $W$:
\begin{equation}
f_{\gamma N}(W)=\left(\frac{\kappa(W)}{\kappa_R}\right)^n\,
\left(\frac{X^2_R+\kappa_R^2}{X^2_R+\kappa^2(W)}\right)\,, \label{eq:fgamman}
\end{equation}
where $\kappa_R=\kappa(M_R)$ and $n$ is obtained from a best fit to the real
photon data. The phase $\phi_R(W,Q^2)$ in Eq.~(\ref{eq:BW}) is introduced to
adjust the total phase such that the Fermi-Watson theorem is fulfilled below
two-pion threshold. For the $S$- and $P$-wave multipoles we extend this
unitarization procedure up to $W=1400$ MeV. Because of a lack of further
information, we assume that the phases $\phi_R$ are constant at the higher
energies. In particular we note that the phase $\phi_R$ for the $P_{33}(1232)$
excitation vanishes at $W=M_R=1232$~MeV for all values of $Q^2$. For this
multipole we may even apply the Fermi-Watson theorem up to $W \approx 1600$~MeV
because the inelasticity parameter $\eta_{\alpha}$ remains close to 1. For the
$D$- and $F$-wave resonances, the phases
$\phi_R$ are set to be constant and determined from the best fit.\\

\begin{table*}[ht]
\caption{The reduced e.m. amplitudes ${\bar{\cal A}}_{\alpha}$
defined by Eq.~(\ref{eq:BW}) in terms of the helicity
amplitudes.}\label{tab:amplitudes}
\begin{center}
\begin{tabular}{c|ccc|ccc|c}
\hline
$N^{\ast}$        &      & $\bar{E}$  & & & $\bar{M}$  &  & $\bar{S}$  \\
\hline
$S_{11}$/$S_{31}$ &  & $-A_{1/2}$  &    &   & ---  &  & $-\sqrt{2}S_{1/2}$\\
$P_{13}$/$P_{33}$ &  & $\frac{1}{2}(\frac{1}{\sqrt{3}}A_{3/2}-A_{1/2})$ & & &
$-\frac{1}{2}(\sqrt{3}A_{3/2}+A_{1/2})$ & &  $-\frac{1}{\sqrt{2}}S_{1/2}$\\
$P_{11}$/$P_{31}$ &  & ---  &    &   & $A_{1/2}$  & &$-\sqrt{2}S_{1/2}$ \\
$D_{13}$/$D_{33}$ &  & $-\frac{1}{2}(\sqrt{3}A_{3/2}+A_{1/2})$ & & &
$-\frac{1}{2}(\frac{1}{\sqrt{3}}A_{3/2}-A_{1/2})$ & & $-\frac{1}{\sqrt{2}}S_{1/2}$\\
$D_{15}$/$D_{35}$&  & $\frac{1}{3}(\frac{1}{\sqrt{2}}A_{3/2}-A_{1/2})$ & & &
$-\frac{1}{3}(\sqrt{2}A_{3/2}+A_{1/2})$ & & $-\frac{\sqrt{2}}{3}S_{1/2}$\\
$F_{15}$/$F_{35}$ &  & $-\frac{1}{3}(\sqrt{2}A_{3/2}+A_{1/2})$ & & &
$-\frac{1}{3}(\frac{1}{\sqrt{2}}A_{3/2}-A_{1/2})$ & & $-\frac{\sqrt{2}}{3}S_{1/2}$\\
$F_{17}$/$F_{37}$ &  & $\frac{1}{4}(\sqrt{\frac{3}{5}}A_{3/2}-A_{1/2})$ & & &
$-\frac{1}{4}(\sqrt{\frac{5}{3}}A_{3/2}+A_{1/2})$ & & $-\frac{1}{2\,\sqrt{2}}S_{1/2}$\\
\hline
\end{tabular}
\end{center}
\end{table*}

The more commonly used helicity amplitudes $A_{1/2}$, $A_{3/2}$, and $S_{1/2}$
are given by linear combinations of the e.m. couplings $\bar{\cal
A}_{\alpha}^R$. These relations take the form
\begin{eqnarray}
A^{\ell +}_{1/2} &=& -\frac{1}{2} [(\ell +2) \bar{E}_{\ell+} + \ell \bar{M}_{\ell+} ]
\,, \\
A^{\ell +}_{3/2} &=& \frac{1}{2}\sqrt{\ell(\ell+2)} (\bar{E}_{\ell+} - \bar{M}_{\ell+})
\,, \label{eq:Alplus}\\
S^{\ell +}_{1/2} &=& -\frac{\ell+1}{\sqrt{2}} \bar{S}_{\ell+}
\end{eqnarray}
for resonances with total spin $j=\ell +\frac {1}{2}$, and
\begin{eqnarray}
A^{\ell -}_{1/2} &=& \frac{1}{2} [(\ell +1) \bar{M}_{\ell -} - (\ell -1) \bar{E}_{\ell -}]\,, \\
A^{\ell -}_{3/2} &=& -\frac{1}{2}\sqrt{(\ell -1)(\ell+1)}(\bar{E}_{\ell -} + \bar{M}_{\ell -})\,, \label{eq:Alminus}\\
S^{\ell -}_{1/2} &=& -\frac{\ell}{\sqrt{2}} \bar{S}_{\ell -}
\end{eqnarray}
for resonances with total spin $j=\ell - \frac {1}{2}$. The inverse relations
for the partial waves are listed in Table~\ref{tab:amplitudes}. The helicity
amplitudes are related to matrix elements of the e.m. current $J_{\mu}$ between
the nucleon and the resonance states, e.g., as obtained in the framework of
quark models,
\begin{eqnarray}
A_{1/2} &=& -\sqrt{\frac{2\pi\alpha_{\rm {em}}}{\kappa_R}}
<N^*,\frac{1}{2}\,|\,J_{+}\,|\,N,-\frac{1}{2}>\, \zeta\,, \\
A_{3/2} &=& -\sqrt{\frac{2\pi\alpha_{\rm {em}}}{\kappa_R}}
<N^*,\frac{3}{2}\,|\,J_{+}\,|\,N,\frac{1}{2}>\,\zeta\,, \label{eq:quark}\\
S_{1/2} &=& +\sqrt{\frac{2\pi\alpha_{\rm {em}}}{\kappa_R}}
<N^*,\frac{1}{2}\,|\,\rho\,|\,N,\frac{1}{2}>\, \zeta \,,
\end{eqnarray}
where $J_{+}=-\frac{1}{\sqrt{2}}(J_x+iJ_y)\,$ and $\alpha_{\rm {em}}=1/137$.
However, these equations define the couplings only up to a phase $\zeta$, which
in principle can be obtained from the pionic decay of the resonance calculated
within the same model. Because this phase is often ignored in the literature,
the comparison of the sign is not always meaningful, especially in critical
cases such as the Roper resonance for which the correct sign is not obvious
from the data. MAID uses the helicity amplitudes $A_{1/2}$, $A_{3/2}$, and
$S_{1/2}$ for photoproduction as input parameters, except for the
$P_{33}(1232)$ resonance which is directly described by the three e.m.
amplitudes $\bar{\cal A}_{\alpha}$.

While the original version of MAID included only the 7 most important nucleon
resonances with only transverse e.m. couplings in most cases, MAID2007
describes all 13 four-star resonances below $W=2$~GeV:
$P_{33}(1232)$,$P_{11}(1440)$,$D_{13}(1520)$,$S_{11}(1535)$,
$S_{31}(1620)$,$S_{11}(1650)$,$D_{15}(1675)$,$F_{15}(1680)$,$D_{33}(1700)$,
$P_{13}(1720)$,$F_{35}(1905)$, $P_{31}(1910)$, and $F_{37}(1950)$.

\begin{table}[htp]
\caption{Masses, widths, single-pion branching ratios, and angles $\phi_R$
included in the MAID analysis. Masses and widths are given in MeV, angles in
degrees, the branching ratios $\beta_\pi$ and $\beta_\gamma$ in \%. The quoted
PDG ranges are from Ref.~\cite{PDG10}.}\label{tab:par_res}
\begin{center}
\begin{tabular}{c|c c c c |cccc}
\hline
$N^{\ast},\Delta$  &$M_{R,PDG}$& $\Gamma_{R,PDG}$& $\beta_{\pi,PDG}$& $\beta^p_{\gamma,PDG}$& $M_R$& $\Gamma_R$& $\beta_{\pi}$ & $\phi_R$\\
\hline
$P_{33}(1232)$ & 1231-1233& 116-120&    100  & 0.52-0.60   &  1232 & 130 &100&  0.0 \\
$P_{11}(1440)$ & 1420-1470& 200-450&   55-75 & 0.035-0.048 &  1440 & 350 & 70&-15   \\
$D_{13}(1520)$ & 1515-1525& 100-125&   55-65 & 0.46-0.56   &  1530 & 130 &60 & 32   \\
$S_{11}(1535)$ & 1525-1545& 125-175&   35-55 & 0.15-0.35   &  1535 & 100 &40 & 8.2  \\
$P_{33}(1600)$ & 1550-1700& 250-460&   10-25 & 0.001-0.020 &       &     &   &      \\
$S_{31}(1620)$ & 1600-1660& 135-150&   20-30 & 0.004-0.044 &  1620 & 150 &25 & 23   \\
$S_{11}(1650)$ & 1645-1670& 145-185&   60-95 & 0.04-0.18   &  1690 & 100 &85 & 7.0  \\
$D_{15}(1675)$ & 1670-1680& 130-165&   35-45 & 0.004-0.023 &  1675 & 150 &45 & 20   \\
$F_{15}(1680)$ & 1680-1690& 120-140&   65-70 & 0.21-0.32   &  1680 & 135 &70 & 10   \\
$D_{13}(1700)$ & 1650-1750& 050-150&   05-15 & 0.01-0.05   &       &     &   &      \\
$P_{11}(1710)$ & 1680-1690& 120-140&   65-70 & 0.002-0.050 &       &     &   &      \\
$D_{33}(1700)$ & 1670-1750& 200-400&   10-20 & 0.12-0.26   &  1740 & 450 &15 & 61   \\
$P_{13}(1720)$ & 1700-1750& 150-300&   10-20 & 0.003-0.100 &  1740 & 250 &20 &  0.0 \\
$F_{35}(1905)$ & 1865-1915& 270-400&   09-15 & 0.01-0.03   &  1905 & 350 &10 & 40   \\
$P_{31}(1910)$ & 1870-1920& 190-270&   15-30 &  0.0-0.2    &  1910 & 250 &25 & 35   \\
$P_{33}(1920)$ & 1900-1970& 150-300&   05-20 &             &       &     &   &      \\
$D_{35}(1930)$ & 1900-2020& 220-500&   05-15 &  0.00-0.02  &       &     &   &      \\
$F_{37}(1950)$ & 1915-1950& 235-335&   35-45 & 0.08-0.13   &  1945 & 280 &40 & 30   \\
\hline
\end{tabular}
\end{center}
\end{table}

In Tables~\ref{tab:helicity_p} and \ref{tab:helicity_n} we compare the helicity
amplitudes obtained from MAID2007 with the results of the PDG~\cite{PDG10},
GWU/SAID~\cite{Dugger:2009pn,Arndt:2002xv},
Bonn-Gatchina~\cite{Anisovich:2009zy}, and Giessen
\cite{Penner:2002md,Shklyar:2006xw} analyses. As is very typical for a global
analysis with about 20,000 data points fitted to a small set of 20-30
parameters, the fit errors appear unrealistically small. However, one should
realize that these errors only reflect the statistical uncertainty of the
experimental error, whereas the model uncertainty can be larger by an order of
magnitude. We therefore do not list our fit errors, which in fact are very
similar to the fits of the SAID group~\cite{Arndt:2002xv,Dugger:2009pn}. The
only realistic error estimate is obtained by comparing different analyses, such
as SAID, MAID, and coupled-channels approaches. A comparison between the
analyses of the different groups with different models and methods, listed in
Table~\ref{tab:helicity_p}, shows a rather clear hierarchy among the resonances
that all have an overall four-star ranking (see also Fig.~\ref{fig:nstars}).
The most reliable e.m. couplings are known from $P_{33}$, $D_{13}$ and
$F_{15}$, whereas even from the latter two only the large $A_{3/2}$ amplitudes
show this quality. The $P_{11}$, $S_{11}$ and $F_{37}$ are reasonably well
known but for the $D_{15}$, $D_{33}$ and $F_{35}$ the couplings show already a
larger spread. Badly known and unreliable are the couplings for the second
$S_{11}$, the $S_{31}$, $P_{13}$ and $P_{31}$. Electromagnetic resonance
couplings for states with overall status of less than four-star should rather
be considered as unknown. In most present analyses and models these states are
neglected. However, with the advent of `complete experiments' and the analysis
of double-polarization observables a big improvement in this field is already
in sight.

\begin{table}[htbp]
\caption{Proton helicity amplitudes at $Q^2=0$ for the major nucleon
resonances, in units $10^{-3}$~GeV$^{-1/2}$. The results with
MAID(MD07)~\cite{MAID07} are compared to the PDG~\cite{PDG10},
GWU/SAID(SP09)~\cite{Dugger:2009pn}, Bonn-Gatchina~\cite{Anisovich:2009zy}, and
Gie{\ss}en (GI02/07)~\cite{Penner:2002md,Shklyar:2006xw} analysis.\protect\\
$^1$~In the SP09 analysis, the $\pi N$ branching ratio
for the $S_{11}(1650)$ is 100\%.} \label{tab:helicity_p}
\begin{center}
\begin{tabular}{l l|r@{$\pm$}l r@{$\pm$}l r@{$\pm$}l r r}
\hline
&  &  \multicolumn{2}{c}{PDG} & \multicolumn{2}{c}{SP09} & \multicolumn{2}{c}{BoGa09} &  GI02/07 &  MD07\\
\hline
$P_{33}(1232)$& $A_{1/2}$ & -135&6 & -139.6&1.8 & -136&5 & -128 & -140\\
              & $A_{3/2}$ & -250&8 & -258.9&2.3 & -267&8 & -247 & -265\\
$P_{11}(1440)$& $A_{1/2}$ & -65 &4 & -56.4 &1.7 & -52 &10 & -84 &  -61\\
$D_{13}(1520)$& $A_{1/2}$ & -24 &9 & -26.0 &1.5 & -32 &6 & -15 & -27\\
              & $A_{3/2}$ & 166 &5 & 141.2 &1.7 & 138 &8 & 146  & 161\\
$S_{11}(1535)$& $A_{1/2}$ &  90 &30 &100.9 &3.0 & 90 &15 & 95 &  66 \\
$S_{31}(1620)$& $A_{1/2}$ &  27 &11 & 47.2&2.3 & 63&12 & -50  & 66 \\
$S_{11}(1650)$& $A_{1/2}$ &  53 & 16 &  9.0 &9.1$^1$ & 60 &20 & 57 & 33 \\
$D_{15}(1675)$& $A_{1/2}$ &  19 &8 &  14.9 &2.1 &  21 &4 &  9  & 15\\
              & $A_{3/2}$ &  15 &9 &  18.4 &2.1 &  24 &8 &  21  & 22\\
$F_{15}(1680)$& $A_{1/2}$ & -15 &6 & -17.6 &1.5 & -12 &6 &  3  & -25\\
              & $A_{3/2}$ & 133 &12& 134.2 &1.6 & 136 &12 & 116  & 134\\
$D_{33}(1700)$& $A_{1/2}$ & 104 &15 & 118.3 &3.3 & 160 &45 & 96 & 226\\
              & $A_{3/2}$ & 85  &22& 110.0  &3.5 & 160 &40 & 154   &  210\\
$P_{13}(1720)$& $A_{1/2}$ &  18 &30&   90.5 &3.3   &  130 &50   & -65  &  73\\
              & $A_{3/2}$ & -19 &20&  -36.0 &3.9  &   100 &50  & 35  &  -11\\
$F_{35}(1905)$& $A_{1/2}$ & 26  &11 & 11.4  &8.0 & 28  &12 &   &  18\\
              & $A_{3/2}$ & -45 &20 & -51.0&8.0 & -42 &15 &    & -28\\
$P_{31}(1910)$& $A_{1/2}$ &  3  &14 & \multicolumn{2}{c}{}& \multicolumn{2}{c}{} & &  18\\
$F_{37}(1950)$& $A_{1/2}$ & -76 &12 & -71.5&1.8 & -83 &8 &     & -94\\
              & $A_{3/2}$ & -97 &10 & -94.7 &1.8 & -92 &8 &   &  -121\\
\hline
\end{tabular}
\end{center}
\end{table}

\begin{table}[htbp]
\caption{Neutron helicity amplitudes at $Q^2=0$ for the major nucleon
resonances, in units $10^{-3}$~GeV$^{-1/2}$. The results with
MAID(MD07)~\cite{MAID07} are compared to the PDG~\cite{PDG10}, GWU/SAID(GW02)~\cite{Arndt:2002xv}
and Gie{\ss}en(GI07)~\cite{Shklyar:2006xw} analysis.}\label{tab:helicity_n}
\begin{center}
\begin{tabular}{l l|r@{$\pm$}l r@{$\pm$}l r r}
\hline
&  &  \multicolumn{2}{c}{PDG} & \multicolumn{2}{c}{GW02} &  GI02/07 &  MD07\\
\hline
$P_{11}(1440)$& $A_{1/2}$ & 40&10 & 47&5 & 138 & 54 \\
$D_{13}(1520)$& $A_{1/2}$ & -59 &9 & -67 &4 & -64& -77\\
              & $A_{3/2}$ &-139 &11&-112 &3 & -136& -154\\
$S_{11}(1535)$& $A_{1/2}$ & -46&27 & -16&5 & -74 & -51 \\
$S_{11}(1650)$& $A_{1/2}$ & -15&21 &-28&4 & -9& 9\\
$D_{15}(1675)$& $A_{1/2}$ & -43 &12& -50 &4 & -56& -62\\
              & $A_{3/2}$ & -58 &13& -71 &5 & -84& -84\\
$F_{15}(1680)$& $A_{1/2}$ &  29 &10&  29 &6 & 30& 28\\
              & $A_{3/2}$ & -33 &9 & -58 &9 &-48&-38\\
$P_{13}(1720)$& $A_{1/2}$ &   1 &15&  \multicolumn{2}{c}{} &  3& -3\\
              & $A_{3/2}$ & -29 &61&  \multicolumn{2}{c}{} & -1& -31\\
\hline
\end{tabular}
\end{center}
\end{table}

\section{Transition form factors}

In most cases, the resonance couplings $\bar{\mathcal A}_{\alpha}^R(W,Q^2)$ are
assumed to be independent of the total energy. However, an energy dependence
may occur if the resonance is parameterized in terms of the virtual photon
three-momentum $k(W,Q^2)$, e.g., in MAID2007 for the $\Delta(1232)$ resonance.
For all other resonances discussed here, we may assume a simple $Q^2$
dependence, $\bar{\mathcal A}_{\alpha}(Q^2)$. These resonance couplings are
taken as constants for a single-Q$^2$ analysis, e.g., for photoproduction
($Q^2=0$) but also at any fixed $Q^2>0$, whenever sufficient data with W and
$\theta$ variation are available, see Table~\ref{tab:database}. Alternatively
the couplings can be parameterized as functions of $Q^2$ by an ansatz like
\begin{equation}
\bar{\mathcal A}_{\alpha}(Q^2) =\bar{\mathcal A}_{\alpha}(0) (1+a_1 Q^2+a_2 Q^4
+a_3 Q^6+a_4 Q^8)\, e^{-b_1 Q^2}\,. \label{eq:ffpar}
\end{equation}
For such an ansatz the parameters $\bar{\mathcal A}_{\alpha}(0)$ are determined
by a fit to the world database of photoproduction, and the parameters $a_i$ and
$b_1$ are obtained from a combined fitting of all the electroproduction data at
different $Q^2$. The latter procedure is called the $Q^2$-dependent fit. In
MAID the photon couplings $\bar{\mathcal A}_{\alpha}(0)$ are input parameters,
directly related to the helicity couplings $A_{1/2},\, A_{3/2}$, and $S_{1/2}$
of nucleon resonance excitation. For further details see Ref.~\cite{MAID07}.

In Tables~\ref{tab:param-p1}, \ref{tab:param-p2}, and \ref{tab:param-n} we list
the parameters obtained from our new $Q^2$-dependent fit to the resonances
above the $\Delta(1232)$. Because we have recently included the 2008 $\pi^+$
data of Park et al.~\cite{Park:2007tn} in our database, our new results differ
from the MAID2007 parametrization for the following six proton transition form
factors: $P_{11}(1440)$,\linebreak $D_{13}(1520)$, $D_{33}(1700)$,
$D_{15}(1675)$, $F_{15}(1680)$, and $P_{13}(1720)$. Our parametrization of the
$\Delta(1232)$ form factors is more complicated, in particular due to built-in
requirements from low-energy theorems in the Siegert limit, as discussed in
Ref.~\cite{MAID07} in further details.

\begin{table}[htb]
\begin{center}
\caption{ \label{tab:database}  Database of pion electroproduction for energies
above the $\Delta$ resonance up to $W=1.7$~GeV, which is used in our
single-$Q^2$ transition form factor analysis.}
\begin{tabular*}{80mm}{@{\extracolsep{\fill}}lccc}
\hline
 Reference & year & reaction & $Q^2\;(GeV^2)$ \\
\hline
Frolov et al.\cite{Fro99} & 1999         & $p\pi^0$ & $ 2.8 - 4.0 $ \\
Gothe et al.\cite{Got00} & 2000          & $p\pi^0$ & $ 0.63 $ \\
Pospischil et al.\cite{Pos01} & 2001     & $p\pi^0$ & $ 0.121 $ \\
Mertz et al.\cite{Mer01} & 2001          & $p\pi^0$ & $ 0.126 $ \\
Joo et al.\cite{Joo02} & 2002            & $p\pi^0$ & $ 0.4 - 1.8 $ \\
Joo et al.\cite{Joo04} & 2004            & $n\pi^+$ & $ 0.4 - 0.65 $ \\
Laveissiere et al.\cite{Lav04} & 2004    & $p\pi^0$ & $ 1.0 $ \\
Kelly et al.\cite{Kelly05} & 2005        & $p\pi^0$ & $ 1.0 $ \\
Elsner et al.\cite{Elsner06} & 2006      & $p\pi^0$ & $ 0.20 $ \\
Stave et al.\cite{Stave06} & 2006        & $p\pi^0$ & $ 0.06 $ \\
Egiyan et al.\cite{Egi06} & 2006         & $n\pi^+$ & $ 0.3 - 0.6 $ \\
Ungaro et al.\cite{Ung06} & 2006         & $p\pi^0$ & $ 3.0 - 6.0 $ \\
Park et al.\cite{Park:2007tn} & 2008     & $n\pi^+$ & $ 1.7 - 4.5 $ \\
Villano et al.\cite{Villano:2009sn}& 2009& $p\pi^0$ & $ 6.4 - 7.7 $ \\
\hline
\end{tabular*}
\end{center}
\end{table}

\begin{table}
\begin{center}
\caption{ \label{tab:param-p1} New MAID2008 parametrization of the transition
form factors, Eq.~(\ref{eq:ffpar}), for proton targets. $\bar{\mathcal
A}_{\alpha}(0)$ is given in units of $10^{-3}\,{\rm {GeV}}^{-1/2}$ and the
coefficients $a_1,\,a_2,\,a_4,\,b_1$ in units of ${\rm {GeV}}^{-2},\,{\rm
{GeV}}^{-4},\,{\rm {GeV}}^{-8},\,{\rm {GeV}}^{-2}$, respectively. For all fits
$a_3=0$.}
\begin{tabular*}{120mm}{@{\extracolsep{\fill}}ccccccc}
\hline
 $N^*,\;\Delta^*$ & & $\bar{\mathcal A}_{\alpha}(0)$ &
$a_1$ & $a_2$ & $a_4$ & $b_1$ \\
\hline
$P_{11}(1440)p$ & $A_{1/2}$ & $-61.4$ & $ 0.871$ & $-3.516$ & $-0.158$ & $1.36$\\
                & $S_{1/2}$ & $  4.2$ & $  40. $ & $  0   $ & $ 1.50 $ & $1.75$\\
$D_{13}(1520)p$ & $A_{1/2}$ & $-27.4$ & $ 8.580$ & $-0.252$ & $ 0.357$ & $1.20$\\
                & $A_{3/2}$ & $160.6$ & $-0.820$ & $ 0.541$ & $-0.016$ & $1.06$\\
                & $S_{1/2}$ & $-63.5$ & $ 4.19 $ & $  0   $ & $   0  $ & $3.40$\\
$D_{15}(1675)p$ & $A_{1/2}$ & $ 15.3$ & $ 0.10 $ & $  0   $ & $   0  $ & $2.00$\\
                & $A_{3/2}$ & $ 21.6$ & $ 1.91 $ & $ 0.18 $ & $   0  $ & $0.69$\\
                & $S_{1/2}$ & $  1.1$ & $  0   $ & $  0   $ & $   0  $ & $2.00$\\
$F_{15}(1680)p$ & $A_{1/2}$ & $-25.1$ & $ 3.780$ & $-0.292$ & $ 0.080$ & $1.25$\\
                & $A_{3/2}$ & $134.3$ & $ 1.016$ & $ 0.222$ & $ 0.237$ & $2.41$\\
                & $S_{1/2}$ & $-44.0$ & $ 3.783$ & $  0   $ & $   0  $ & $1.85$\\
$D_{33}(1700) $ & $A_{1/2}$ & $226. $ & $ 1.91 $ & $  0   $ & $   0  $ & $1.77$\\
                & $A_{3/2}$ & $210. $ & $ 0.88 $ & $ 1.71 $ & $   0  $ & $2.02$\\
                & $S_{1/2}$ & $  2.1$ & $   0  $ & $  0   $ & $   0  $ & $2.00$\\
$P_{13}(1720)p$ & $A_{1/2}$ & $ 73.0$ & $ 1.89 $ & $  0   $ & $   0  $ & $1.55$\\
                & $A_{3/2}$ & $-11.5$ & $10.83 $ & $-0.66 $ & $   0  $ & $0.43$\\
                & $S_{1/2}$ & $-53.0$ & $ 2.46 $ & $  0   $ & $   0  $ & $1.55$\\
\hline
\end{tabular*}
\end{center}
\end{table}

Above the third resonance region there is an energy gap between $1800-1900$~MeV
where no four-star resonances have been found. Beyond this gap and up to 2~GeV,
three more four-star resonances are listed in
PDG,$F_{35}(1905)$,$P_{31}(1910)$, and $F_{37}(1950)$, which are also included
in MAID. There is essentially nothing known about these states in
electroproduction, and we have just introduced their reported photon couplings,
multiplied with a simple gaussian form factor, exp$(-2.0\,Q^2/{\rm {GeV}}^2)$.
The main role of these resonances in MAID is to define a global high-energy
behavior that is needed for applications with dispersion relations and sum
rules. Future experiments in this region will give
us the necessary information to map out these form factors in more detail.\\

\begin{table}
\begin{center}
\caption{ \label{tab:param-p2} Maid2007 parameterizations,
Eq.~(\ref{eq:ffpar}), for proton targets ($a_{2,3,4}=0)$.}
\begin{tabular*}{80mm}{@{\extracolsep{\fill}}ccccc}
\hline
 $N^*,\;\Delta^*$ & & $\bar{\mathcal A}_{\alpha}(0)$ & $a_1$ & $b_1$ \\
\hline
$S_{11}(1535)p$ & $A_{1/2}$ & $ 66.4$ & $ 1.608$ & $0.70$\\
                & $S_{1/2}$ & $ -2.0$ & $  23.9$ & $0.81$\\
$S_{31}(1620) $ & $A_{1/2}$ & $ 65.6$ & $ 1.86 $ & $2.50$\\
                & $S_{1/2}$ & $ 16.2$ & $ 2.83 $ & $2.00$\\
$S_{11}(1650)p$ & $A_{1/2}$ & $ 33.3$ & $ 1.45 $ & $0.62$\\
                & $S_{1/2}$ & $ -3.5$ & $ 2.88 $ & $0.76$\\
\hline
\end{tabular*}
\end{center}
\end{table}
\vspace*{0.5cm}

\begin{table}
\begin{center}
\caption{ \label{tab:param-n} Same as Table~\ref{tab:param-p2}, for neutron
targets.}
\begin{tabular*}{80mm}{@{\extracolsep{\fill}}ccccc}
\hline $N^*$ & & $\bar{\mathcal A}_{\alpha}(0)$ & $a_1$ & $b_1$ \\
\hline
$P_{11}(1440)n$ & $A_{1/2}$ & $ 54.1$ & $ 0.95 $ & $1.77$\\
                & $S_{1/2}$ & $-41.5$ & $ 2.98 $ & $1.55$\\
$D_{13}(1520)n$ & $A_{1/2}$ & $-76.5$ & $-0.53 $ & $1.55$\\
                & $A_{3/2}$ & $-154.$ & $ 0.58 $ & $1.75$\\
                & $S_{1/2}$ & $ 13.6$ & $ 15.7 $ & $1.57$\\
$S_{11}(1535)n$ & $A_{1/2}$ & $-50.7$ & $ 4.75 $ & $1.69$\\
                & $S_{1/2}$ & $ 28.5$ & $ 0.36 $ & $1.55$\\
$S_{11}(1650)n$ & $A_{1/2}$ & $  9.3$ & $ 0.13 $ & $1.55$\\
                & $S_{1/2}$ & $ 10. $ & $-0.50 $ & $1.55$\\
$D_{15}(1675)n$ & $A_{1/2}$ & $-61.7$ & $ 0.01 $ & $2.00$\\
                & $A_{3/2}$ & $-83.7$ & $ 0.01 $ & $2.00$\\
                & $S_{1/2}$ & $  0  $ & $  0   $ & $ 0  $\\
$F_{15}(1680)n$ & $A_{1/2}$ & $ 27.9$ & $  0   $ & $1.20$\\
                & $A_{3/2}$ & $-38.4$ & $ 4.09 $ & $1.75$\\
                & $S_{1/2}$ & $  0  $ & $  0   $ & $  0 $\\
$P_{13}(1720)n$ & $A_{1/2}$ & $ -2.9$ & $ 12.70$ & $1.55$\\
                & $A_{3/2}$ & $-31.0$ & $ 5.00 $ & $1.55$\\
                & $S_{1/2}$ & $  0  $ & $   0  $ & $  0 $\\
\hline
\end{tabular*}
\end{center}
\end{table}

\subsection{First Resonance Region}

The $\Delta(1232)P_{33}$ is the only nucleon resonance with a well-defined
Breit-Wigner resonance position, $M_R=1232$~MeV. It is an ideal single-channel
resonance, the Watson theorem applies, and the Breit-Wigner position coincides
with the K-matrix pole position. For these reasons, the $N\to\Delta(1232)$ form
factors can be determined in an essentially model independent way.

\begin{figure}
\begin{center}
\includegraphics[width=13cm]{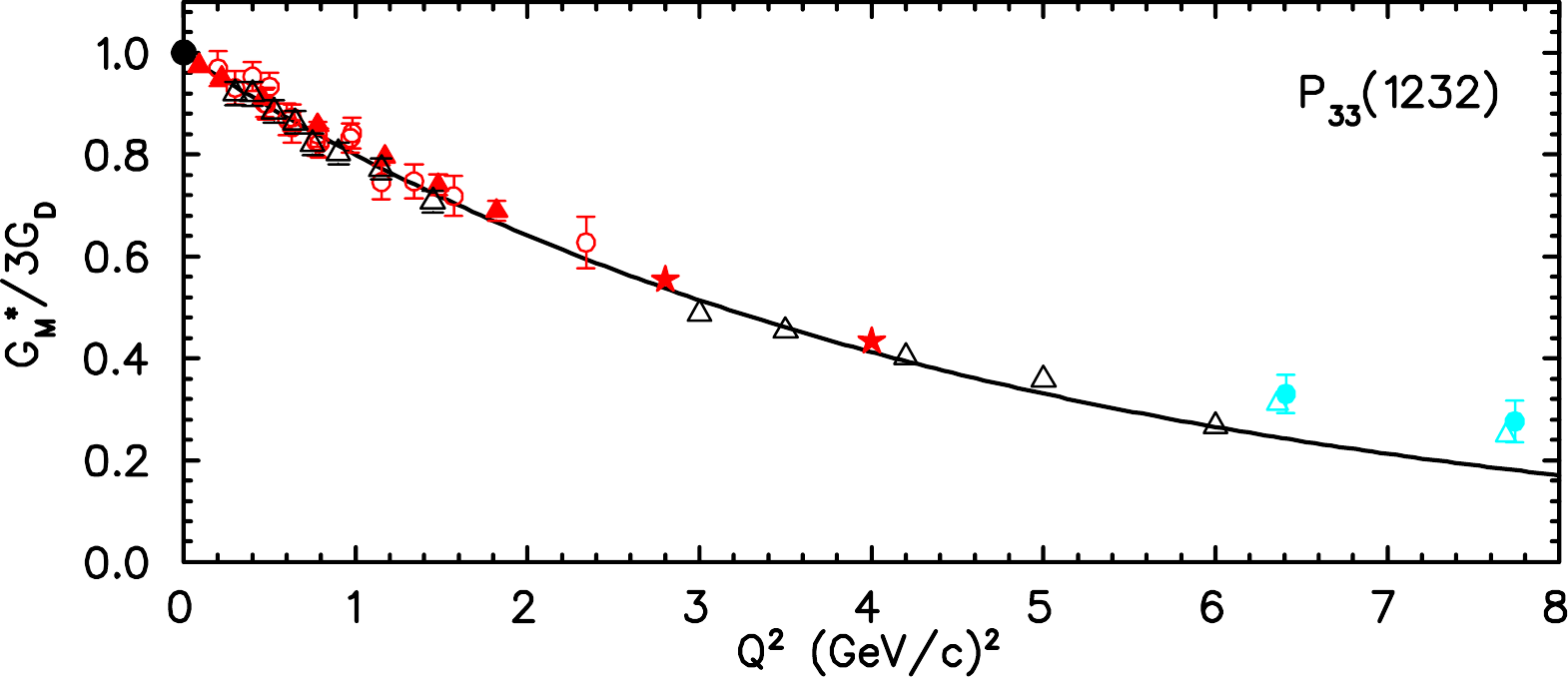}
\vspace{3mm} \caption{The $Q^2$ dependence of the magnetic form factor $G_M^*$
for the $N\Delta (1232)$ transition divided by $3\,G_D(Q^2)$ in the definition
of Ash~\cite{Ash67}. The curve shows the result of our MAID2007 fit with
Eq.~(\ref{eq:GEMstar}). The black solid circle at $Q^2=0$ shows the result of the
Mainz photoproduction experiment~\cite{Beck00} and is practically equal to 1.0,
the red data points are from Refs.~\cite{Stein} for solid triangles,
\cite{Bartel} for open circles and \cite{Fro99} for solid stars. The black open
triangles show the new JLab analysis of Aznauryan et al.~\cite{Azn09}. The cyan
open triangles and solid circles show the isobar analysis of
Ref.~\cite{Villano:2009sn} and our own new analysis, respectively.}
\label{fig:gmstar}
\end{center}
\end{figure}

The magnetic form factor shown in Fig.~\ref{fig:gmstar} is very well known up
to high momentum transfer, $Q^2=10$~GeV$^2$, and can be parameterized in a
surprisingly simple form found in our previous MAID analysis,
\begin{equation}
G_M^*(Q^2)=3\,G_D(Q^2) e^{-0.21 Q^2/{\rm {GeV}}^2}\,,
\end{equation}
with $G_D$ the standard dipole form factor. The electric and Coulomb form
factors are much smaller and are usually given as ratios to the magnetic form
factor.

In the literature the e.m. properties of the $N \Delta(1232)$ transition are
described by either the magnetic ($G_M^*$), electric ($G_E^*$), and Coulomb
($G_C^*$) form factors or the helicity amplitudes $A_{1/2}$, $A_{3/2}$, and
$S_{1/2}$, which can be derived from the reduced e.m. amplitudes $\bar{\mathcal
A}_\alpha$ defined by Eq.~(\ref{eq:BW}). It is worthwhile pointing out that the
latter amplitudes are related to the multipoles over the full energy region,
that is, they are the primary target of the fitting procedure. The form factors
and helicity amplitudes are then obtained by evaluating the reduced e.m.
amplitudes at the resonance position $W=M_\Delta$=1232~MeV. The respective
relations take the form
\begin{eqnarray}
G_M^*(Q^2) &=&  -c_\Delta (A_{1/2}+\sqrt{3} A_{3/2})=2 c_\Delta \,
\bar{\mathcal A}_M^\Delta(M_\Delta,Q^2)\,,\\
G_E^*(Q^2) &=&  \,\,\, c_\Delta (A_{1/2}-\frac{1}{\sqrt{3}} A_{3/2})=-2
c_\Delta \,\bar{\mathcal A}_E^\Delta(M_\Delta,Q^2)\,,\\
G_C^*(Q^2) &=& \sqrt{2} c_\Delta \frac{2M_{\Delta}}{k_\Delta} S_{1/2}=-2
c_\Delta \, \frac{2M_{\Delta}}{k_\Delta}\bar{\mathcal A}_S^\Delta(M_\Delta,Q^2)
\,,\\
\mbox{with} \;\; c_\Delta &=& \left( \frac{M_N^3 \kappa_\Delta}{4\pi\alpha_{\rm
{em}} M_\Delta k_\Delta^2} \right)^{1/2}\,, \label{eq:GstarToA}
\end{eqnarray}
where  $k_\Delta=k_\Delta (Q^2)=k(M_{\Delta},Q^2)$ and
$\kappa_\Delta=\kappa(M_\Delta)=k(M_\Delta,0)$ are the virtual photon momentum
and the photon equivalent energy at resonance. Because the $\Delta(1232)$ is
very close to an ideal resonance, the real parts of the amplitudes vanish at
$W=M_{\Delta}$, and the form factors can be directly expressed by the imaginary
parts of the corresponding multipoles at the resonance position,
\begin{eqnarray}
G^{\ast}_M(Q^2) &=& b_\Delta\, {\rm {Im}}\{M_{1+}^{(3/2)}(M_\Delta,Q^2)\}
\,,\label{eq:GstarMtoMult}\\
G^{\ast}_E(Q^2) &=& -b_\Delta\, {\rm {Im}}\{E_{1+}^{(3/2)}(M_\Delta,Q^2)\}
\,,\label{eq:GstarEtoMult}\\
G^{\ast}_C(Q^2) &=& -b_\Delta \frac{2M_{\Delta}}{k_{\Delta}}\,
{\rm {Im}}\{S_{1+}^{(3/2)}(M_\Delta, Q^2)\}\,,\label{eq:GstarCtoMult}\\
\mbox{where}\;\; b_\Delta &=& \left( \frac{8\, M_N^2\, q_\Delta\,
\Gamma_\Delta}{3\,\alpha_{\rm {em}}\, k_\Delta^2} \right)^{1/2} \,,
\end{eqnarray}
and with $\Gamma_{\Delta}=115$ MeV and $q_{\Delta}=q(M_{\Delta})$ the pion
momentum at resonance. The above definition of the form factors is due to
Ash~\cite{Ash67}. The form factors of Jones and Scadron~\cite{Jon73} are
obtained by multiplying our form factors with $\sqrt{1+Q^2/(M_N +M_\Delta)^2}$.
We note that the form factor $G_C^{\ast}$ differs from our previous
work~\cite{Tiator:2003xr} by the factor $2M_\Delta/k_{\Delta}$ in
Eq.~(\ref{eq:GstarCtoMult}). With these definitions all 3 transition form
factors remain finite at pseudo-threshold (Siegert limit). In the literature,
the following ratios of multipoles have been defined:
\begin{eqnarray}
R_{EM} &=& \frac{E_{1+}^{3/2}}{M_{1+}^{3/2}} =
-\frac{G_E^{\ast}}{G_M^{\ast}}=\frac{A_{1/2}-\frac{1}{\sqrt{3}}
A_{3/2}}{A_{1/2}+\sqrt{3} A_{3/2}}\,,\label{eq:REM}\\
R_{SM} &=& \frac{S_{1+}^{3/2}}{M_{1+}^{3/2}} =
-\frac{k_\Delta}{2M_{\Delta}}\frac{G_C^{\ast}}{G_M^{\ast}}=\frac{\sqrt{2}
S_{1/2}}{A_{1/2}+\sqrt{3} A_{3/2}}\,. \label{eq:RSM}
\end{eqnarray}
In MAID2007 the $Q^2$ dependence of the e.m. $N\Delta$ transition form factors
is parameterized as follows:
\begin{eqnarray}
G_{E,M}^*(Q^2)&=&g_{E,M}^0 (1 + \beta_{E,M}
Q^{2})  e^{-\gamma_{E,M} Q^2}G_D(Q^2)\,\label{eq:GEMstar} \,,\\
G_{C}^*(Q^2)&=&g_{C}^0 \frac{1 + \beta_{C}Q^{2}}{1+d_C Q^2/(4M_N^2)} \frac{2
M_\Delta}{\kappa_\Delta} e^{-\gamma_{C} Q^2}G_D(Q^2)\,,\label{eq:GCstar}
\end{eqnarray}
where $G_D(Q^2)=1/(1+Q^2/0.71\,{\rm {GeV}}^2)^2$ is the dipole form factor. In
order to fulfill the Siegert theorem, we have changed the
parametrization of the Coulomb amplitude accordingly~\cite{MAID07}.
\begin{table}[htbp]
\caption{Parameters for the $N \Delta$ transition form factors
$G_M^*,G_E^*,G_C^*$ given by Eqs.~(\ref{eq:GEMstar}-\ref{eq:GCstar}). The
normalization values at the photon point $(Q^2=0)$, $g_\alpha^0$ and $d_\alpha$
are dimensionless, the parameters $\beta$ and $\gamma$ in GeV$^{-2}$. }
\label{tab:GMEC}
\begin{center}
\begin{tabular}{c|ccc}
\hline
  & M1 & E2 & C2 \\
\hline
  $g_\alpha^0$  & 3.00 & 0.0637 & 0.1240    \\
 $\beta_\alpha$ & 0.0095 & -0.0206& 0.120 \\
$\gamma_\alpha$ &  0.23  & 0.16 &  0.23  \\
$d_\alpha$      &  0     &  0  &  4.9  \\
\hline
\end{tabular}
\end{center}
\end{table}
The result of MAID2007 for $G_M^{\ast}(Q^2)$ is compared to the data in
Fig.~\ref{fig:gmstar}. Except for the latest data by Villano et
al.~\cite{Villano:2009sn} at the highest $Q^2$ shown, our single-$Q^2$ analysis
for all other data follows the global fit closely and is not shown in the
figure. We further note that $G_M^{\ast}(0)/3$ takes the value of 1.0 to an
accuracy of 1\%. This value is related to the $N\rightarrow\Delta$ magnetic
transition moment, $\mu_{N\Delta}=(3.46\pm 0.03$) nuclear magnetons, see
Eq.~(\ref{eq:muNDelta}), by the following equation:
\begin{equation}
G_M^*(0)=\sqrt{\frac{M_N}{M_\Delta}}\,\mu_{N\Delta}\,.
\end{equation}

\begin{figure}
\begin{center}
\includegraphics[height=4.8cm]{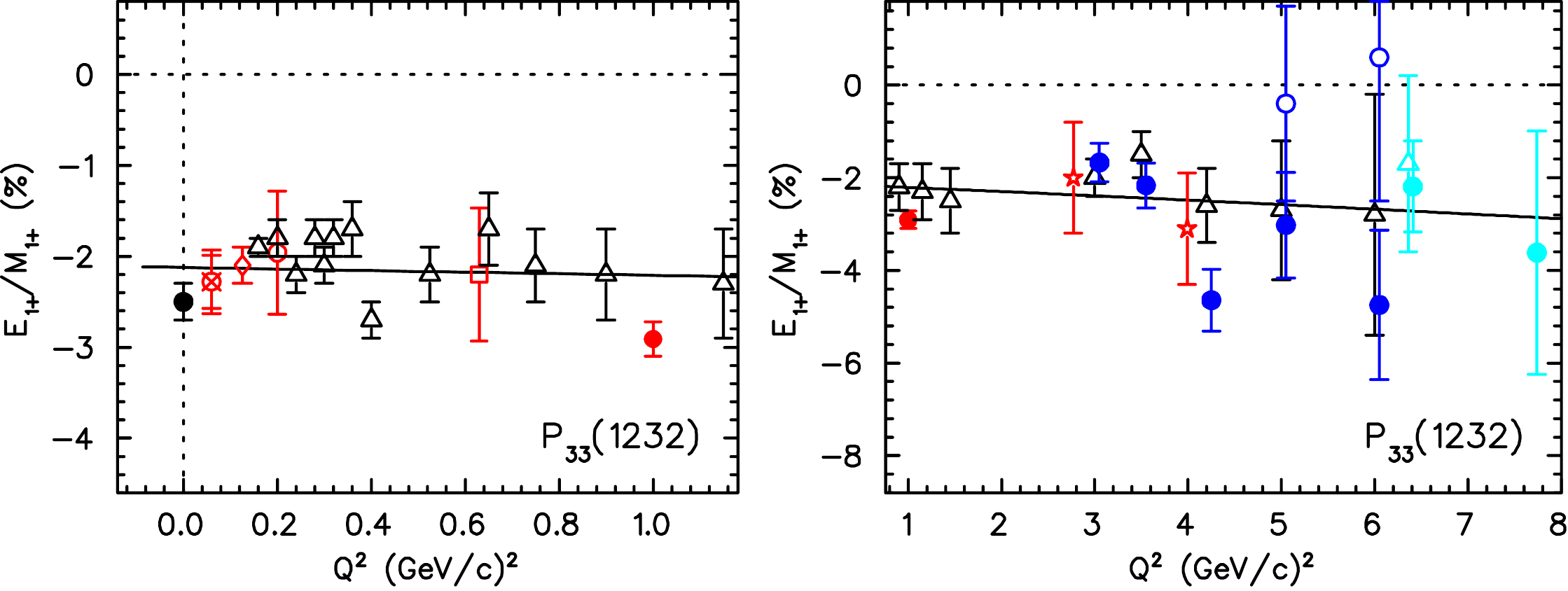}
\vspace{3mm} \caption{The $Q^2$ dependence of the ratio $R_{EM}$ at the
$\Delta(1232)$ resonance. The curve shows the result of our MAID2007 fit with
Eq.~(\ref{eq:GEMstar}). The black circle at $Q^2=0$ represents the Mainz
photoproduction experiment~\cite{Beck00}. The red data points are from
Refs.~\cite{Stave06} (cross), \cite{Stave:2008tv} (open circles), \cite{Mer01}
(open triangle), \cite{Got00,Ban03} (open square), \cite{Kelly05} (solid
circle), and \cite{Fro99} (solid stars). The black open triangles show the new
JLab analysis of Aznauryan et al.~\cite{Azn09}. The blue solid circles are our
2007 analysis~\cite{MAID07} and the blue open circles from our new work
discussed in the text. The cyan open triangles and cyan solid circles show the
isobar model analysis of Ref.~\cite{Villano:2009sn} and our own new analysis,
respectively.} \label{fig:emratios}
\end{center}
\end{figure}

\begin{figure}
\begin{center}
\includegraphics[height=4.8cm]{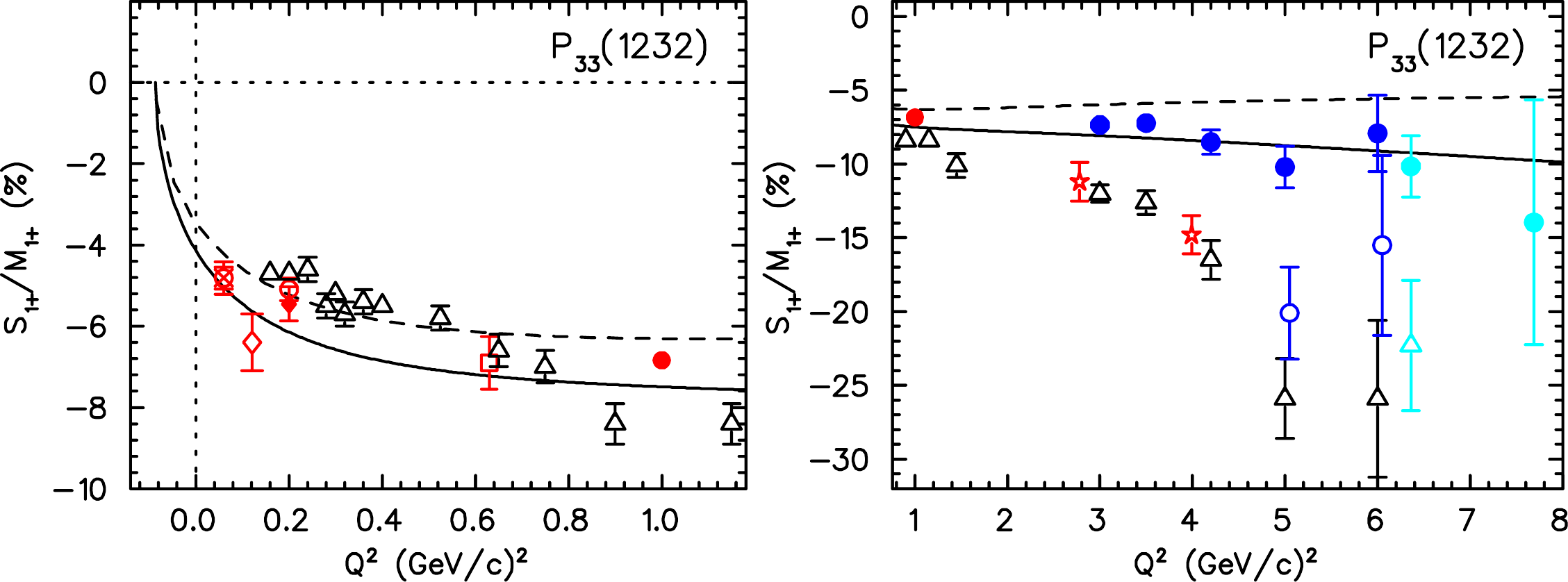}
\vspace{3mm} \caption{The $Q^2$ dependence of the ratio $R_{SM}$ at the
$\Delta(1232)$ resonance position. The solid line shows our MAID2007 fit with
Eq.~(\ref{eq:GCstar}) and the dashed curve is the Buchmann prediction relating
$R_{SM}$ to the elastic neutron form factors~\cite{Buchmann:2004ia}. The red
data points are from Refs.~\cite{Stave06} (cross), \cite{Pos01} (open diamond),
\cite{Elsner06} (solid diamond), \cite{Stave:2008tv} (open circles),
\cite{Got00,Ban03} (open square), \cite{Kelly05} (solid circle), and
\cite{Fro99} (solid stars). The black open triangles show the new JLab analysis
of Aznauryan et al.~\cite{Azn09}, the blue solid circles are our 2007
analysis~\cite{MAID07}, and the blue open circles result from our new
investigation discussed in the text. The cyan open triangles and cyan solid
circles show the isobar analysis of Ref.~\cite{Villano:2009sn} and our new
analysis, respectively.} \label{fig:smratios}
\end{center}
\end{figure}

Figures~\ref{fig:emratios} and \ref{fig:smratios} compare the MAID2007
solutions (solid lines) for the ratios $R_{EM}$ and $R_{SM}$ with other
analyses. The ratio $R_{EM}$ from MAID2007 stays always below the zero line, in
agreement with the original analysis of the data~\cite{Ung06,Fro99} and also
with the dynamical model of Sato and Lee~\cite{SL01} who concluded that
$R_{EM}$ remains negative and tends towards more negative values with
increasing $Q^2$. This indicates that the predicted helicity conservation at
the quark level is irrelevant for the present experimental $Q^2$ range. We also
analyzed the new data of Ref.~\cite{Ung06} in the range of 3~GeV$^2\leq Q^2
\leq 6$~GeV$^2$ and found slightly decreasing values of $R_{EM}$ from our
single-$Q^2$ analysis. For the ratio $R_{SM}$ both the $Q^2$-dependent and the
single-$Q^2$ fits approach a negative constant for large $Q^2$. This result is
in good agreement with the prediction of Ji et al.~\cite{Ji:2003fw} and of
Buchmann~\cite{Buchmann:2004ia} (dashed curve in Fig.~\ref{fig:smratios}) who
derived the following relation between the ratio $R_{SM}$ and the ratio of the
electric and magnetic neutron form factors:
\begin{equation}
R_{SM}(Q^2)=\frac{M_N \, k_\Delta(Q^2)\,G_E^n(Q^2)} {2\, Q^2 \, G_M^n(Q^2)}\,.
\label{eq:RsmBuch}
\end{equation}

For $Q^2>1~{\rm {GeV}}^2$ our Maid2007 analysis for $R_{SM}$  disagrees with
the JLab analysis of Aznauryan et al.~\cite{Azn09}. Whereas our analysis stays
almost constant, the JLab analysis suggests a much larger negative slope. We
note that we had obtained a similar slope in our previous MAID2003
analysis~\cite{Tiator:2003xr}. However, while repeating the data analysis at
$Q^2=5$ and $6~GeV^2$ in different energy ranges, we found a strong dependence
of the fit on the energy interval used. Whereas our results of 2007 were
obtained with the full energy range of $W=(1110-1390)$~MeV,
Fig.~\ref{fig:smratios} also shows an analysis (blue open circles) in the
energy range of $W=1200-1260$~MeV, much closer to the resonance energy. If we
choose the energy interval even closer to resonance, $W=1220-1240~$MeV, the
errors increase further by a factor of 2 and the $E/M$ ratio becomes large and
positive, while the $S/M$ ratio remains the same. We conclude that the analysis
in this $Q^2$ range strongly depends on the energy interval and the
parametrization of the background used in the analysis and requires further
studies. In order to solve this problem it may be necessary to obtain higher
statistics in the data or to measure additional polarization observables, such
as recoil polarization. The same conclusion is also likely for the highest
$Q^2$ values that were analyzed with the data of Villano et
al.~\cite{Villano:2009sn} at $Q^2$ values of 6.3 and 7.6~GeV$^2$. For these
data we obtain a $G_M^*$ form factor consistent with the JLab analysis.
However, these fits yield values about 30-50\% higher than our global fit to
the world data.

\subsection{Second Resonance Region}

Above the two-pion threshold, we can no longer apply the two-channel unitarity
and consequently the Watson theorem does not hold anymore. Therefore, the
background amplitude of the partial waves does not vanish at resonance as is
the case for the $\Delta(1232)$. As an immediate consequence the
resonance-background separation becomes model-dependent. In MAID2007 we choose
to separate the background and resonance contributions according to the
K-matrix approximation, Eqs.~(\ref{eq:Kmatrix},\ref{eq:BW}). Furthermore, we
recall that the absolute values of the helicity amplitudes are correlated with
the input used for the total resonance width $\Gamma_R$ and the single-pion
branching ratio $\beta_\pi$, which gives rise to additional uncertainties from
these hadronic resonance parameters. On the experimental side, the data at the
higher energies are no longer as abundant as in the $\Delta$ region. However,
the large data set recently obtained mainly by the CLAS collaboration (see
Table~\ref{tab:database}) enabled us to determine the transverse and
longitudinal helicity couplings as functions of $Q^2$ for all the four-star
resonances below 1800~MeV. These data are available in the kinematical regions
of $1100~\mbox{MeV}<W<1680~\mbox{MeV}$ and
$0.4~\mbox{GeV}^2<Q^2<4.5~\mbox{GeV}^2$.

\begin{figure}
\begin{center}
\includegraphics[height=4.0cm]{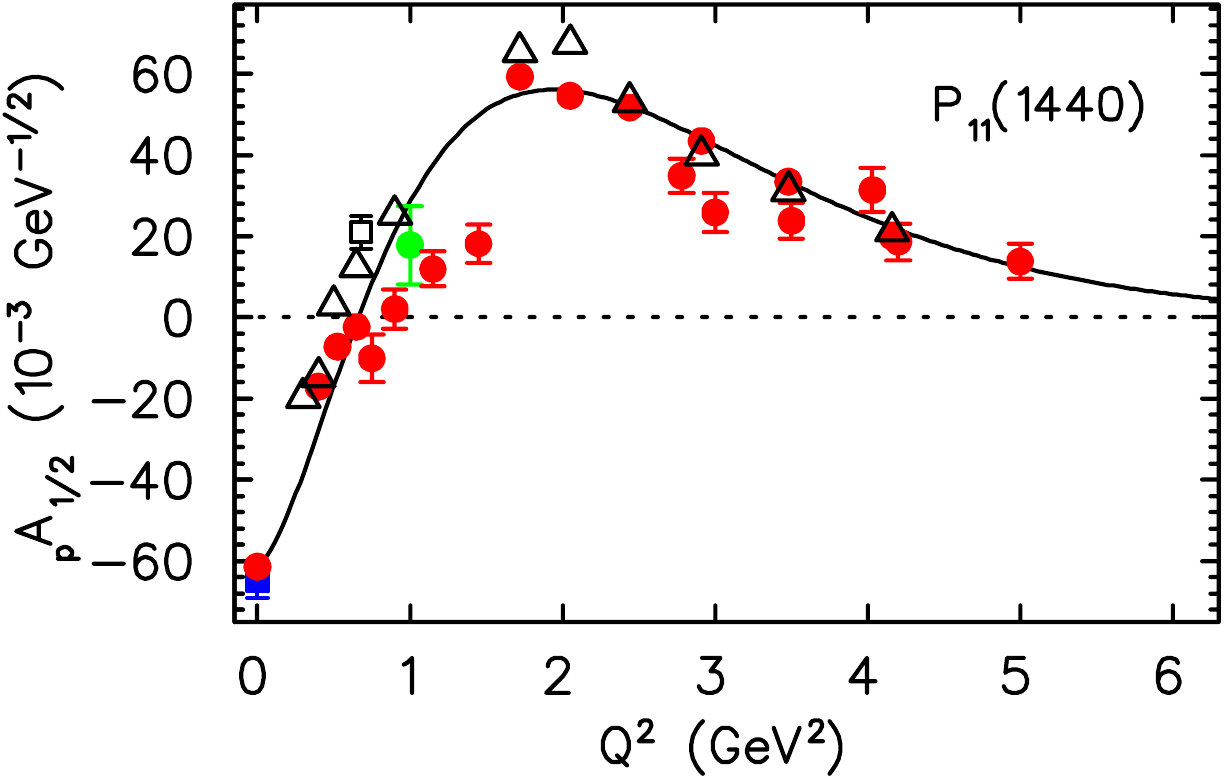}\hspace*{0.2cm}
\includegraphics[height=4.0cm]{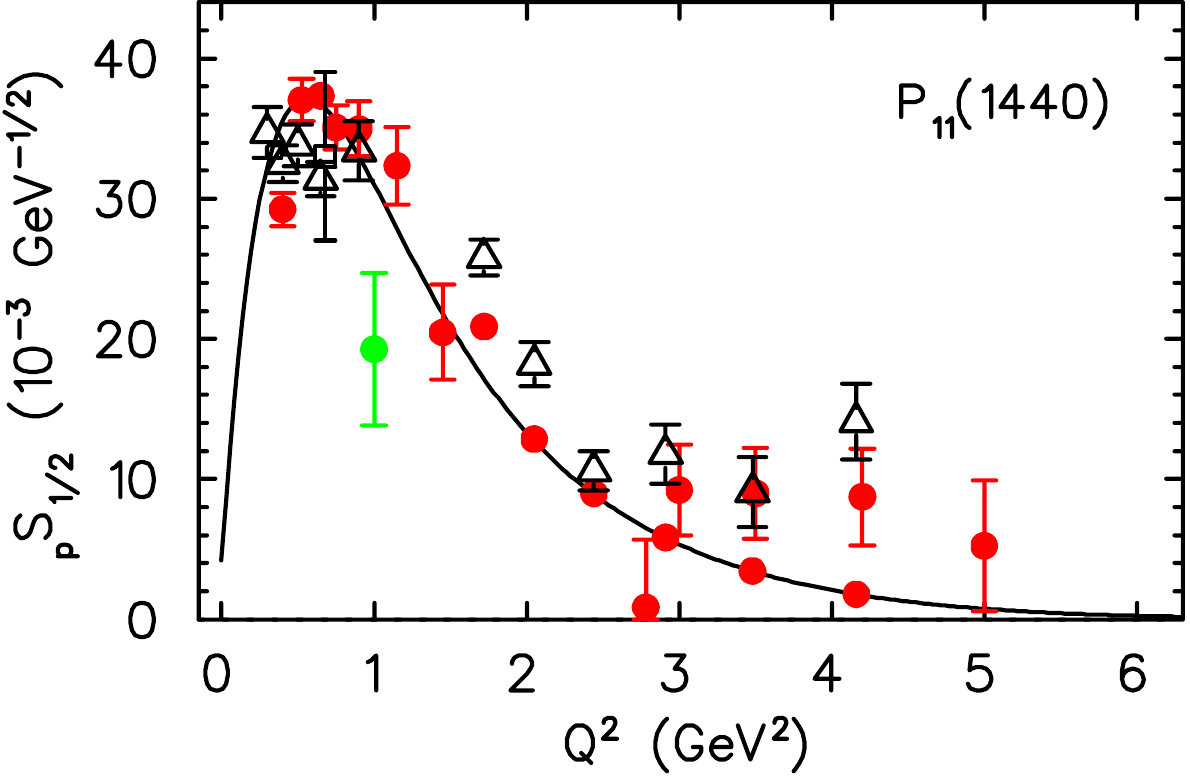}
\vspace{3mm} \caption{\label{fig:p11phel} Transverse and longitudinal form
factors of the $P_{11}(1440)$ Roper resonance. The red circles are the MAID
analysis of 2007~\protect\cite{MAID07} and 2008, the open black triangles are
the 2009 JLab analysis~\protect\cite{Azn09} and the open squares are from
Ref.~\cite{Aznauryan:2005tp}. For the transverse form factor at $Q^2=0$ we show
the PDG average~\protect\cite{PDG10} (blue square) and our MAID2007 result of
photoproduction (red circle), which partly overlap here. The green point at
$Q^2=1.0$~GeV$^2$ shows our analysis of the JLab/Hall A data of Laveissiere et
al.~\cite{Lav04}. The curves show the MAID2008 parametrization. All FF data
points analyzed by the MAID group can be downloaded from the MAID
website~\cite{MAID07}.}
\end{center}
\end{figure}

The helicity amplitudes for the Roper resonance $P_{11}(1440)$ are displayed in
Fig.~\ref{fig:p11phel}. Our latest $Q^2$-dependent solution (solid lines) is in
reasonable agreement with the single-$Q^2$ analysis (red circles). The figure
shows a zero crossing of the transverse helicity amplitude at $Q^2\approx
0.7$~GeV$^2$ and a maximum at the relatively large momentum transfer
$Q^2\approx 2.0$~GeV$^2$. The longitudinal Roper excitation rises to large
values around $Q^2\approx 0.5$~GeV$^2$ and, in fact, produces one of the
strongest longitudinal amplitudes we find in our analysis. This answers the
question raised by Li and Burkert\cite{Burk92} whether the Roper resonance is a
radially excited 3-quark state or a quark-gluon hybrid, because in the latter
case the longitudinal coupling should vanish completely.

\begin{figure}
\begin{center}
\includegraphics[height=4.0cm]{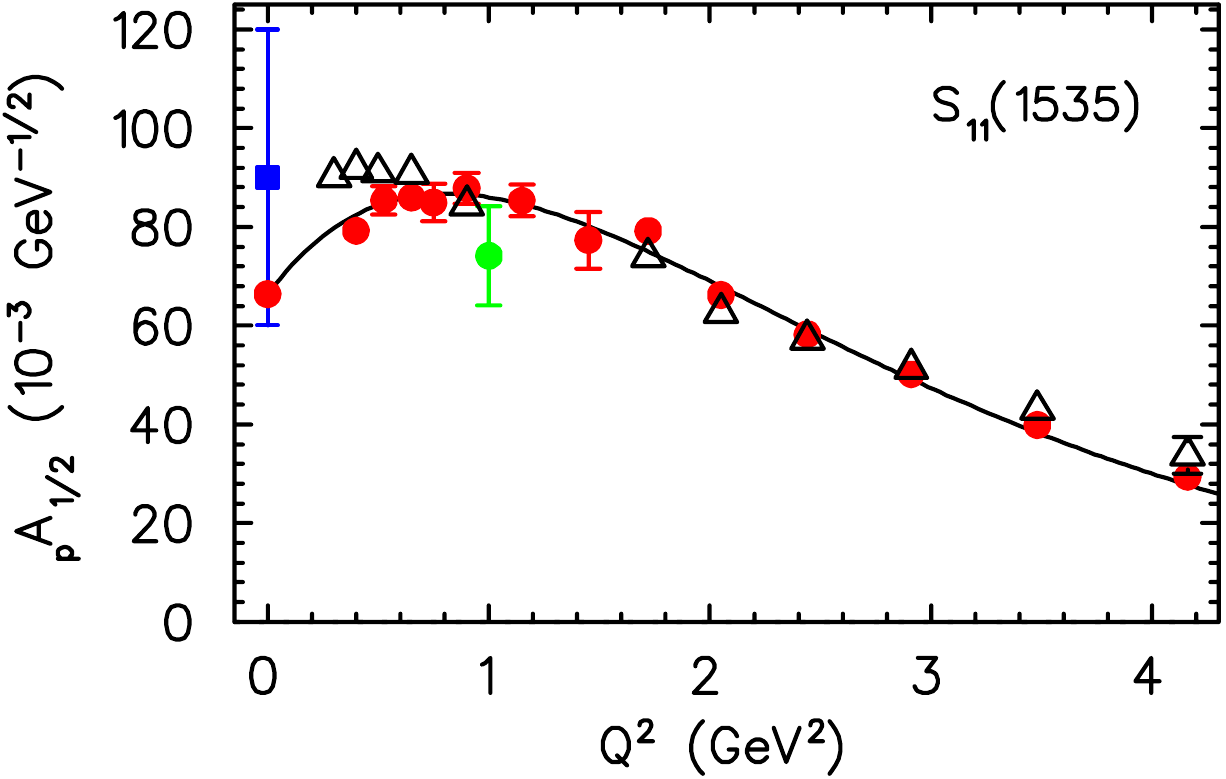}\hspace*{0.2cm}
\includegraphics[height=4.0cm]{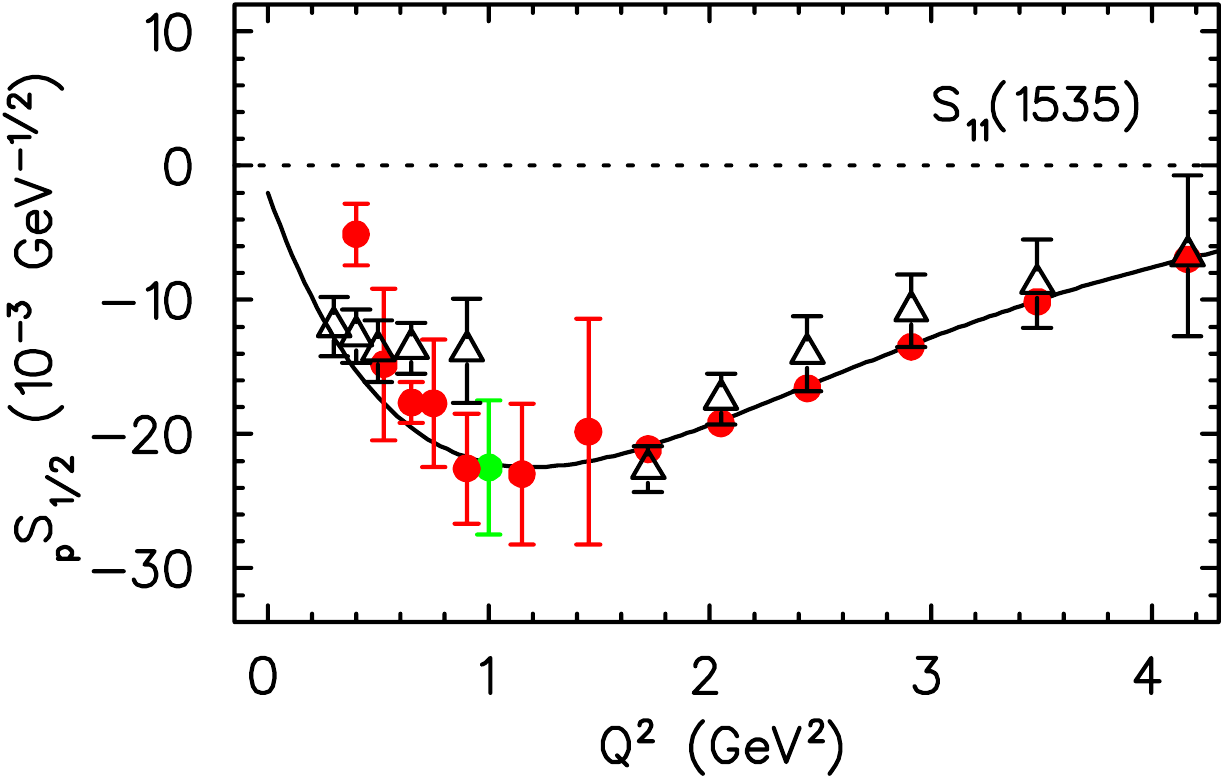}
\vspace{3mm} \caption{\label{fig:s11phel} Transverse and longitudinal form
factors of the $S_{11}(1535)$ resonance. The curves show the MAID2007
parametrization. Further notation as in Fig.~\ref{fig:p11phel}.}
\end{center}
\end{figure}

Figure~\ref{fig:s11phel} displays the results for the $S_{11}(1535)$ resonance.
The red single-$Q^2$ data points show our results of 2007, which are in good
agreement with our $Q^2$-dependent analysis (solid lines). The black triangles
are the 2009 results of Ref.~\cite{Azn09}. The blue data point at $Q^2=0$
represents the PDG average over several $\gamma,\pi$ and $\gamma,\eta$
analyses. Whereas we find values around 65 in all MAID analyses, the JLab
analysis obtains values around 90 for $\gamma,\pi$ and 110 for
$\gamma,\eta$~\cite{Aznauryan:2011ub}. A peculiarity is worth mentioning for
the longitudinal form factor. In the long-wavelength limit near
pseudothreshold, at $Q^2=-(M_R-M_N)^2$ or $-0.36$~GeV$^2$ for the $S_{11}$, the
Siegert theorem predicts a positive value for $S_{1/2}(Q^2)$ with a positive
slope, whereas both the MAID and the JLab groups find negative values for all
physical values, $Q^2>0$.

\begin{figure}
\begin{center}
\includegraphics[height=4.0cm]{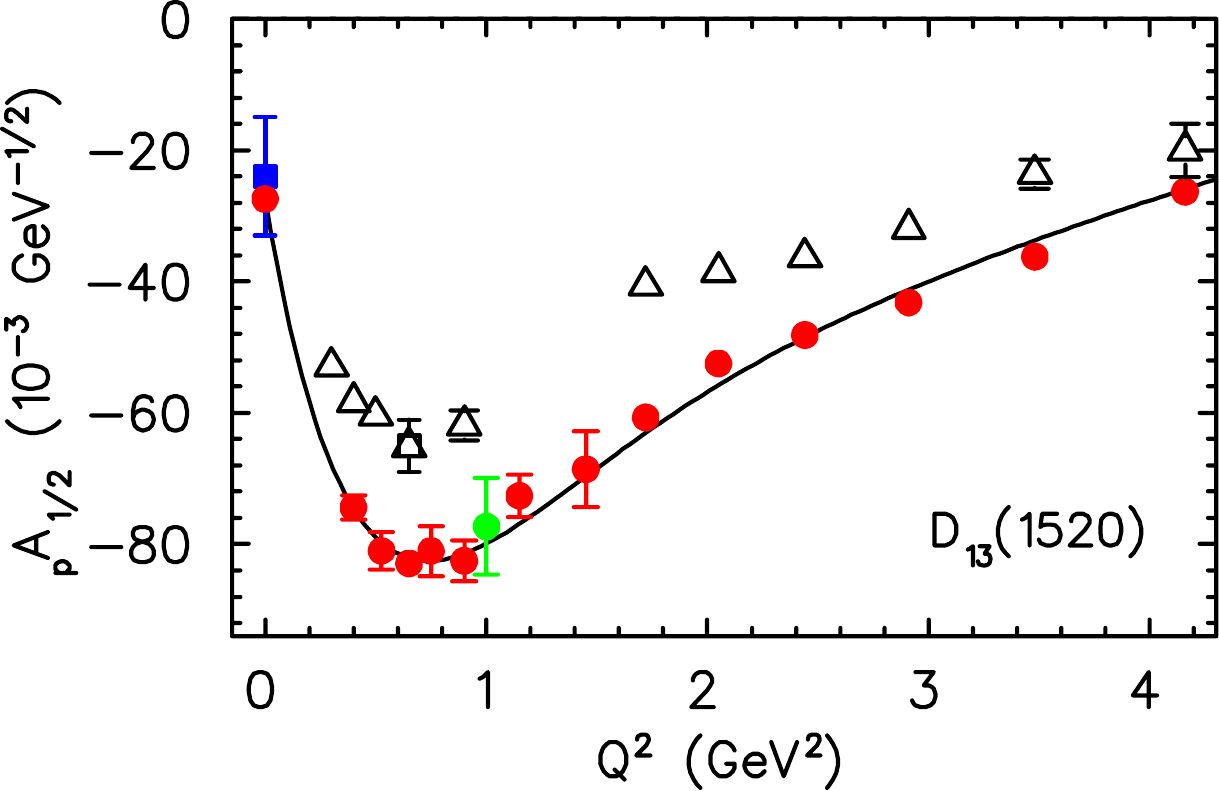}\hspace*{0.2cm}\includegraphics[height=4.0cm]{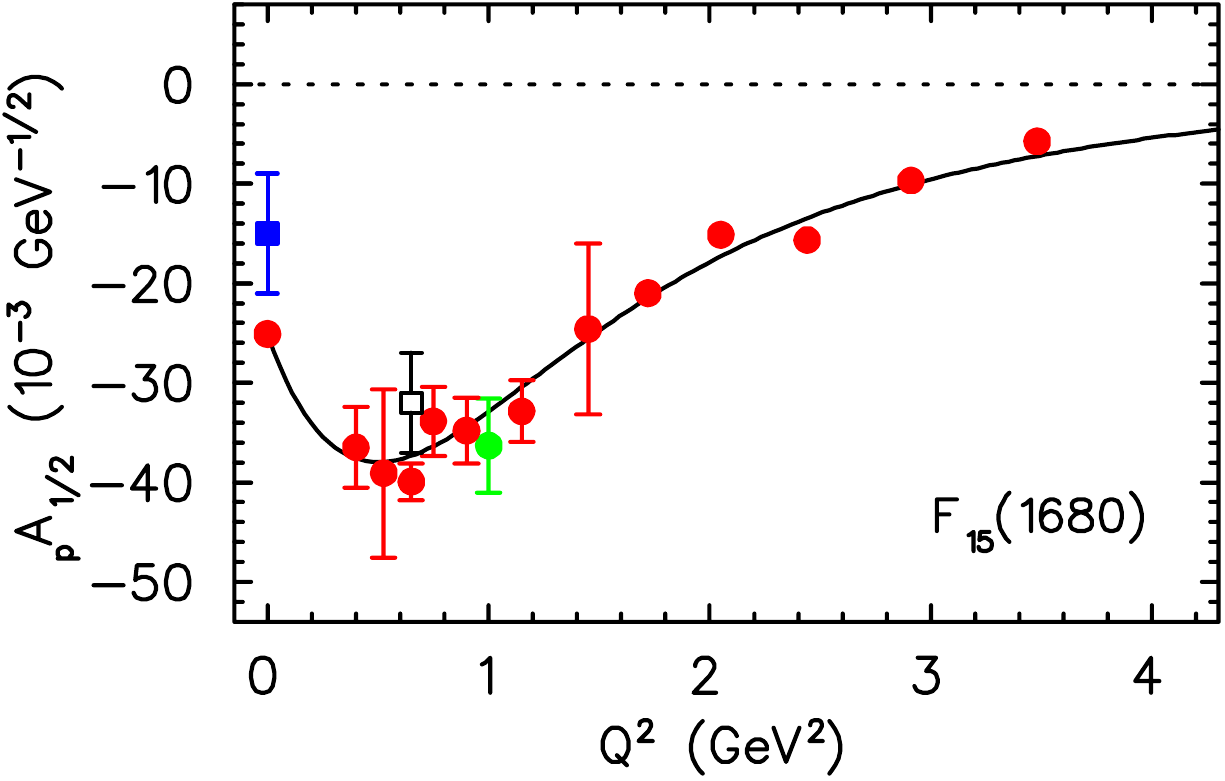}

\vspace*{0.2cm}
\includegraphics[height=4.0cm]{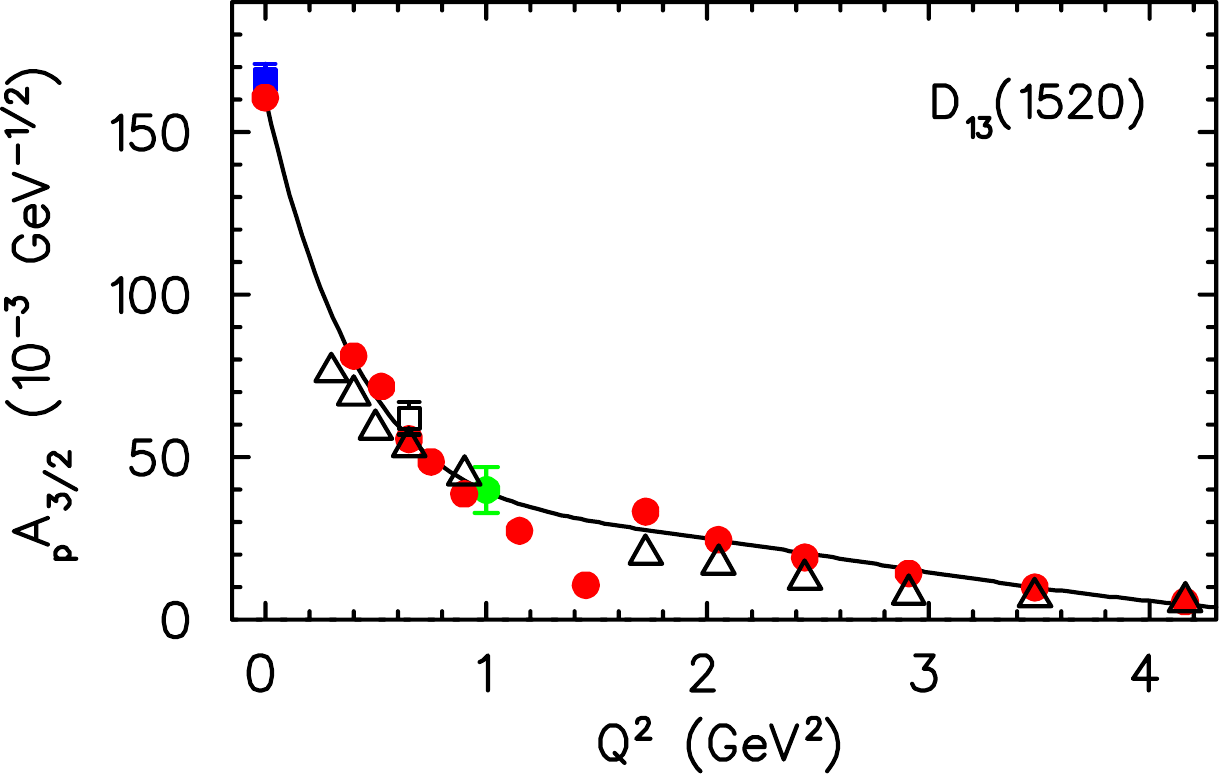}\hspace*{0.2cm}\includegraphics[height=4.0cm]{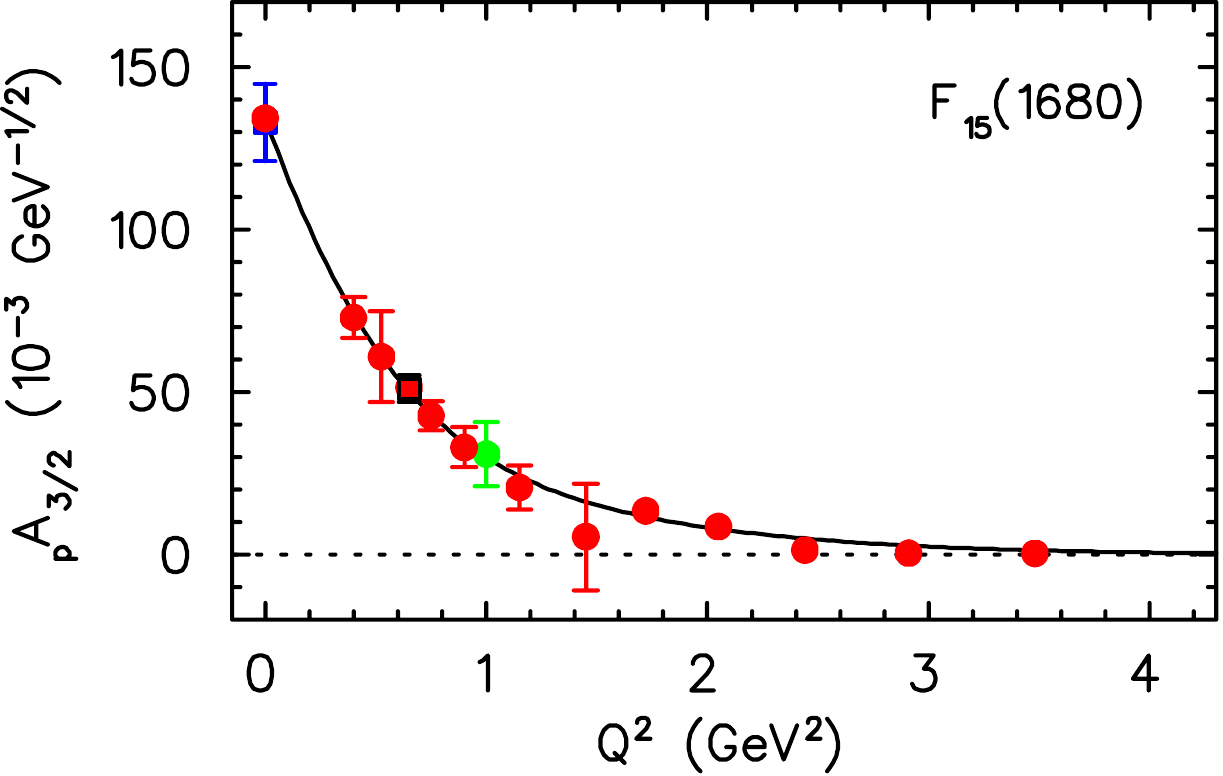}

\vspace*{0.2cm}
\includegraphics[height=4.0cm]{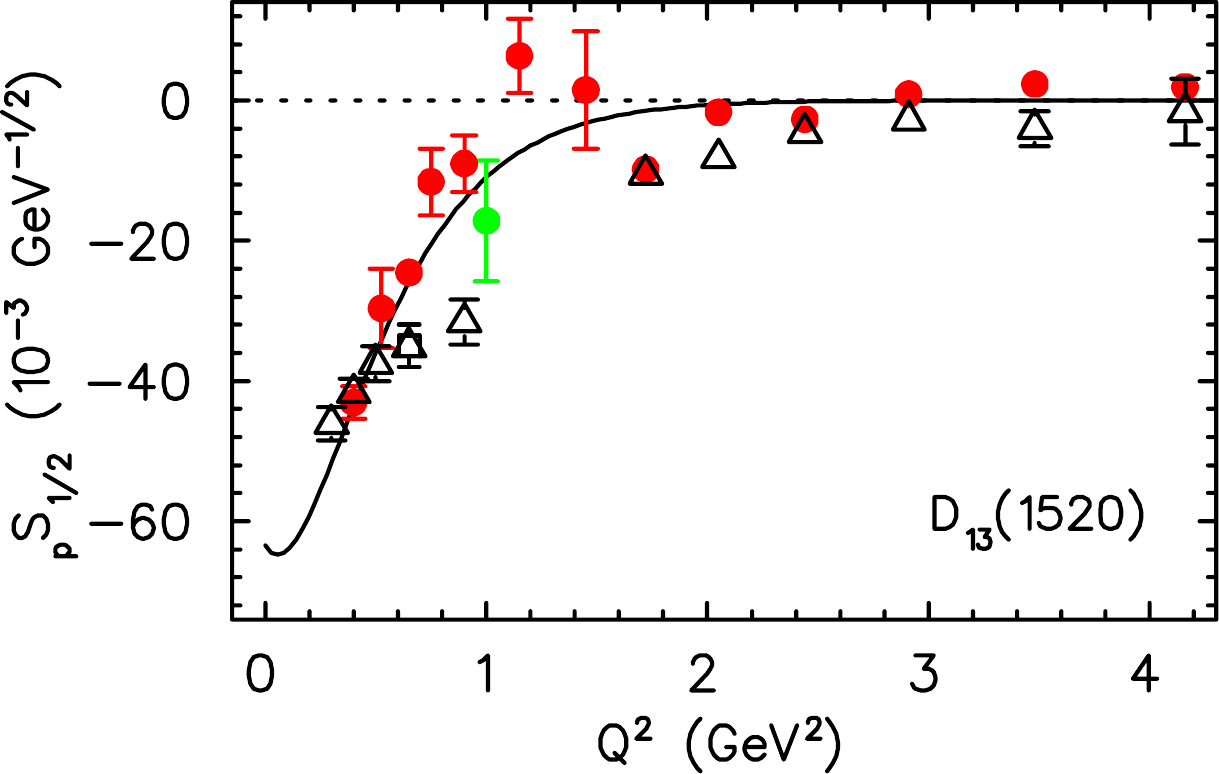}\hspace*{0.2cm}\includegraphics[height=4.0cm]{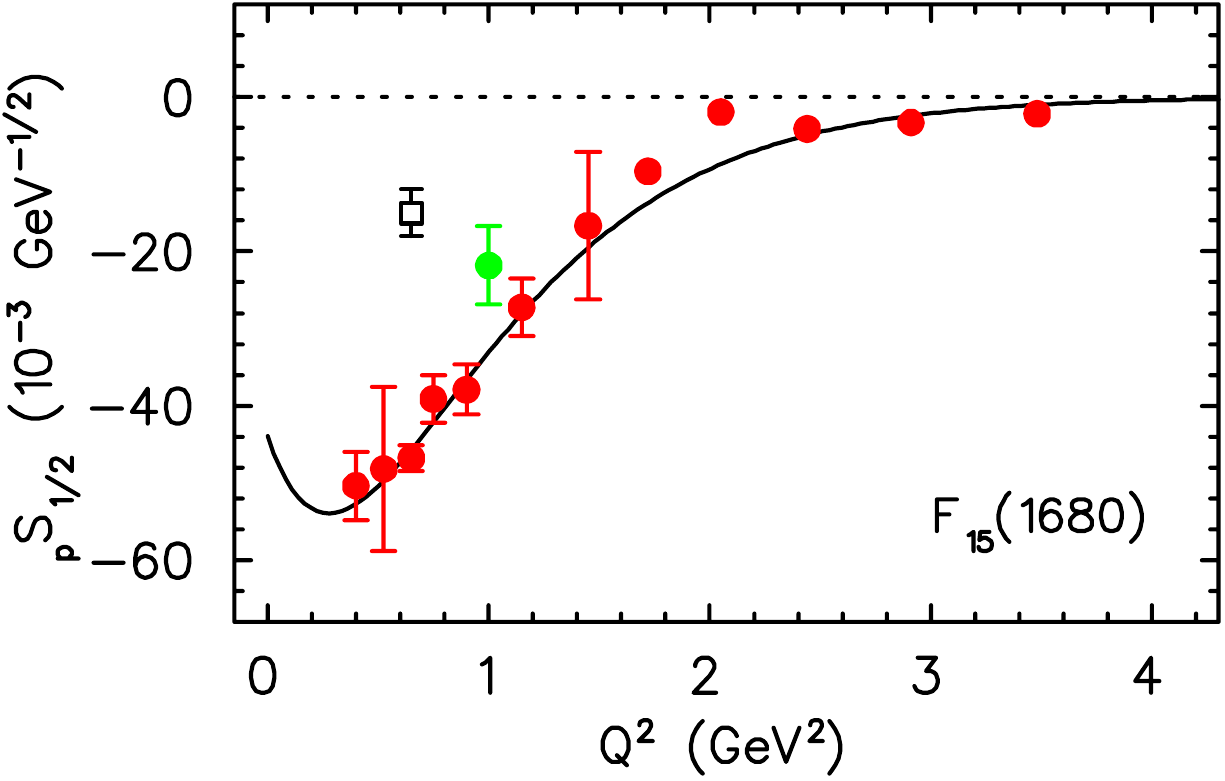}
\vspace{2mm} \caption{\label{fig:d13f15phel} Transverse and longitudinal form
factors of the $D_{13}(1520)$ (left panels) and $F_{15}(1680)$ (right panels)
resonances. The curves show the MAID2008 parametrization. Further notation as
in Fig.~\ref{fig:p11phel}.}
\end{center}
\end{figure}

Whereas the inclusion of the 2008 Park $\pi^+$ data~\cite{Park:2007tn} did not
modify our 2007 solution for the $S_{11}$ resonance, some modifications were
necessary for the $D_{13}(1520)$ resonance shown by the left panels of
Fig.~\ref{fig:d13f15phel}. For this resonance we find significant deviations in
the comparison with the JLab partial wave analysis of 2009, especially for
$A_{1/2}$ over the whole range of $Q^2$.

\begin{figure}
\begin{center}
\includegraphics[height=4.0cm]{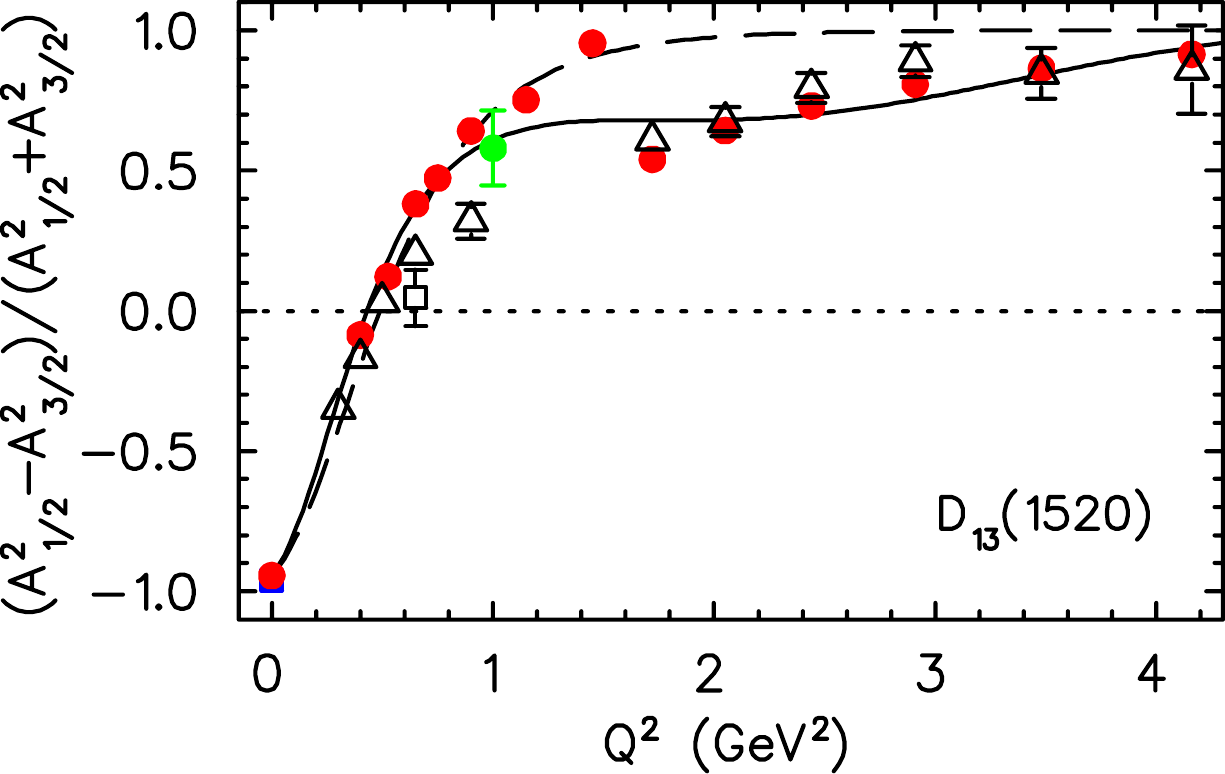}\hspace*{0.2cm}
\includegraphics[height=4.0cm]{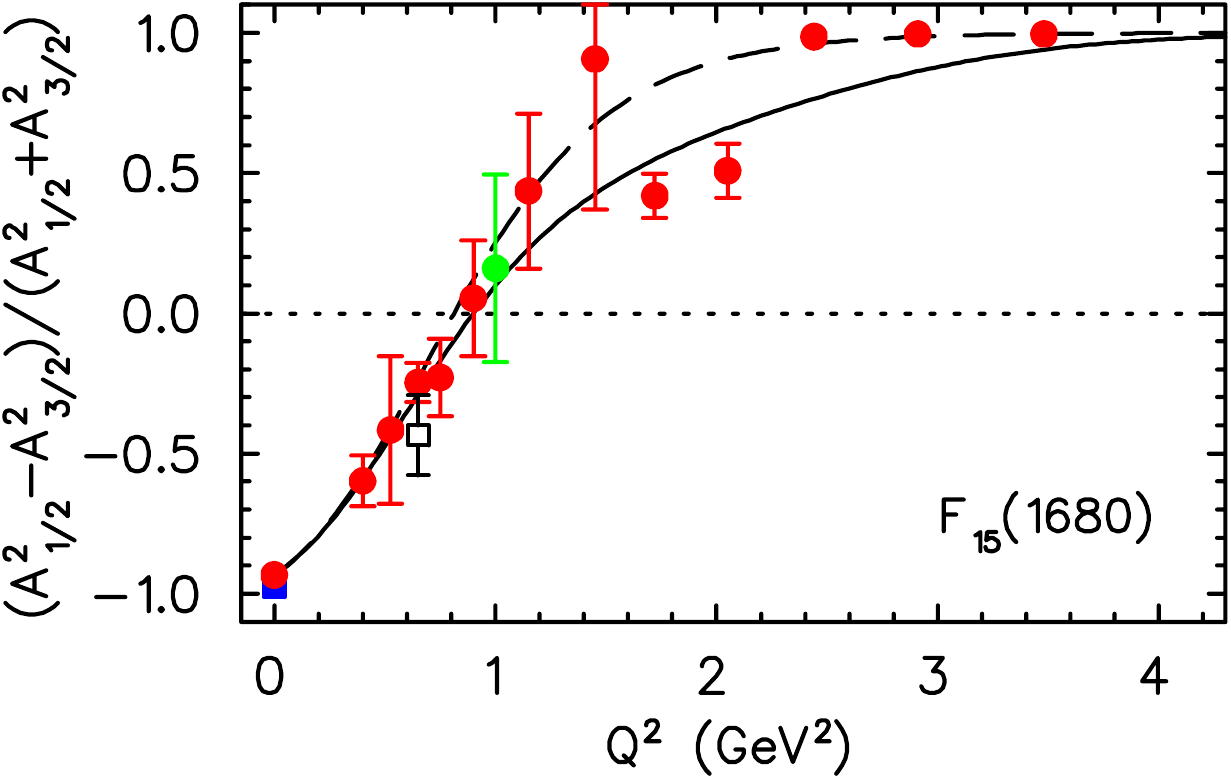}
\vspace{2mm} \caption{The helicity asymmetry ${\mathcal {A}}(Q^2)$ of
Eq.~(\ref{eq:HelAsym}) for the $D_{13}(1520)$ and $F_{15}(1680)$ resonances of
the proton. The dashed and solid curves are the MAID2007 and most recent
MAID2008 solutions, respectively. Further notation as in
Fig.~\ref{fig:p11phel}. } \label{fig:asym}
\end{center}
\end{figure}
\begin{figure}
\begin{center}
\includegraphics[height=4.0cm]{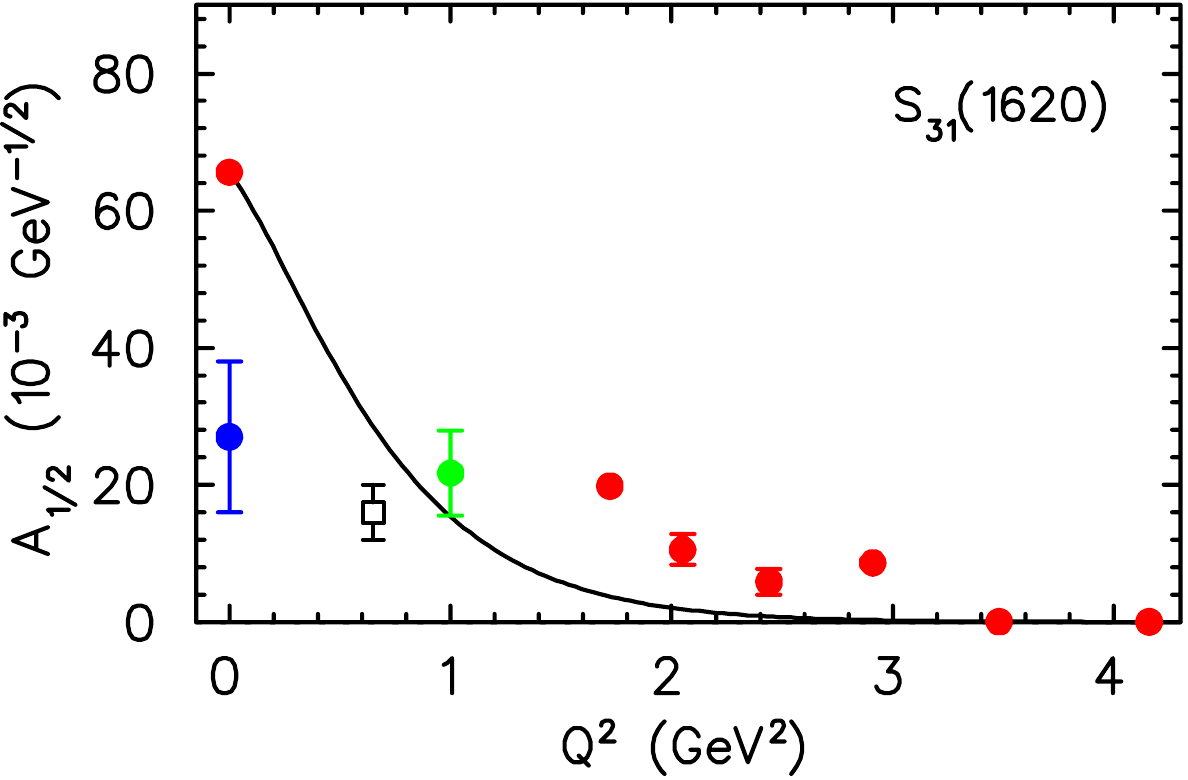}\hspace*{0.2cm}
\includegraphics[height=4.0cm]{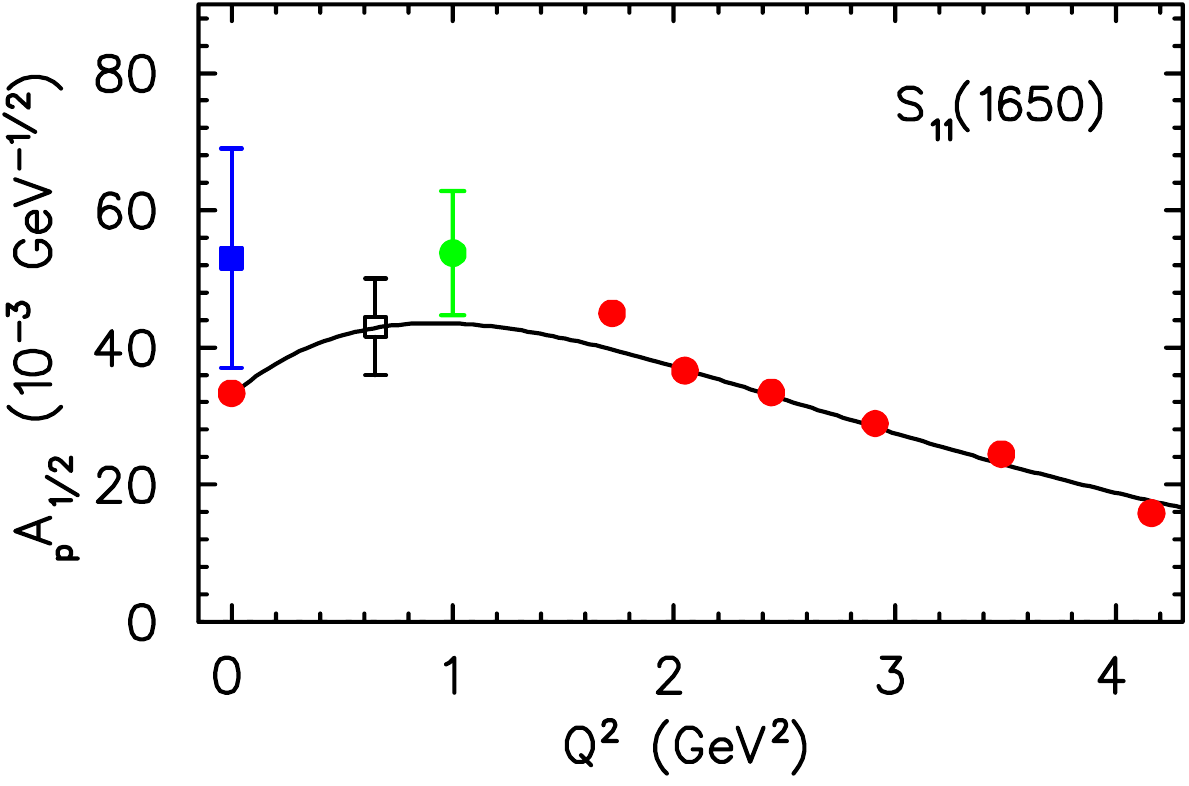}
\vspace{3mm} \caption{\label{fig:s31s11sphel} Transverse form factors of the
$S_{31}(1620)$ and $S_{11}(1650)$ resonances. The curves show the MAID2007
parametrization. Further notation as in Fig.~\ref{fig:p11phel}.}
\end{center}
\end{figure}
\begin{figure}
\begin{center}
\includegraphics[height=4.0cm]{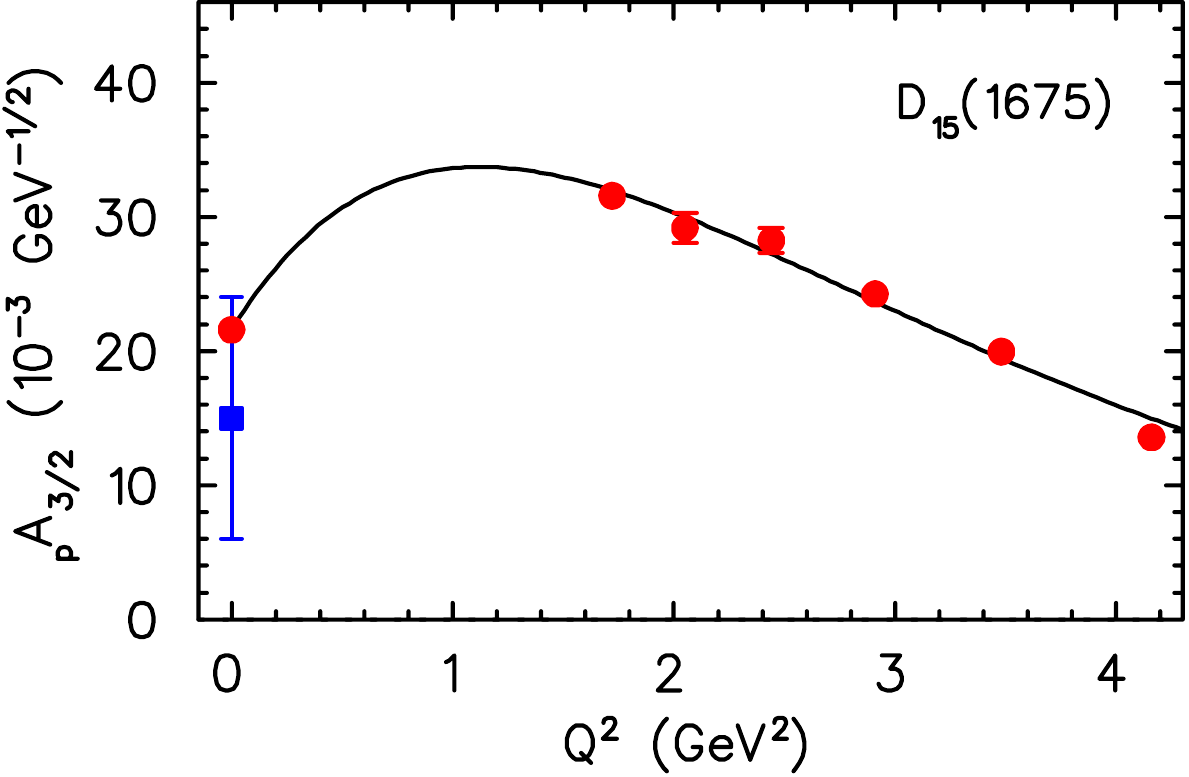}\hspace*{0.2cm}
\includegraphics[height=4.0cm]{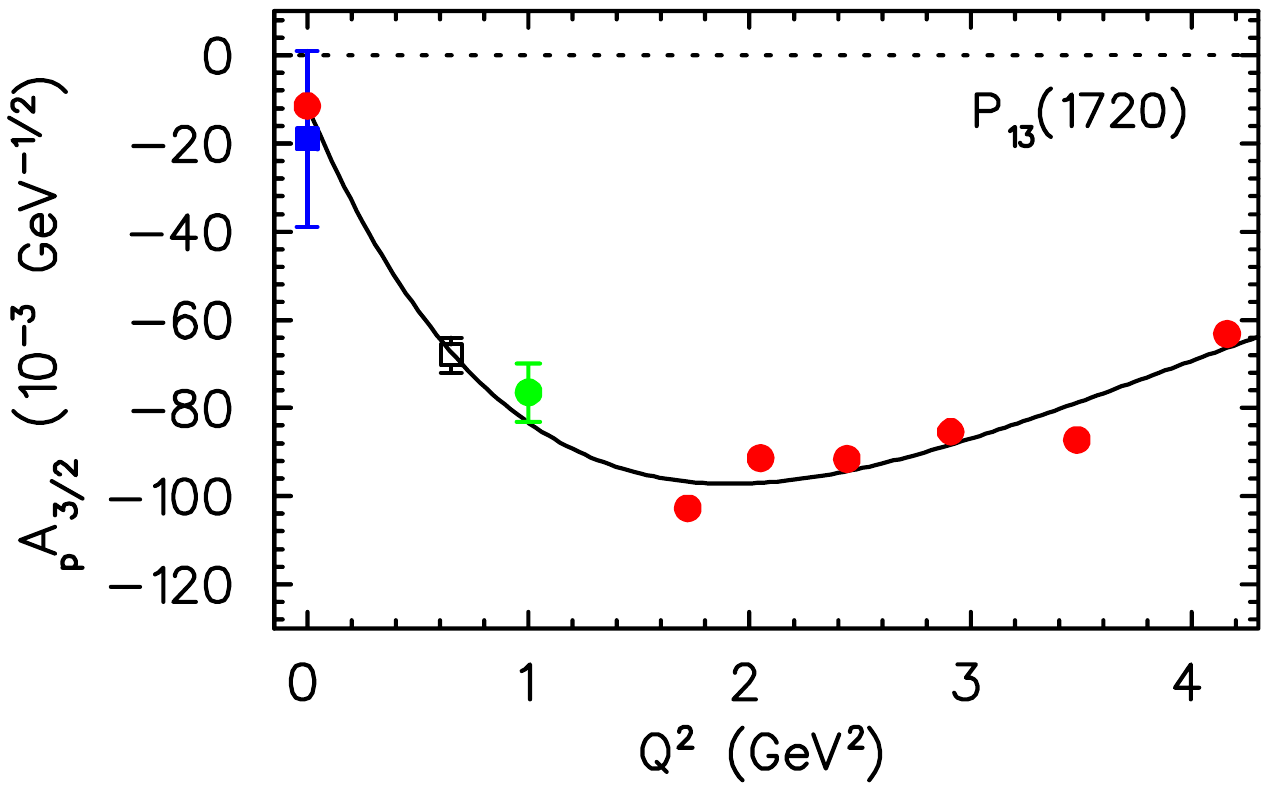}
\vspace{3mm} \caption{\label{fig:d15p13phel} Transverse form factor of the
$D_{15}(1675)$ (left) and $P_{13}(1720)$ (right) resonance. For these two
states the $A_{1/2}$ form factors are practically zero in the MAID analysis.
The curves show the MAID2008 parametrization. Further notation as in
Fig.~\ref{fig:p11phel}.}
\end{center}
\end{figure}

\subsection{Third Resonance Region}

A similar situation as for the $D_{13}$ resonance is obtained for the
$F_{15}(1680)$ shown in Fig.~\ref{fig:d13f15phel} (right panels). However,
there is only one data point at $Q^2=0.65$~GeV$^2$ from the JLab analysis of
$1\pi-2\pi$ analysis~\cite{Aznauryan:2005tp} to be compared which agrees quite
well for the transverse form factors but differs significantly for $S_{1/2}$.
For both of these proton resonances the helicity non-conserving amplitude
$A_{3/2}$ dominates for real photons but, with increasing values of $Q^2$,
drops much faster than the helicity conserving amplitude $A_{1/2}$. As a
consequence the asymmetry
\begin{equation}
{\mathcal {A}}(Q^2)=\frac{\mid A_{1/2} \mid^2 - \mid A_{3/2} \mid^2} {\mid
A_{1/2} \mid^2 + \mid A_{3/2} \mid^2} \label{eq:HelAsym}
\end{equation}
changes rapidly from values close to $-1$ to values near $+1$ over a very small
$Q^2$ range. This behavior has dramatic consequences for $Q^2$-dependent sum
rules. Whereas we found significant differences between the MAID and JLab
analyses for individual form factors, these differences almost disappear in the
asymmetries. For comparison, the asymmetry $\mathcal {A}$ for the
$\Delta(1232)$ resonance takes the value $\approx -0.5$, practically constant
over the plotted $Q^2$ range. This is one more indication for the special role
of the $\Delta$ resonance, which `does not care' about the helicity
conservation required by asymptotic QCD, at least in the currently available
$Q^2$ range.

The resonances $S_{31}(1620)$ and $S_{11}(1650)$ are shown in
Fig.~\ref{fig:s31s11sphel}. The photon couplings of the $S_{31}$ resonance show
a very large spread, see Table~\ref{tab:helicity_p}. Whereas the PDG $A_{1/2}$
value at the photon point is only $(27\pm 11 )\cdot 10^{-3}$ GeV$^{-1/2}$, our
MAID2007 value is 66 in the same units. In the range $Q^2=1-2$~GeV$^2$ we
obtain very small values, and for higher $Q^2$ this amplitude tends towards
zero. For the transverse FF of the $S_{11}(1650)$ resonance we find a solution
close to the shape of the first $S_{11}$ resonance, albeit with a much reduced
overall strength. The analysis of the longitudinal FFs leads to very small
values for both resonances.

Finally, in Fig.~\ref{fig:d15p13phel} we show the situation for the
$D_{15}(1675)$ and $P_{13}(1720)$ re\-sonances, both without a significant
longitudinal coupling. Unlike the situation discussed before, these two
resonances have dominantly helicity 3/2 transitions, whereas the $A_{1/2}$
transition is consistent with zero. As for the $\Delta(1232)$, these are
further examples for which the pQCD prediction for helicity conservation does
not hold in the $Q^2$ region below 5~GeV$^2$. However, for these resonances our
data analysis is at its kinematical limit, we can hardly reach the resonance
positions and therefore only the low-energy tails were analyzed.

\section{Empirical transverse charge transition densities}

In this section we consider the e.m. $N \to N^*$ transition when viewed from a
light front moving towards the baryon. Equivalently, this corresponds to a
frame in which the baryons have large momentum components along the $z$-axis
chosen along the direction of $P = (p + p^\prime)/2$, where $p$ ($p^\prime$)
are the initial (final) baryon four-momenta. We indicate the baryon light-front
``plus'' component by $P^+$ (defining $a^\pm \equiv a^0 \pm a^3$). We can
furthermore choose a symmetric frame in which the virtual photon four-momentum
$q$ has $q^+ = 0$ and a transverse component (in the $xy$-plane) indicated by
the transverse vector $\vec q_\perp$, satisfying $q^2 = - {\vec q_\perp}^{\, 2}
\equiv - Q^2$. In such a symmetric frame, the virtual photon only couples to
forward moving partons and the component $J^+$ of the electromagnetic current
can be interpreted as the quark charge density operator. Considering only $u$
and $d$ quarks, we have $J^+(0) = +2/3 \, \bar u(0) \gamma^+ u(0) - 1/3 \, \bar
d(0) \gamma^+ d(0)$. Each term in this expression is a positive operator since
$\bar q \gamma^+ q \propto | \gamma^+ q |^2$.

We define a transition charge density for the unpolarized $N \to N^\ast$
transition by the Fourier transform
\begin{equation}
\rho_0^{N N^\ast}(\vec b) \equiv \int \frac{d^2 \vec q_\perp}{(2 \pi)^2} \,
e^{- i \, \vec q_\perp \cdot \vec b} \, \frac{1}{2 P^+} \langle P^+, \frac{\vec
q_\perp}{2}, \lambda \,|\, J^+(0) \,|\, P^+, -\frac{\vec q_\perp}{2}, \lambda
\rangle , \label{eq:ndens0}
\end{equation}
where $\lambda$ denotes the nucleon and resonance light-front helicities, $\vec
q_\perp = Q ( \cos \phi_q \hat e_x + \sin \phi_q \hat e_y )$, and the
2-dimensional vector $\vec b$ points to a position in the $xy$-plane.

We will consider the cases of $j=1/2$ resonances, as $P_{11}$ and $S_{11}$, and
of $j=3/2$ resonances, as $P_{33}$ and $D_{13}$. For the spin $1/2$ resonances
we can observe monopole and dipole patterns as for the nucleon, but for the
spin $3/2$ resonances, we also obtain quadrupole patterns.

The above unpolarized transition charge density involves only one of the two or
three independent $N \to N^\ast$ e.m. form factors and leads to the monopole
pattern. To extract the full information of the $N N^*$ transition, we consider
the transition charge densities for transversely polarized nucleons and
resonances.

We denote this transverse polarization direction by $\vec S_\perp = \cos \phi_S
\hat e_x + \sin \phi_S \hat e_y$. The transverse spin state can be expressed in
terms of the light front helicity spinors as $| s_\perp = + \frac{1}{2} \rangle
= \left( | \lambda = + \frac{1}{2} \rangle + e^{i \phi_S } \, | \lambda = -
\frac{1}{2} \rangle \right) / \sqrt{2}$, with $s_\perp$ the nucleon spin
projection along the direction of $\vec S_\perp$.

We can then define a transition charge density for transversely polarized $N$
and $N^\ast$, both along the direction of $\vec S_\perp$ as
\begin{equation}
\rho_T^{N N^\ast}(\vec b) \equiv \int \frac{d^2 \vec q_\perp}{(2 \pi)^2} \,
e^{-i \, \vec q_\perp \cdot \vec b} \, \frac{1}{2 P^+} \langle P^+, \frac{\vec
q_\perp}{2}, s'_\perp \,|\, J^+(0) \,|\, P^+, -\frac{\vec q_\perp}{2}, s_\perp
\rangle .\label{eq:ndens2}
\end{equation}

The nonsymmetric pattern, which describes the deviation from the circular
symmetric unpolarized charge density, depends on the orientation of $\vec b = b
( \cos \phi_b \hat e_x + \sin \phi_b \hat e_y )$. In the following we choose
the transverse spin along the $x$-axis ($\phi_S = 0$).

In general we can write the transition densities in the following way for
$j=1/2$ and $j=3/2$ states:
\begin{equation}
\rho_T^{N N^\ast}(\vec b) = \rho_0^{ N N^\ast}(b) + \sin (\phi_b - \phi_S)
\,\rho_1^{ N N^\ast}(b) + \cos 2(\phi_b - \phi_S) \,\rho_2^{ N N^\ast}(b)\,.
\label{eq:generaldens}
\end{equation}
For a comparison we first consider the transverse charge densities in a
nucleon~\cite{Carlson:2007xd},
$(j^\pi=1/2^+,\, s_\perp = s'_\perp = +\frac{1}{2},\, \rho_2\equiv 0)$,
\begin{eqnarray}
\rho_0^{N}(b) &=& \int_0^\infty \frac{d Q}{2 \pi} Q \, J_0(b \, Q)
F_1^{N}(Q^2)\,,\\
\rho_1^{N}(b) &=& \int_0^\infty \frac{d Q}{2 \pi} Q \, J_1(b \, Q)
\frac{Q}{2M_N}F_2^{N}(Q^2)\,, \label{eq:ndensities}
\end{eqnarray}
where $F_1,F_2$ are the usual Dirac form factors of the nucleon.

For the transverse charge densities in the nucleon to Roper transition~\cite{Tiator2009},\\
$(j^\pi=1/2^+,\, s_\perp = s'_\perp = +\frac{1}{2},\, \rho_2\equiv 0)$, we get
\begin{eqnarray}
\rho_0^{N P_{11}}(b) &=& \int_0^\infty \frac{d Q}{2 \pi} Q \, J_0(b \, Q)
F_1^{N P_{11}}(Q^2)\,,\\
\rho_1^{N P_{11}}(b) &=& \int_0^\infty \frac{d Q}{2 \pi} Q \, J_1(b \, Q)
\frac{Q}{(M_R+M_N)}F_2^{N P_{11}}(Q^2)\,, \label{eq:P11densities}
\end{eqnarray}
where $F_1,F_2$ are related to the helicity transition form factors
$A_{1/2}(Q^2)$ and $S_{1/2}(Q^2)$ in the following way
\begin{eqnarray}
F_1^{N P_{11}}(Q^2) &=& \frac{1}{N_-} \frac{Q^2}{Q_+^2} (A_{1/2}
+\frac{\sqrt{2}(M_R+M_N)}{k_R}S_{1/2})\,,\\
F_2^{N P_{11}}(Q^2) &=& \frac{1}{N_-} \frac{Q^2}{Q_+^2}
(\frac{(M_R+M_N)^2}{Q^2}A_{1/2} -\frac{\sqrt{2}(M_R+M_N)}{k_R}S_{1/2})\,.
\label{eq:P11ff}
\end{eqnarray}

For the transverse charge densities in the nucleon to $S_{11}(1535)$ transition,\\
$(j^\pi=1/2^-,\, s_\perp = -s'_\perp = +\frac{1}{2},\, \rho_2^{N}\equiv 0)$, we get
\begin{eqnarray}
\rho_0^{N S_{11}}(b) &=& \quad\int_0^\infty \frac{d Q}{2 \pi} Q \, J_0(b \, Q)
F_1^{N S_{11}}(Q^2)\,,\\
\rho_1^{N S_{11}}(b) &=& -\int_0^\infty \frac{d Q}{2 \pi} Q \, J_1(b \, Q)
\frac{Q}{(M_R+M_N)}F_2^{N S_{11}}(Q^2)\,, \label{eq:S11densities}
\end{eqnarray}
where $F_1,F_2$ are related to the transition form factors $A_{1/2}(Q^2)$ and
$S_{1/2}(Q^2)$ in the following way
\begin{eqnarray}
F_1^{N S_{11}}(Q^2) &=& \frac{1}{N_+} \frac{Q^2}{Q_-^2} (A_{1/2}
-\frac{\sqrt{2}(M_R-M_N)}{k_R}S_{1/2})\,,\\
F_2^{N S_{11}}(Q^2) &=& \frac{1}{N_+} \frac{Q^2}{Q_-^2}
(\frac{(M_R^2-M_N^2)}{Q^2}A_{1/2} +\frac{\sqrt{2}(M_R+M_N)}{k_R}S_{1/2})\,.
\label{eq:S11ff}
\end{eqnarray}

For the transverse charge densities in the nucleon to $\Delta(1232)P_{33}$ transition~\cite{Carlson:2007xd},
$(j^\pi=3/2^+,\, s_\perp = s'_\perp = +\frac{1}{2})$, we get
\begin{eqnarray}
\rho_0^{N \Delta}(b) &=& \quad\int_0^\infty \frac{d Q}{2 \pi} Q \, J_0(b \, Q)
\frac{1}{2}G^+_{+\frac{1}{2} \, +\frac{1}{2}} (Q^2)\,,\\
\rho_1^{N \Delta}(b) &=& -\int_0^\infty \frac{d Q}{2 \pi} Q \, J_1(b \, Q)
\frac{1}{2}\left[ \sqrt{3} G^+_{+\frac{3}{2} \, +\frac{1}{2}}(Q^2)
+ G^+_{+\frac{1}{2} \, -\frac{1}{2}}(Q^2)\right]\,,\\
\rho_2^{N \Delta}(b) &=& -\int_0^\infty \frac{d Q}{2 \pi} Q \, J_2(b \, Q)
\frac{1}{2}\sqrt{3} \, G^+_{+\frac{3}{2} \, -\frac{1}{2}}(Q^2)\,,
\label{eq:P33densities}
\end{eqnarray}
where
\begin{eqnarray}
G^+_{+\frac{1}{2} \, +\frac{1}{2}} &=& \frac{-1}{4} \, \frac{(M_\Delta + M_N
)}{M_N}
    \frac{Q^2}{Q_+^2} \left\{ G_M^\ast \right. \nonumber \\
&&+G_E^\ast \, \frac{3}{Q_-^2} \left[ (3 M_\Delta + M_N)(M_\Delta - M_N) -Q^2\right] \nonumber \\
&&+\left.\,2\,G_C^\ast \left[ -\frac{(M_\Delta + M_N)}{M_\Delta}+\frac{3\,Q^2}{Q_-^2}\right]\right\}\,,\\
\sqrt{3} \, G^+_{+\frac{3}{2} \, +\frac{1}{2}} + G^+_{+\frac{1}{2} \,
-\frac{1}{2}} &=&
 \frac{(M_\Delta + M_N )}{M_N} \, \frac{Q}{Q_+^2}
 \left\{ G_M^\ast (M_\Delta + M_N) + \, G_C^\ast \, \frac{Q^2}{2 M_\Delta} \right\}, \\
G^+_{+\frac{3}{2} \, -\frac{1}{2}} &=& \frac{\sqrt{3}}{4} \frac{(M_\Delta + M_N
)}{M_N}
\frac{Q^2}{Q_+^2} \left\{ G_M^\ast \right. \nonumber\\
&& \left. +G_E^\ast \left[ 1 - \frac{ 4 M_\Delta (M_\Delta - M_N)}{Q_-^2}
  \right] - \, G_C^\ast \, \frac{2 Q^2}{Q_-^2} \right\}.
\label{eq:P33ff}
\end{eqnarray}

Finally, we obtain for the transverse charge densities in the nucleon to
$D_{13}(1520)$ transition, $(j^\pi=3/2^-,\, s_\perp = -s'_\perp = +\frac{1}{2})$,
\begin{eqnarray}
\rho_0^{N D_{13}}(b) &=& \int_0^\infty \frac{d Q}{2 \pi} Q \, J_0(b \, Q)
\frac{1}{2}G^+_{+\frac{1}{2} \, +\frac{1}{2}} (Q^2)\,,\\
\rho_1^{N D_{13}}(b) &=& \int_0^\infty \frac{d Q}{2 \pi} Q \, J_1(b \, Q)
\frac{1}{2}\left[ \sqrt{3} G^+_{+\frac{3}{2} \, +\frac{1}{2}}(Q^2)
+ G^+_{-\frac{1}{2} \, +\frac{1}{2}}(Q^2) \right]\,,\\
\rho_2^{N D_{13}}(b) &=& \int_0^\infty \frac{d Q}{2 \pi} Q \, J_2(b \, Q)
\frac{1}{2}\sqrt{3} \, G^+_{-\frac{3}{2} \, +\frac{1}{2}}(Q^2)\,,
\label{eq:D13densities}
\end{eqnarray}
where
\begin{eqnarray}
G^+_{+\frac{1}{2} \, +\frac{1}{2}} &=& \frac{1}{\sqrt{6}} \, \frac{Q^2}{Q_+^2}
\left\{ - F_1^{N D_{13}} - \frac{1}{2} F_2^{N D_{13}} + \frac{(M_R + M_N)}{M_R}
F_3^{N D_{13}}  \right\}\,,
\\
G^+_{+\frac{3}{2} \, +\frac{1}{2}} &=& \frac{1}{\sqrt{2}} \, \frac{Q
M_R}{Q_+^2} \left\{ F_1^{N D_{13}} + \frac{(M_R + M_N)}{2 M_R} F_2^{N D_{13}}
\right\},
 \\
G^+_{-\frac{1}{2} \, +\frac{1}{2}} &=& \frac{1}{\sqrt{6}} \, \frac{Q
M_N}{Q_+^2} \left\{ - F_1^{N D_{13}} - \frac{(M_R + M_N)}{2 M_N} F_2^{N D_{13}}
- \frac{Q^2}{M_R M_N} F_3^{N D_{13}}  \right\}\,,
 \\
G^+_{-\frac{3}{2} \, +\frac{1}{2}} &=& \frac{1}{2 \sqrt{2}} \,
\frac{Q^2}{Q_+^2} \, F_2^{N D_{13}}\,, \label{eq:D13ff}
\end{eqnarray}
and
\begin{eqnarray}
F_1^{N D_{13}}  &=& \frac{\sqrt{2}}{N_+} \frac{Q_+^2}{Q_-^2}
\left\{  A_{3/2} - \sqrt{3} A_{1/2}  \right\},  \\
F_2^{N D_{13}}  &=& \frac{\sqrt{2}}{N_+} \frac{1}{Q_-^2}
\left\{ (M_R^2 - M_N^2 - Q^2)  \left[ A_{3/2} + \sqrt{3} A_{1/2} \right]\right.\nonumber \\
&&-\left. \sqrt{\frac{3}{2}}\, \frac{4\, M_R\, Q^2}{k_R} S_{1/2} \right\} - F_1^{N D_{13}},  \\
F_3^{N D_{13}}  &=& -\frac{\sqrt{2}}{N_+} \frac{2 M_R^2}{Q_-^2} \left\{ A_{3/2}
+ \sqrt{3} A_{1/2} + \frac{\sqrt{6}}{2M_R k_R} (M_R^2 - M_N^2 - Q^2) S_{1/2}
\right\}. \label{eq:D13ffhel}
\end{eqnarray}
Throughout these definitions we have used the abbreviations
\begin{eqnarray}
Q_\pm^2 &=& (M_R \pm M_N)^2 + Q^2\,, \\
N_\pm   &=& \sqrt{\frac{\pi\,\alpha\, Q_\pm^2}{M_R\, M_N\, \kappa_R}}\,,
\end{eqnarray}
and $k_R=Q_+ Q_-/(2M_R)$ for the $Q^2$-dependent virtual photon momentum at
$W=M_R$, which is equal to the equivalent photon energy $\kappa_R$ at $Q^2=0$.

\begin{figure}[htbp]
\begin{center}
\includegraphics[width=6.0cm]{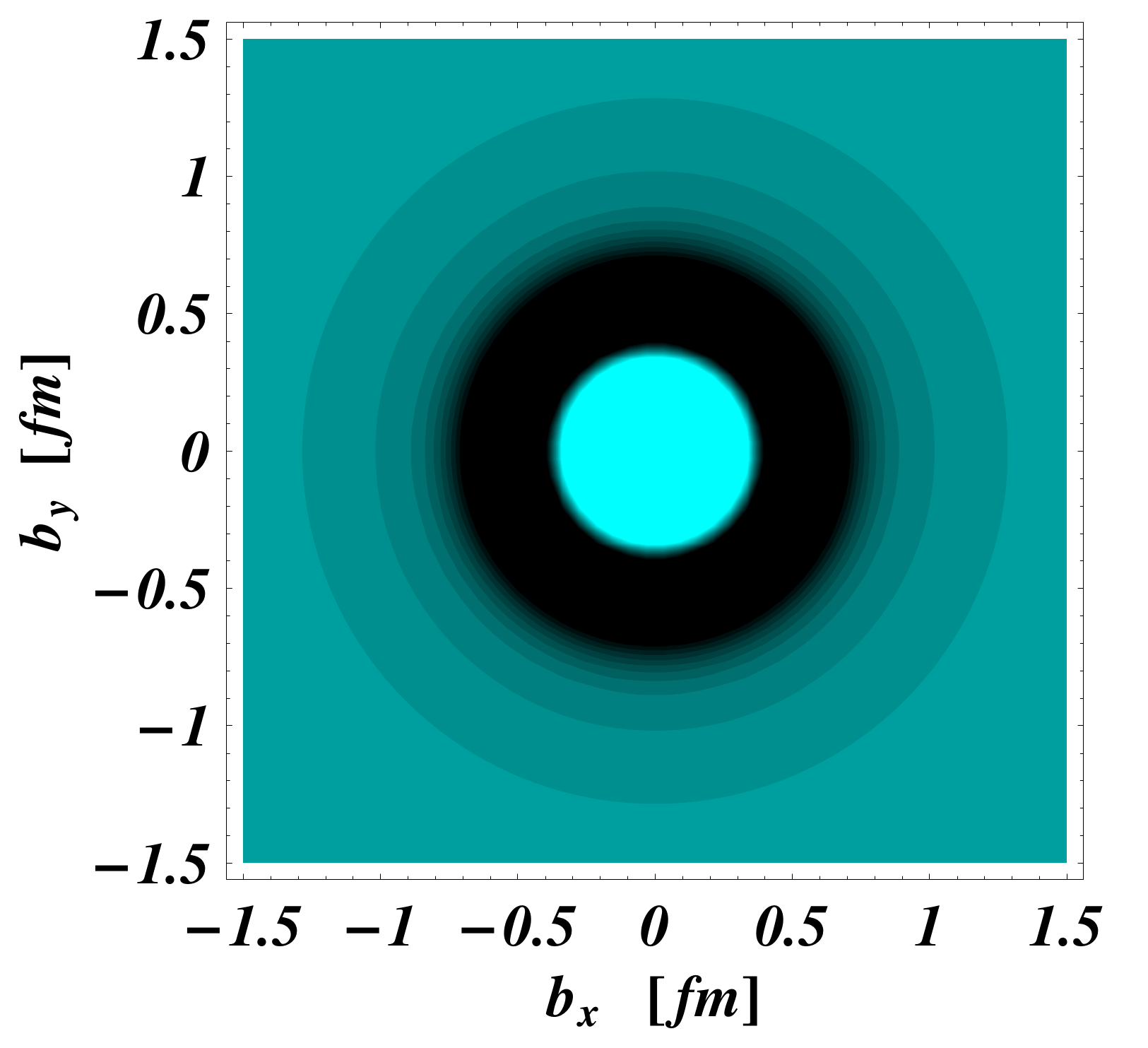}
\hspace{0.3cm}
\includegraphics[width=6.0cm]{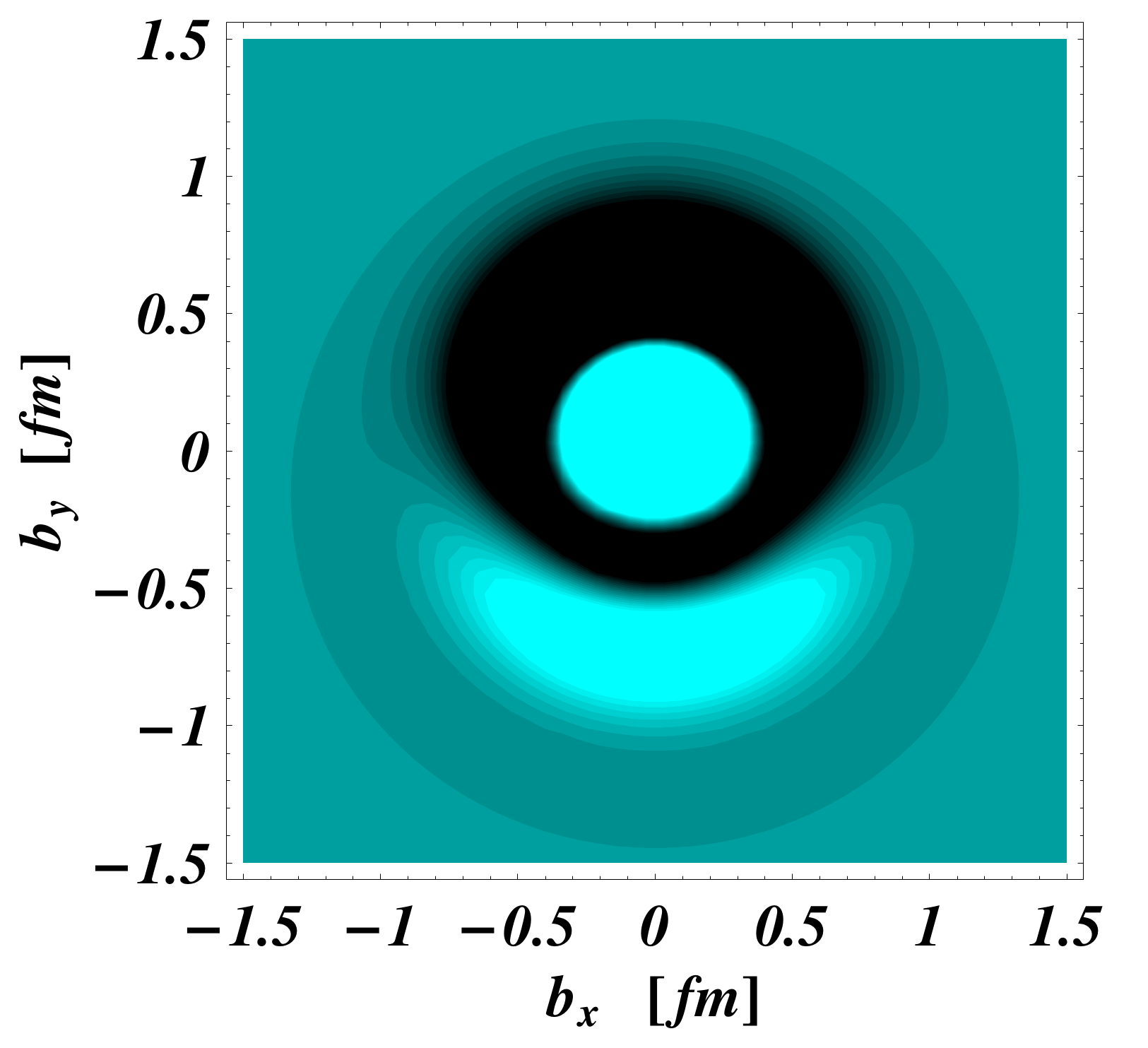}
\vspace{3mm} \caption{\label{fig:p11dens1p} Quark transverse charge density
corresponding to the $p \to P_{11}(1440)$ e.m. transition. Left panel: $p$ and
$P_{11}$ are unpolarized ($\rho_0^{p P_{11}}$). Right panel: $p$ and $P_{11}$
are polarized along the $x$-axis ($\rho_T^{p P_{11}}$). The light (dark)
regions correspond to positive (negative) densities. For the $p \to
P_{11}(1440)$ e.m. transition FFs, we use the MAID2008 parametrization.}
\end{center}
\begin{center}
\includegraphics[width=6.0cm]{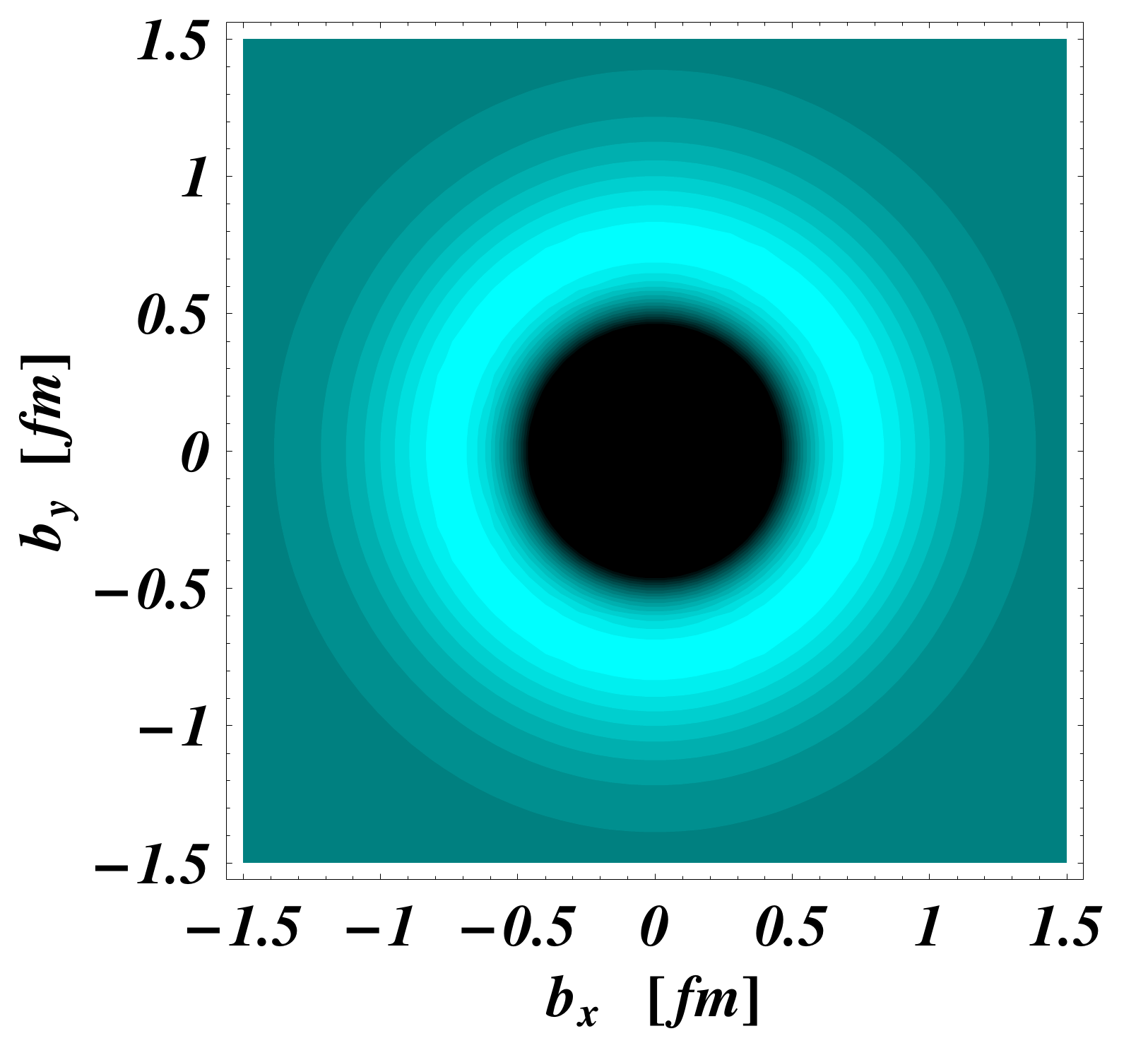}
\hspace{0.3cm}
\includegraphics[width=6.0cm]{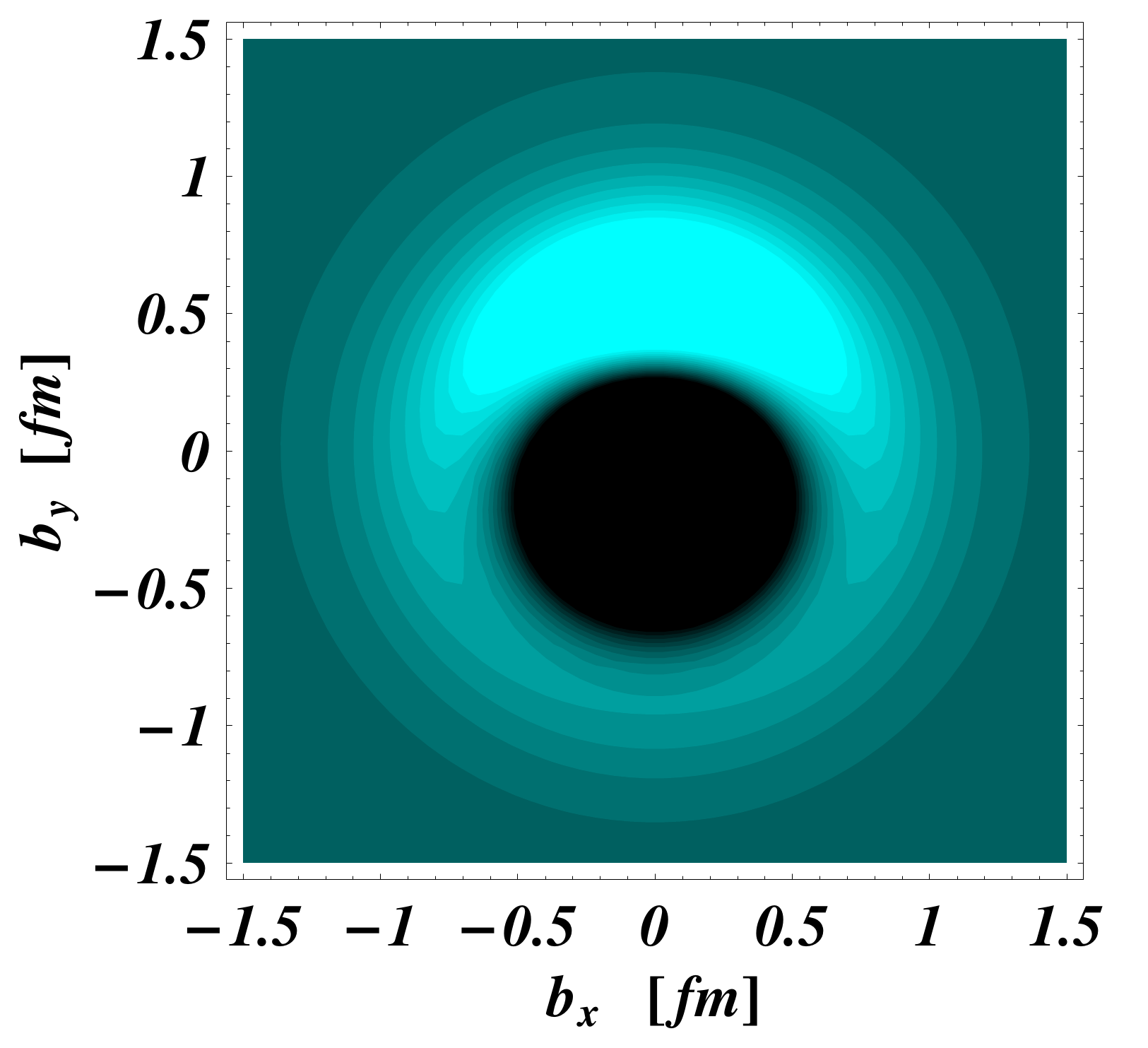}
\vspace{3mm} \caption{\label{fig:p11dens1n}Quark transverse charge density
corresponding to the $n \to P_{11}(1440)$ e.m. transition. Left panel: $n$ and
$P_{11}$ are unpolarized ($\rho_0^{n P_{11}}$). Right panel: $n$ and $P_{11}$
are polarized along the $x$-axis ($\rho_T^{n P_{11}}$). The light (dark)
regions correspond to positive (negative) densities. For the $n \to
P_{11}(1440)$ e.m. transition FFs, we use the MAID2007 parametrization.}
\end{center}
\end{figure}

In Figs.~\ref{fig:p11dens1p} and \ref{fig:p11dens1n} we map the results for the
$N \to P_{11}(1440)$ transition charge densities for protons and neutrons,
respectively. The left panels show the unpolarized case, the right panels are
obtained for transverse polarization of the nucleon and the Roper. For the
transition on a proton, which is well constrained by the data, we use the
MAID2008 parametrization and find an inner region of positive quark charge
concentrated within 0.4~fm, accompanied by a relatively broad band of negative
charge extending out to about 0.8~fm.

For transversely polarized baryons, the large magnetic transition strength at
the real photon point yields a sizeable shift of the charge distribution which
induces an electric dipole moment. For the neutron, which is not so very well
constrained by the data, the MAID2007 analysis yields charge distributions of
opposite sign compared to the proton, with the active quarks spreading out over
even larger spatial distances.

\begin{figure}[ht]
\begin{center}
\includegraphics[width=6.0cm]{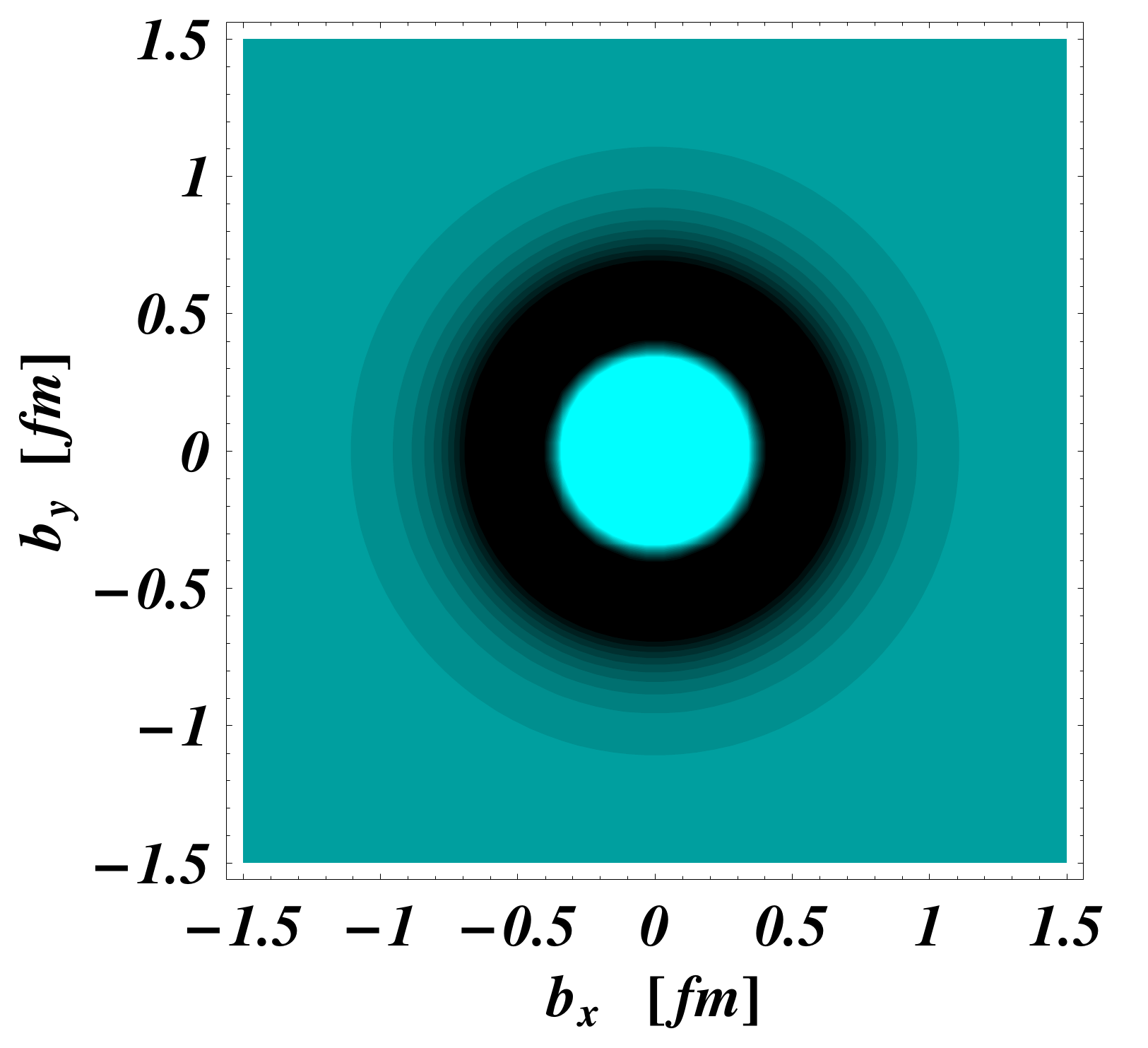}
\hspace{0.3cm}
\includegraphics[width=6.0cm]{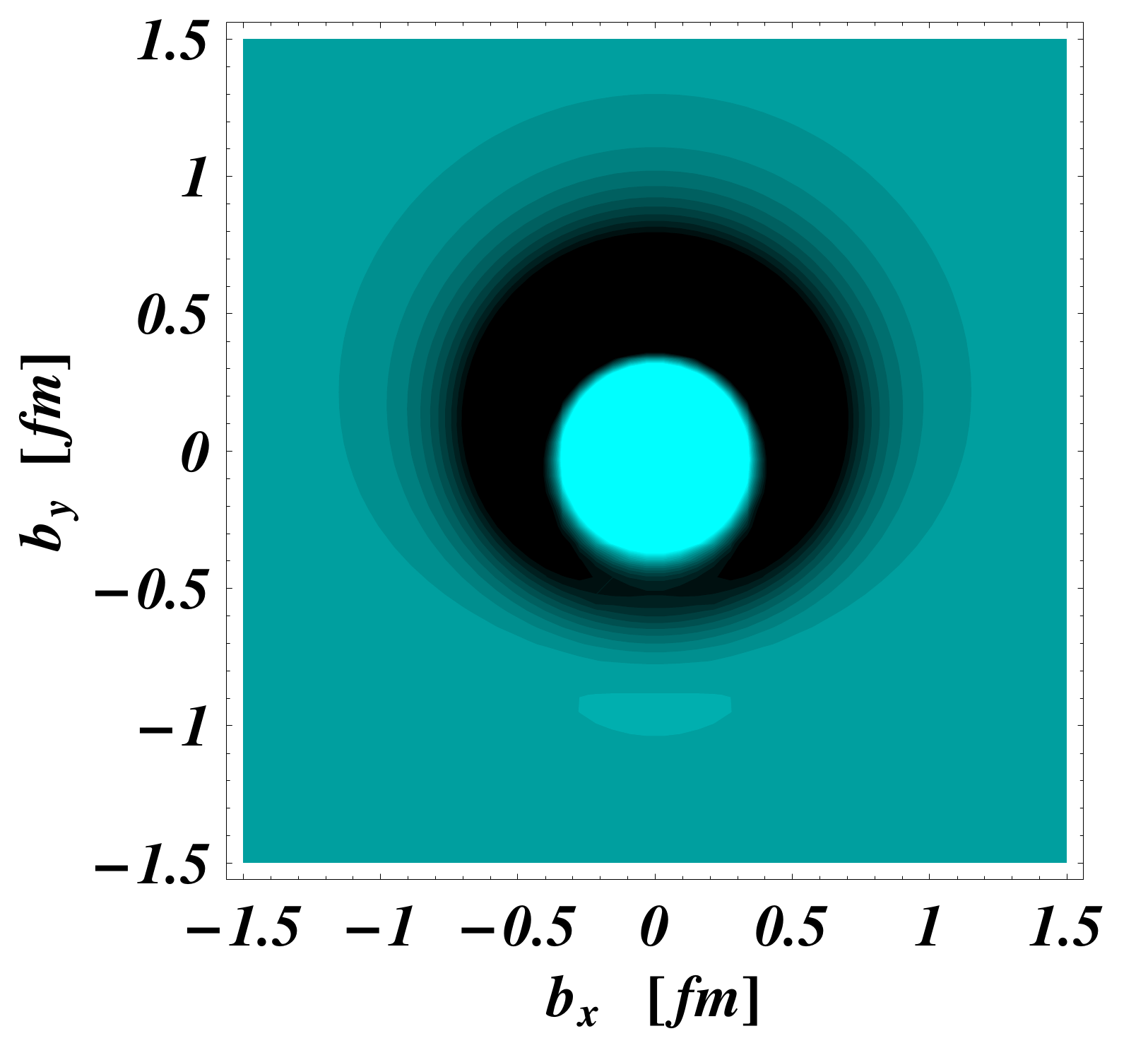}
\vspace{3mm} \caption{\label{fig:s11dens1p} Quark transverse charge density
corresponding to the $p \to S_{11}(1535)$ e.m. transition. Left panel: $p$ and
$S_{11}$ are in a light-front helicity +1/2 state ($\rho_0^{p S_{11}}$). Right
panel: $p$ and $S_{11}$ are polarized along the $x$-axis with opposite spin
projections ($\rho_T^{p S_{11}}$), i.e., $s_\perp=-s'_\perp=+1/2$. The light
(dark) regions correspond to positive (negative) densities. For the $p \to
S_{11}(1535)$ e.m. transition FFs, we use the MAID2007 parametrization. }
\end{center}
\end{figure}

Figure~\ref{fig:s11dens1p} shows the unpolarized and polarized transition
charge densities from the proton to the $S_{11}(1535)$ resonance. Comparing
these results to the corresponding Fig.~\ref{fig:p11dens1p} for the Roper, we
find that the u and d quarks are similarly distributed in the unpolarized
densities but the dipole contribution to the polarized densites is much less
pronounced for the $S_{11}$ due to the much smaller $F_2^{NN^*}/F_1^{NN^*}$
form factor ratio.

\begin{figure}[ht]
\begin{center}
\includegraphics[width=6.0cm]{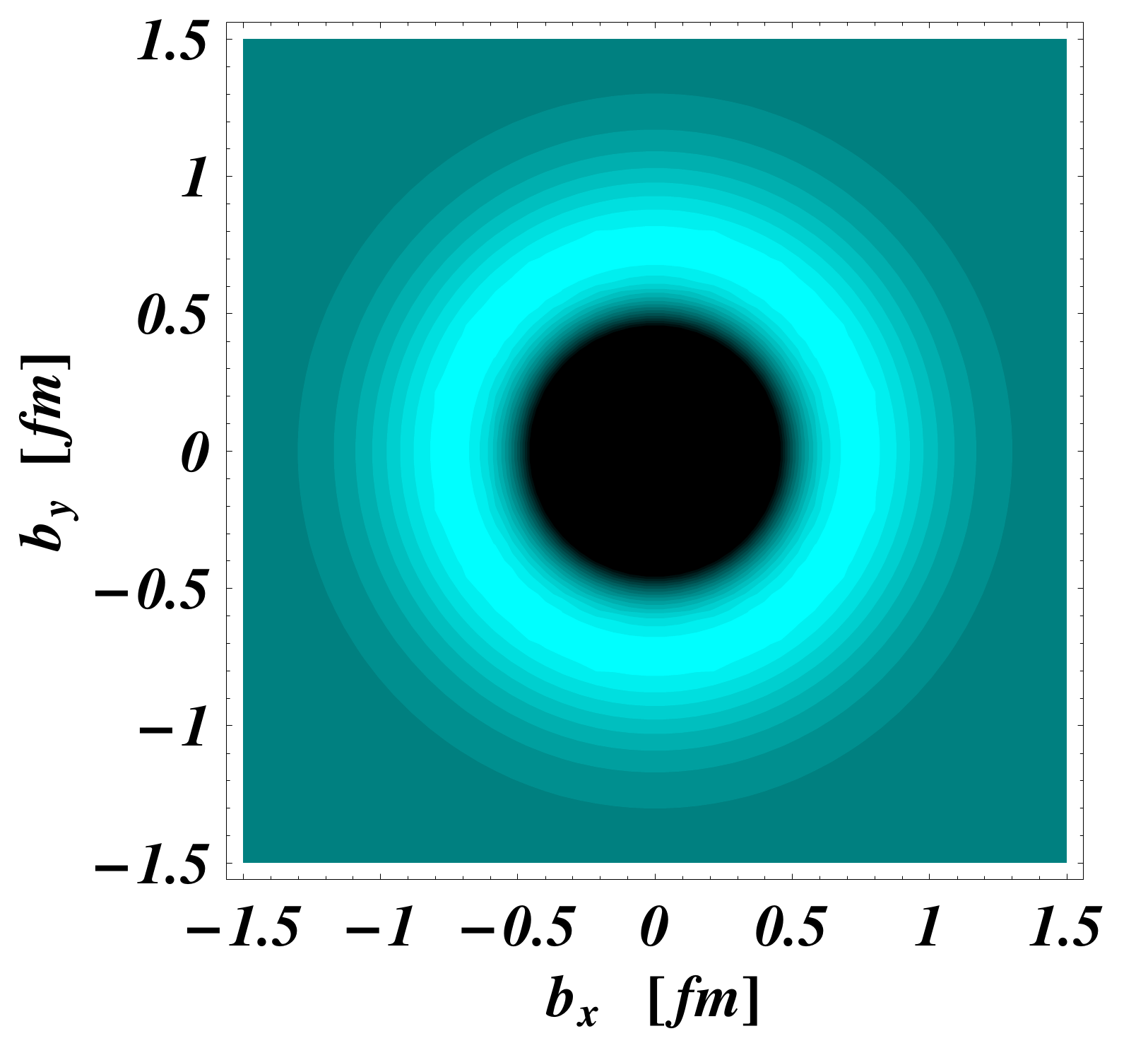}
\hspace{0.3cm}
\includegraphics[width=6.0cm]{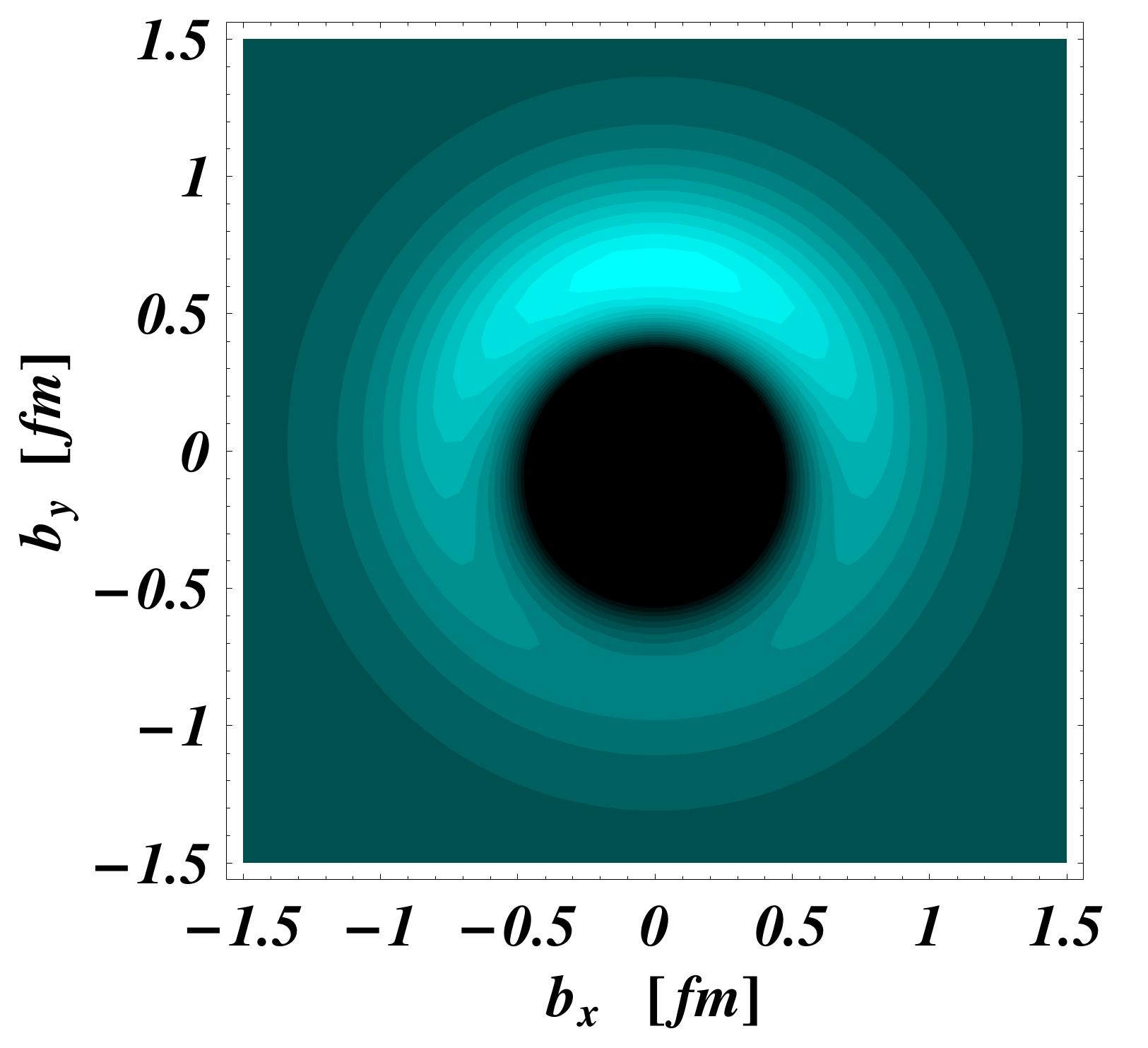}
\vspace{3mm} \caption{\label{fig:s11dens1n} Quark transverse charge density
corresponding to the $n \to S_{11}(1535)$ e.m. transition. Notation as in
Fig.~\ref{fig:s11dens1p}. }
\end{center}
\end{figure}

\begin{figure}
\begin{center}
\includegraphics[width=6.0cm]{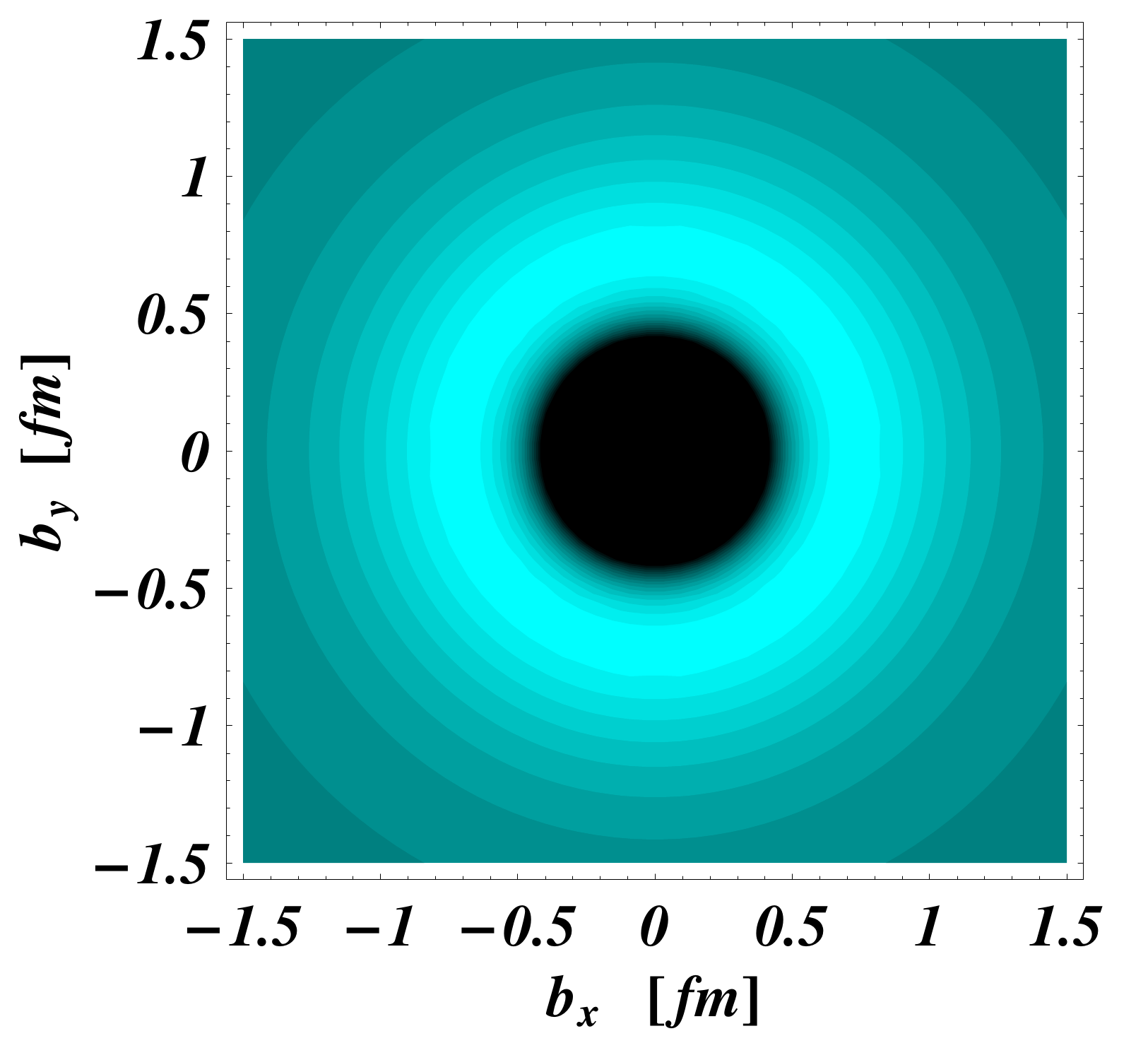}
\hspace{0.3cm}
\includegraphics[width=6.0cm]{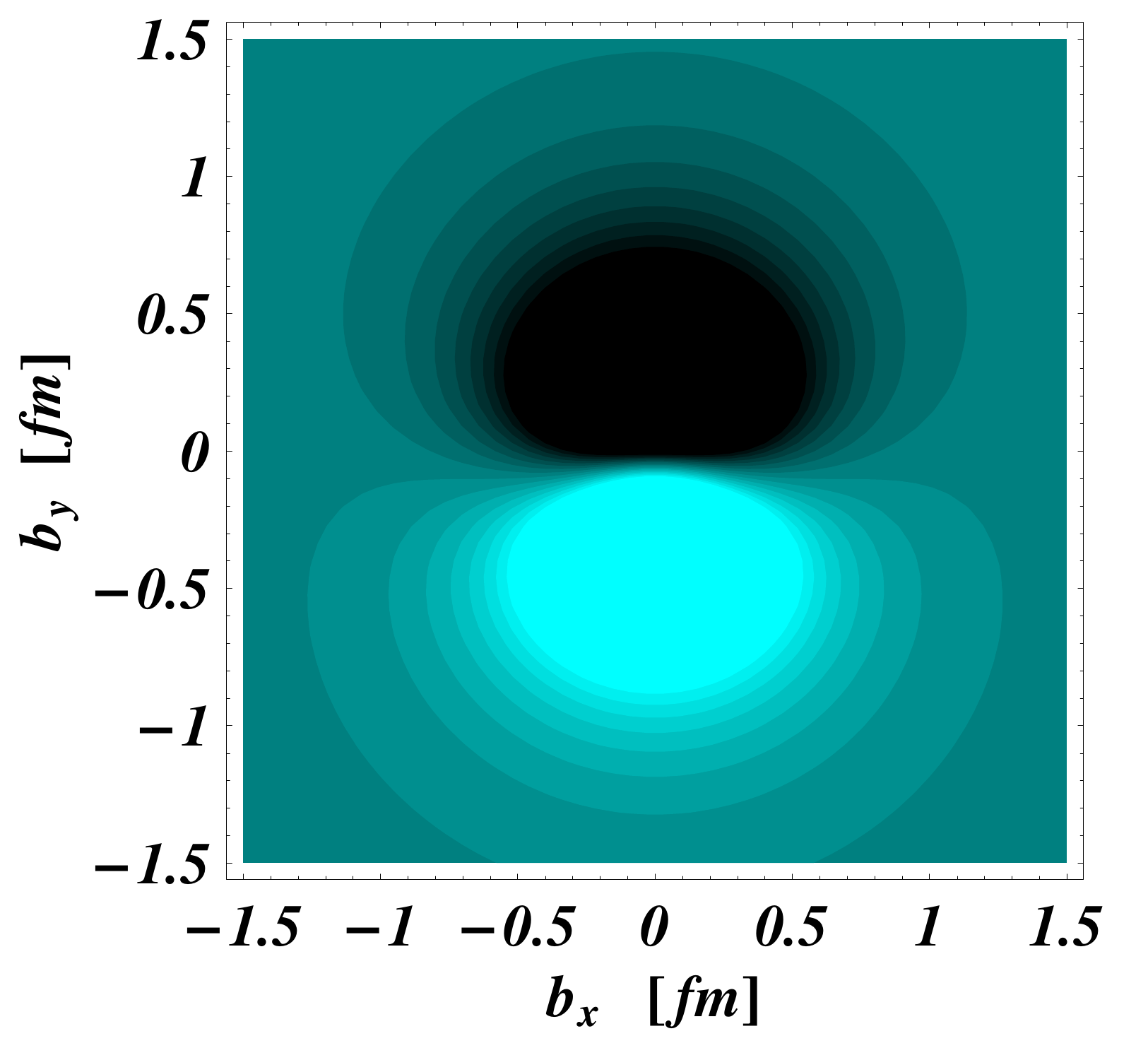}
\includegraphics[width=6.0cm]{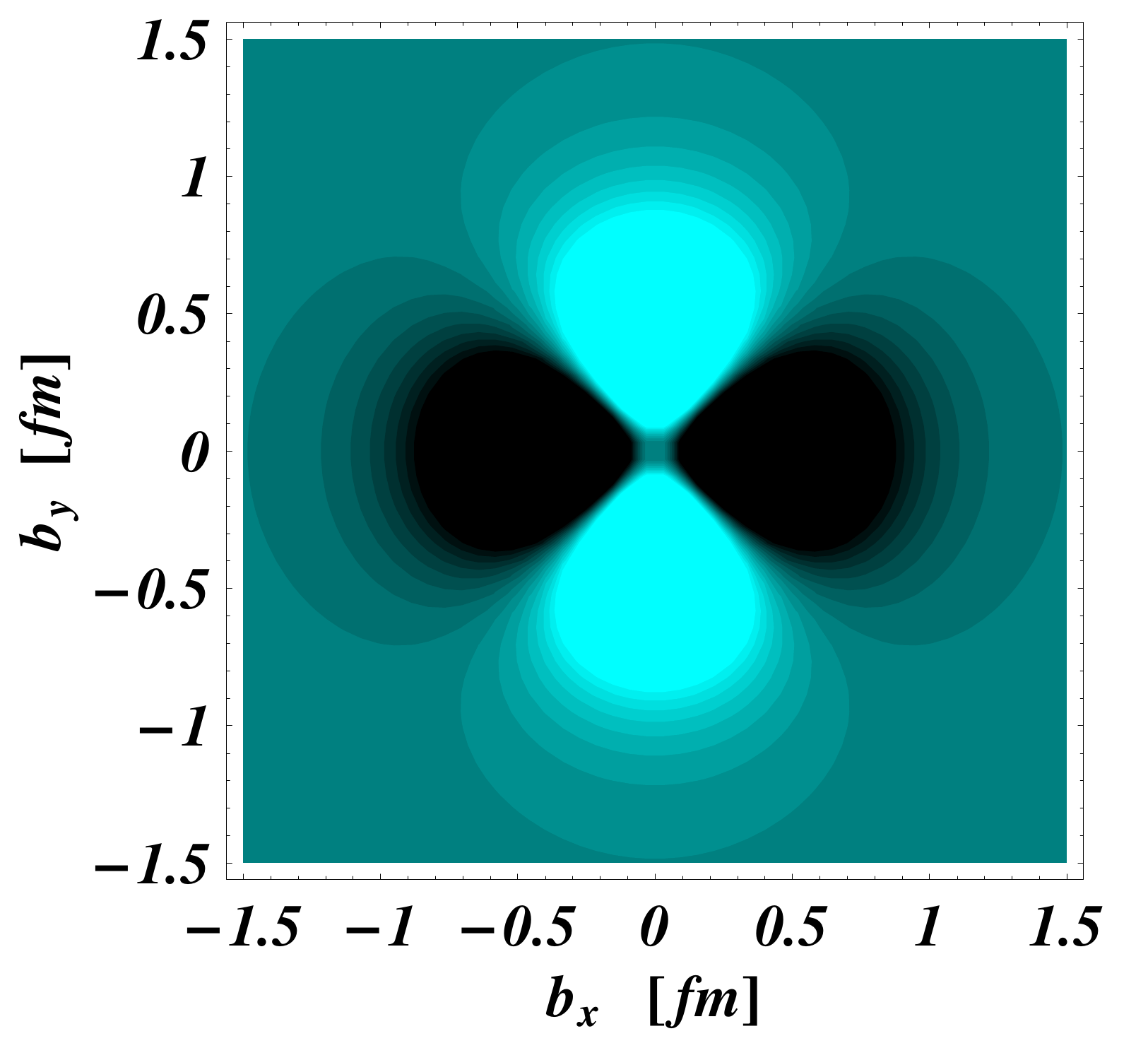}
\vspace{3mm} \caption{\label{fig:p33dens1p} Quark transverse charge density
corresponding to the $p \to \Delta(1232)P_{33}$ e.m. transition. Upper left
panel: p and $\Delta$ are in a light-front helicity +1/2 state ($\rho_0^{p
P_{33}}$). Upper right panel: $p$ and $\Delta$ are polarized along the $x$-axis
($\rho_T^{p P_{33}}$) as in Fig.~\ref{fig:p11dens1p}. The lower panel shows the
quadrupole pattern, whose contribution to the polarized transition density is
very small due to the weak $E2/C2$ admixtures in the $N\Delta$ transition and
practically invisible in the upper right panel. The light (dark) regions
correspond to positive (negative) densities. For the $p \to P_{33}(1232)$ e.m.
transition FFs, we use the MAID2007 parametrization.}
\end{center}
\end{figure}

In the last two figures, we show the transition densities to the $j=3/2$
resonances $\Delta(1232)$ and $D_{13}(1520)$. For both transitions we obtain
three multipole patterns, monopole, dipole, and quadrupole, which add to the
polarized transition density, according Eq.~(\ref{eq:generaldens}). For the $p
\rightarrow \Delta(1232)$ transition in Fig.~\ref{fig:p33dens1p} we show the
unpolarized and polarized transition densities as for the previous cases. In
addition we also present the quadrupole pattern alone, which has a very small
magnitude due to the suppressed $E2,C2$ admixtures in the $N\Delta$ transition.
Therefore it cannot be seen in the combined polarized density.

\begin{figure}
\begin{center}
\includegraphics[width=6.0cm]{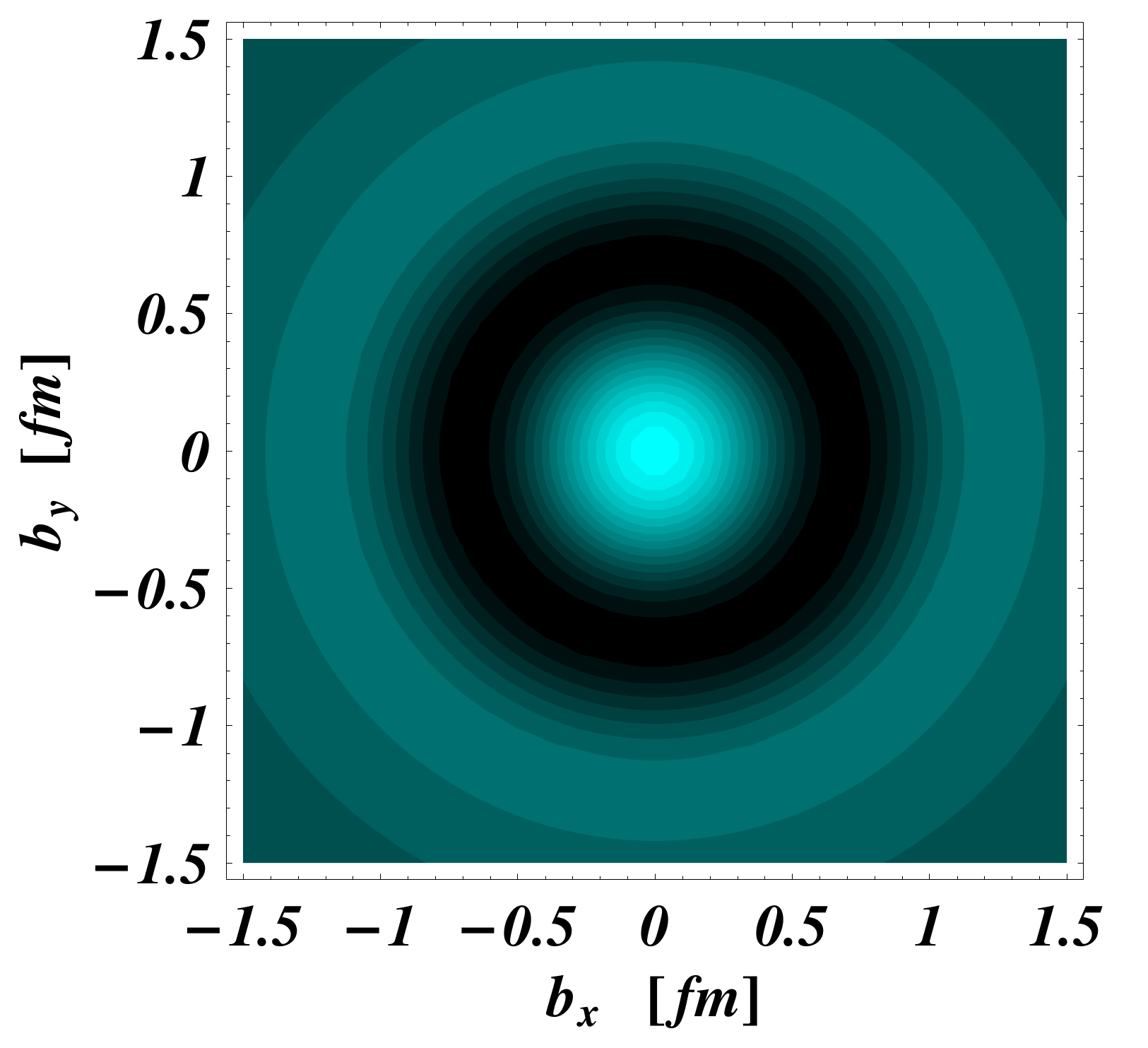}
\hspace{0.3cm}
\includegraphics[width=6.0cm]{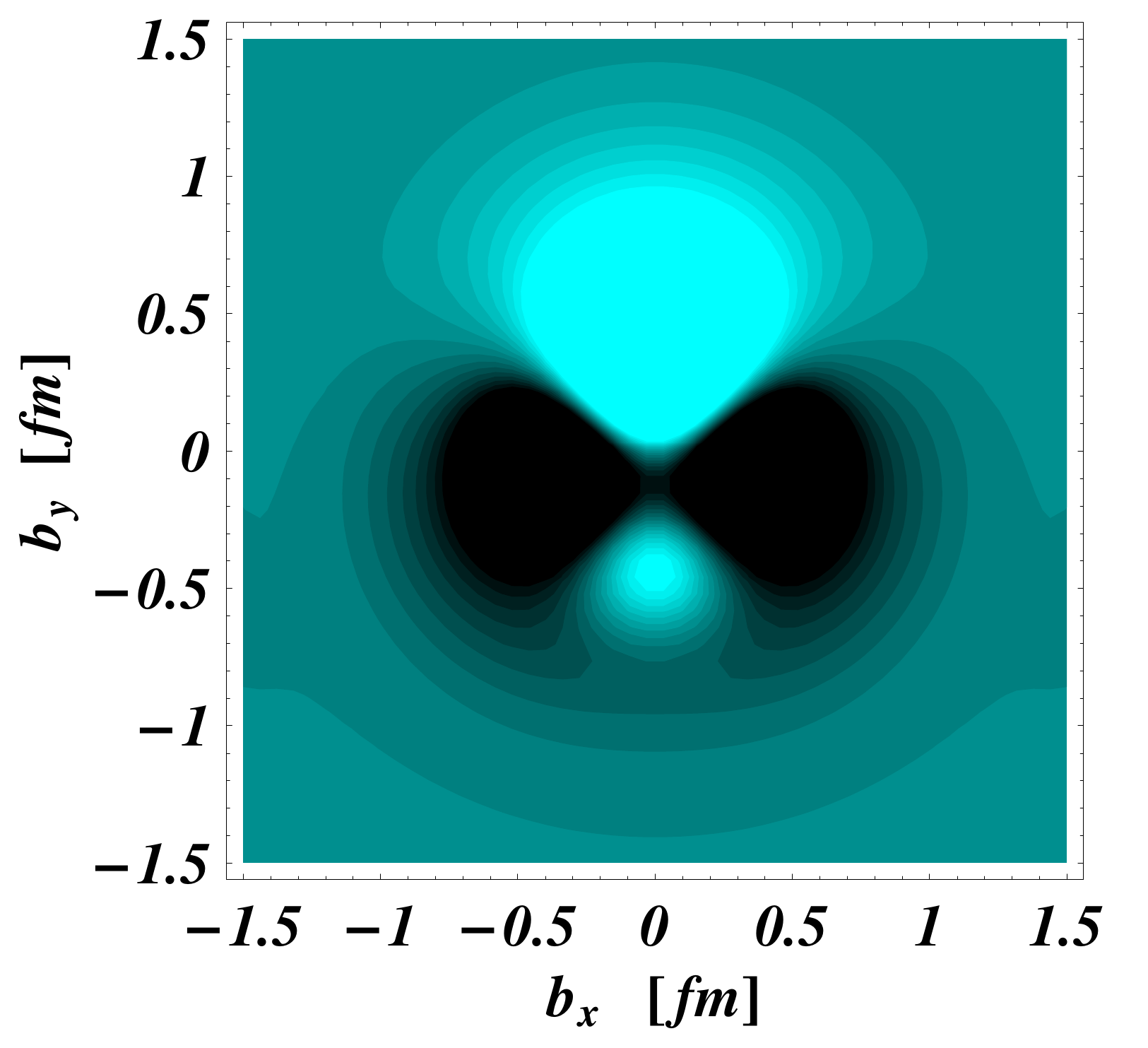}
\vspace{3mm} \caption{\label{fig:d13dens1p} Quark transverse charge density
corresponding to the $p \to D_{13}(1520)$ e.m. transition. Left panel: $p$ and
$D_{13}$ are in a light-front helicity +1/2 state ($\rho_0^{p D_{13}}$). Right
panel: $p$ and $D_{13}$ are polarized along the $x$-axis with spin projections
($\rho_T^{p D_{13}}$) as in Fig.~\ref{fig:s11dens1p}. The light (dark) regions
correspond to positive (negative) densities. For the $p \to D_{13}(1520)$ e.m.
transition FFs, we use the MAID2008 parametrization.}
\end{center}
\end{figure}

Finally, for the $p \rightarrow D_{13}(1520)$ transition in
Fig.~\ref{fig:d13dens1p} we show again the unpolarized density and the
polarized density containing all 3 patterns. The unpolarized or monopole
density of $p \rightarrow D_{13}$ shows a more concentrated central positive
charge for $b< 0.3$~fm and a broad ring of negative charges up to $b\approx
1$~fm. Compared to the Roper and $S_{11}$ monopole densities the boundaries
between u and d quarks are more diffuse for the $D_{13}(1520)$. In contrast to
the $N\Delta$ transition, here the quadrupole pattern is much more dominant and
shows up clearly in the polarized density, where also the dipole pattern has a
significant influence.

\section{Summary and Conclusions}

For a complete understanding of the nucleon structure it is prerequisite to
know both the ground state form factors and the excitation spectrum, in
particular, the transition form factors from the ground state to all the
excited states. Among all $N^*$ and $\Delta$ resonances listed in PDG, the
four-star nucleon resonances are the relevant states that determine the
response of the nucleon to electromagnetic probes as photons and electrons.

Using the world database of pion photo- and electroproduction as well as recent
data from Mainz, Bonn, Bates, and JLab we have extracted all longitudinal and
transverse helicity amplitudes of nucleon resonance excitation for the
four-star resonances below $W=1.8$~GeV. For this purpose we have extended our
unitary isobar model MAID, in particular, by parameterizing  the $Q^2$
dependence of the transition amplitudes in more detail. The comparison between
single-$Q^2$ fits and overall $Q^2$-dependent fits gives us confidence in the
determination of the transverse and longitudinal heli\-ci\-ty form factors for
the resonances $P_{33}(1232),\, P_{11}(1440),\, S_{11}(1535),\, D_{13}(1520)$,
and $F_{15}(1680)$, even though the model uncertainty of the longitudinal
amplitudes can be as large as 50\% for the $D_{13}$ and $F_{15}$. Similar model
uncertainties can be expected for transverse transition form factors of the
resonances  $S_{31}(1620)$, $S_{11}(1650)$, $D_{15}(1675)$, and $P_{31}(1720)$,
which we analyzed for the first time.

Sufficient reliable form factors exist for the excitation of the resonances
$P_{33}(1232)$, $P_{11}(1440)$, $S_{11}(1535)$, and $D_{13}(1520)$, from which
we extract the quark transverse charge densities inducing these transitions.
The rings of up and down quarks in these two-dimensional representations show
distinctly different angular (multipole) and radial structures which
characterize the individual nucleon resonances in a rather lucid way.

For further improvement a model-independent partial wave analysis for pion, eta
and kaon photoproduction is currently in preparation. First data using beam,
target and recoil polarization in various combinations are already taken.
Similar to the analysis of the $P_{33}$ partial waves that allowed us to get
model-independent $\Delta(1232)$ resonance parameters, we will obtain values
for all transverse form factors at $Q^2=0$ with high statistical precision and
small model dependence. For longitudinal form factors we can not get
experimental information at $Q^2=0$, therefore it is important to perform
dedicated experiments in $(e,e'\pi)$ and $(e,e'\eta)$ for very low $Q^2$. At
MAMI such experiments are proposed for the next few years, where longitudinal
transition form factors will be studied for $Q^2\approx0.05$~GeV$^2$. Without
this knowledge the longitudinal form factors will always remain very uncertain
in the low-$Q^2$ region. At large $Q^2$ both longitudinal and transverse form
factors will be further investigated at JLab, where photon virtualities of
$Q^2$ up to 12~GeV$^2$ will soon be available.

\begin{acknowledgement}
We want to thank the Deutsche Forschungsgemeinschaft for the support by the
Collaborative Research Center 443 (SFB 443).
\end{acknowledgement}


\begin{thebibliography}{}
\bibitem{PDG10} K. Nakamura et al. (Particle Data Group),
   J. Phys. G \textbf{37} (2010) 075021.

\bibitem{Chen:2007cy}
  G.~Y.~Chen, S.~S.~Kamalov, S.~N.~Yang, D.~Drechsel, L.~Tiator,
  Phys.\ Rev.\  {\bf C76 } (2007)  035206.

\bibitem{Tiator:2010rp}
  L.~Tiator, S.~S.~Kamalov, S.~Ceci, G.~Y.~Chen, D.~Drechsel, A.~Svarc, S.~N.~Yang,
  Phys.\ Rev.\  {\bf C82 } (2010)  055203.

\bibitem{Devenish:1975jd}
  R.~C.~E.~Devenish, T.~S.~Eisenschitz, J.~G.~K\"orner,
  Phys.\ Rev.\  {\bf D14 } (1976)  3063.

\bibitem{Ash67} W.~W. Ash et al., Phys. Lett. B {\bf 24} (1967) 165.

\bibitem{Bartel} W. Bartel {et al.}, Phys. Lett. {\bf 28B} (1968) 148

\bibitem{Baetzner}
  K.~Baetzner, U.~Beck, K.~H.~Becks, C.~Berger, J.~Drees, G.~Knop, M.~Leenen, K.~Moser et al.,
  Phys.\ Lett.\  {\bf B39 } (1972)  575-578.

\bibitem{Alder:1972di}
  J.~C.~Alder, F.~W.~Brasse, E.~Chazelas, W.~Fehrenbach, W.~Flauger, K.~H.~Frank, E.~Ganssauge, J.~Gayler et al.,
  Nucl.\ Phys.\  {\bf B46 } (1972)  573-592.

\bibitem{Stein} S. Stein et al., Phys. Rev. {\bf D12} (1975) 1884.

\bibitem{Jon73} H.~F. Jones and M.~D. Scadron, Annals Phys. {\bf 81} (1973) 1.

\bibitem{Stoler:1993yk}
  P.~Stoler,
  Phys.\ Rept.\  {\bf 226 } (1993)  103-171.

\bibitem{Stuart:1996zs}
  L.~M.~Stuart, P.~E.~Bosted, L.~Andivahis, A.~Lung, J.~Alster, R.~G.~Arnold, C.~C.~Chang, F.~S.~Dietrich et al.,
  Phys.\ Rev.\  {\bf D58 } (1998)  032003.

\bibitem{Pos01} Th.~Pospischil {et al.}, Phys. Rev. Lett. {\bf 86} (2001) 2959.

\bibitem{Elsner06} D.~Elsner {et al.}, Eur. Phys. J. A {\bf 27} (2006) 91.

\bibitem{Stave06} S.~Stave {et al.}, Eur. Phys. J. A {\bf 30} (2006) 471.

\bibitem{Got00} R.~W.~Gothe, Prog. Part. Nucl. Phys. {\bf 44} (2000) 185 and
 T.~Bantes and R.~W.~Gothe, private communication.

\bibitem{Ban03} T.~Bantes, PhD thesis, Bonn 2003.

\bibitem{Mer01} C.~Mertz {et al.}, Phys. Rev. Lett. {\bf 86} (2001) 2963.

\bibitem{Sparveris:2004jn}
  N.~F.~Sparveris et al. [ OOPS Collaboration ],
  Phys.\ Rev.\ Lett.\  {\bf 94 } (2005)  022003.

\bibitem{Beck97}
  R.~Beck, H.~P.~Krahn, J.~Ahrens, H.~J.~Arends, G.~Audit, A.~Braghieri, N.~d'Hose, S.~J.~Hall et al.,
  Phys.\ Rev.\ Lett.\  {\bf 78 } (1997)  606-609.

\bibitem{Blanpied:1997zz}
  G.~Blanpied, M.~Blecher, A.~Caracappa, C.~Djalali, G.~Giordano, K.~Hicks, S.~Hoblit, M.~Khandaker et al.,
  Phys.\ Rev.\ Lett.\  {\bf 79 } (1997)  4337-4340.

\bibitem{HDT97} O. Hanstein, D. Drechsel, and L. Tiator,
  Phys. Lett. {\bf B} 399 (1997) 13.

\bibitem{HDT98} O. Hanstein, D. Drechsel, and L. Tiator,
  Nucl. Phys. {\bf A632} (1998) 561.

\bibitem{Beck00}
  R.~Beck et al.,
  Phys.\ Rev.\  C {\bf 61} (2000) 035204.

\bibitem{Fro99} V.~V.~Frolov {et al.}, Phys. Rev. Lett. {\bf 82} (1999) 45.

\bibitem{Lav04} G.~Laveissiere {et al.}, Phys. Rev. C {\bf 69} (2004) 045202
 and Proc. of NSTAR2001, Mainz, World Scientific 2001, p271.

\bibitem{Joo02} K.~Joo {et al.} [ CLAS Collaboration ], Phys. Rev. Lett. {\bf 88} (2002) 122001-1.

\bibitem{Joo03} K.~Joo {et al.} [ CLAS Collaboration ], Phys. Rev. C {\bf 68} (2003) 032201.

\bibitem{Joo04} K.~Joo {et al.} [ CLAS Collaboration ], Phys. Rev. C {\bf 70} (2004) 042201.

\bibitem{Kelly05}J.~J.~Kelly {et al.}, Phys. Rev. Lett. {\bf 95} (2005) 102001
and Phys. Rev. C {\bf 75} (2007), 025201.  

\bibitem{Egi06} H.~Egiyan {et al.} [ CLAS Collaboration ], Phys. Rev. C {\bf 73} (2006) 025204.

\bibitem{Ung06} M. Ungaro et al. [ CLAS Collaboration ], {Phys. Rev. Lett.} \textbf{97} (2006) 112003.

\bibitem{Park:2007tn}
  K.~Park et al. [ CLAS Collaboration ],
  Phys.\ Rev.\  {\bf C77 } (2008)  015208.

\bibitem{Villano:2009sn}
  A.~N.~Villano, P.~Stoler, P.~E.~Bosted, S.~H.~Connell, M.~M.~Dalton, M.~K.~Jones, V.~Kubarovsky, G.~S.~Adams et al.,
  Phys.\ Rev.\  {\bf C80 } (2009)  035203.

\bibitem{Arndt:1990ej}
  R.~A.~Arndt, R.~L.~Workman, Z.~Li, L.~D.~Roper,
  Phys.\ Rev.\  {\bf C42 } (1990)  1864-1866.

\bibitem{Arndt:1995ak}
  R.~A.~Arndt, I.~I.~Strakovsky, R.~L.~Workman,
  Phys.\ Rev.\  {\bf C53 } (1996)  430-440; (SP99 solution of the GW/SAID
  analysis); http://gwdac.phys.gwu.edu/.

\bibitem{Arndt:2006ym}
  R.~A.~Arndt, W.~J.~Briscoe, I.~I.~Strakovsky, R.~L.~Workman,
  AIP Conf.\ Proc.\  {\bf 904 } (2007)  269-275.

\bibitem{Penner:2002md}
  G.~Penner, U.~Mosel,
  Phys.\ Rev.\  {\bf C66 } (2002)  055212.

\bibitem{Shklyar:2006xw}
  V.~Shklyar, H.~Lenske, U.~Mosel,
  Phys.\ Lett.\  {\bf B650 } (2007)  172-178.

\bibitem{Anisovich:2005tf}
  A.~V.~Anisovich, A.~Sarantsev, O.~Bartholomy, E.~Klempt, V.~A.~Nikonov, U.~Thoma,
  Eur.\ Phys.\ J.\  {\bf A25 } (2005)  427-439.

\bibitem{Anisovich:2009zy}
  A.~V.~Anisovich, E.~Klempt, V.~A.~Nikonov, M.~A.~Matveev, A.~V.~Sarantsev, U.~Thoma,
  Eur.\ Phys.\ J.\  {\bf A44 } (2010)  203-220; http://pwa.hiskp.uni-bonn.de/.

\bibitem{Doring:2009yv}
  M.~D\"oring, C.~Hanhart, F.~Huang, S.~Krewald, U.~-G.~Meissner,
  Nucl.\ Phys.\  {\bf A829 } (2009)  170-209.

\bibitem{Doring:2009uc}
  M.~D\"oring, K.~Nakayama,
  Eur.\ Phys.\ J.\  {\bf A43 } (2010)  83-105.

\bibitem{Huang:2010vb}
  F.~Huang, M.~D\"oring, H.~Haberzettl, S.~Krewald, K.~Nakayama,

\bibitem{Djukanovic:2007bw}
  D.~Djukanovic, J.~Gegelia, S.~Scherer,
  Phys.\ Rev.\  {\bf D76 } (2007)  037501.

\bibitem{Capstick:2007tv}
  S.~Capstick, A.~Svarc, L.~Tiator, J.~Gegelia, M.~M.~Giannini, E.~Santopinto, C.~Hanhart, S.~Scherer et al.,
  Eur.\ Phys.\ J.\  {\bf A35 } (2008)  253-266.

\bibitem{Gegelia:2009py}
  J.~Gegelia, S.~Scherer,
  Eur.\ Phys.\ J.\  {\bf A44 } (2010)  425-430.

\bibitem{Hemmert:1997ye}
  T.~R.~Hemmert, B.~R.~Holstein, J.~Kambor,
  J.\ Phys.\ G {\bf G24 } (1998)  1831-1859.

\bibitem{Gail:2005gz}
  T.~A.~Gail, T.~R.~Hemmert,
  Eur.\ Phys.\ J.\  {\bf A28 } (2006)  91-105.

\bibitem{Pascalutsa:2005vq}
  V.~Pascalutsa, M.~Vanderhaeghen,
  Phys.\ Rev.\  {\bf D73 } (2006)  034003.

\bibitem{Pascalutsa:2006up}
  V.~Pascalutsa, M.~Vanderhaeghen, S.~N.~Yang,
  Phys.\ Rept.\  {\bf 437 } (2007)  125-232.

\bibitem{DMT:2001}
  S.~S.~Kamalov, S.~N.~Yang, D.~Drechsel, and L.~Tiator
  Phys.\ Rev.\ C {\bf 64}, (2001) 032201.

\bibitem{KY99} S.~S.~Kamalov, and S.~N.~Yang, Phys. Rev. Lett. {\bf 83}, (1999) 4494.

\bibitem{SL01} T.~Sato and T.-S.~H.~Lee, Phys. Rev. C {\bf 63}, (2001) 055201.

\bibitem{JuliaDiaz:2009ww}
  B.~Julia-Diaz, H.~Kamano, T.-S.~H.~Lee, A.~Matsuyama, T.~Sato, N.~Suzuki,
  Phys.\ Rev.\  {\bf C80 } (2009)  025207.

\bibitem{Suzuki:2010yn}
  N.~Suzuki, T.~Sato, T.-S.~H.~Lee,
  Phys.\ Rev.\  {\bf C82 } (2010)  045206.

\bibitem{Maid98} D. Drechsel, O. Hanstein, S.~S. Kamalov, and L. Tiator,
     {Nucl. Phys. A} \textbf{645} (1999) 145.

\bibitem{MAID07}
  D.~Drechsel, S.~S.~Kamalov, and L.~Tiator,
  Eur.\ Phys.\ J.\  A {\bf 34} (2007) 69;
  http://www.kph.uni-mainz.de/MAID/.

\bibitem{Tiator:2003uu}
  L.~Tiator, D.~Drechsel, S.~S.~Kamalov, M.~M.~Giannini, E.~Santopinto, A.~Vassallo,
  Eur.\ Phys.\ J.\  {\bf A19 } (2004)  55-60.

\bibitem{Tiator:2003xr}
  L.~Tiator, D.~Drechsel, S.~S.~Kamalov, S.~N.~Yang,
  Eur.\ Phys.\ J.\  {\bf A17 } (2003)  357-363.

\bibitem{Tiator:2009mt}
  L.~Tiator, D.~Drechsel, S.~S.~Kamalov, and M.~Vanderhaeghen,
 {CPC(HEP \& NP)} \textbf{33} (2009) 1051.

\bibitem{Aznauryan:2008pe}
  I.~G.~Aznauryan et al. [ CLAS Collaboration ],
  Phys.\ Rev.\  {\bf C78 } (2008)  045209.

\bibitem{Azn09}
  I.~G.~Aznauryan et al. [ CLAS Collaboration ],
  Phys.\ Rev.\  {\bf C80 } (2009)  055203.

\bibitem{Aznauryan:2011ub}
  I.~Aznauryan, V.~D.~Burkert, T.-S.~H.~Lee, V.~Mokeev,
  [arXiv:1102.0597 [nucl-ex]].

\bibitem{Alexandrou:2010uk}
  C.~Alexandrou, G.~Koutsou, J.~W.~Negele, Y.~Proestos, A.~Tsapalis,
  Phys.\ Rev.\  {\bf D83 } (2011)  014501.

\bibitem{Alexandrou:2009hs}
  C.~Alexandrou, T.~Korzec, G.~Koutsou, C.~Lorce, J.~W.~Negele, V.~Pascalutsa, A.~Tsapalis, M.~Vanderhaeghen,
  Nucl.\ Phys.\  {\bf A825 } (2009)  115-144.

\bibitem{Lin:2008qv}
  H.-W.~Lin, S.~D.~Cohen, R.~G.~Edwards, D.~G.~Richards,
  Phys.\ Rev.\  {\bf D78 } (2008)  114508.

\bibitem{Braun:2009jy}
  V.~M.~Braun, M.~G\"ockeler, R.~Horsley, T.~Kaltenbrunner, A.~Lenz, Y.~Nakamura, D.~Pleiter, P.~E.~L.~Rakow et al.,
  Phys.\ Rev.\ Lett.\  {\bf 103 } (2009)  072001.

\bibitem{Drechsel:2004ki}
  D.~Drechsel, L.~Tiator,
  Ann.\ Rev.\ Nucl.\ Part.\ Sci.\  {\bf 54 } (2004)  69-114.

\bibitem{Pasquini:2004nq}
  B.~Pasquini, D.~Drechsel, L.~Tiator,
  Eur.\ Phys.\ J.\  {\bf A23 } (2005)  279-289.

\bibitem{Pasquini:2006yi}
  B.~Pasquini, D.~Drechsel, L.~Tiator,
  Eur.\ Phys.\ J.\  {\bf A27 } (2006)  231-242.

\bibitem{Pasquini:2007fw}
  B.~Pasquini, D.~Drechsel, L.~Tiator,
  Eur.\ Phys.\ J.\  {\bf A34 } (2007)  387-403.

\bibitem{Pasquini:2011ek}
  B.~Pasquini, D.~Drechsel, M.~Vanderhaeghen,
  [arXiv:1105.4454 [hep-ph]].

\bibitem{Miller:2007uy}
  G.~A.~Miller,
  {Phys. Rev. Lett.} \textbf{99} (2007) 112001.

\bibitem{Carlson:2007xd}
  C.~E.~Carlson and M.~Vanderhaeghen,
  {Phys. Rev. Lett.}  \textbf{100} (2008) 032004.

\bibitem{Carlson:2008zc}
  C.~E.~Carlson and M.~Vanderhaeghen,
  {Eur.\ Phys.\ J.\ A} \textbf{41} (2009) 1.

\bibitem{PDG92} K.~I. Hikasa et al. (Particle Data Group),
   Phys. Rev. D \textbf{45}, (1992) S1.

\bibitem{Workman:1998tv}
  R.~Workman,
  Phys.\ Rev.\  {\bf C59 } (1999)  3441-3443.

\bibitem{Bernard:1991rt}
  V.~Bernard, N.~Kaiser, J.~Gasser, U.~G.~Meissner,
  Phys.\ Lett.\  {\bf B268 } (1991)  291-295.

\bibitem{Kamalov:2001qg}
  S.~S.~Kamalov, G.-Y.~Chen, S.-N.~Yang, D.~Drechsel, L.~Tiator,
  Phys.\ Lett.\  {\bf B522 } (2001)  27-36.

\bibitem{Dugger:2009pn}
  M.~Dugger et al. [ CLAS Collaboration ],
  Phys.\ Rev.\  {\bf C79 } (2009)  065206.

\bibitem{Arndt:2002xv}
  R.~A.~Arndt, W.~J.~Briscoe, I.~I.~Strakovsky, R.~L.~Workman,
  Phys.\ Rev.\  {\bf C66 } (2002)  055213.

\bibitem{Arndt:2006bf}
  R.~A.~Arndt, W.~J.~Briscoe, I.~I.~Strakovsky, R.~L.~Workman,
  Phys.\ Rev.\  {\bf C74 } (2006)  045205.

\bibitem{Ji:2003fw}
  X.-d.~Ji, J.-P.~Ma, F.~Yuan,
  Phys.\ Rev.\ Lett.\  {\bf 90 } (2003)  241601.

\bibitem{Buchmann:2004ia}
  A.~J.~Buchmann,
  Phys.\ Rev.\ Lett.\  {\bf 93 } (2004)  212301.

\bibitem{Stave:2008tv}
  S.~Stave et al. [ A1 Collaboration ],
  Phys.\ Rev.\  {\bf C78 } (2008)  025209.

\bibitem{Aznauryan:2005tp}
  I.~G.~Aznauryan, V.~D.~Burkert, G.~V.~Fedotov, B.~S.~Ishkhanov, V.~I.~Mokeev,
  Phys.\ Rev.\  {\bf C72 } (2005)  045201.

\bibitem{Burk92} Z.~P. Li, V. Burkert, and Z. Li,  {J Phys. Rev. D} \textbf{46} (1992) 70.

\bibitem{Tiator2009} L. Tiator, M. Vanderhaeghen,
   {Phys. Lett. B} \textbf{642} (2009) 344.

\end{thebibliography}
\end{document}